\documentclass[iop,twocolappendix]{emulateapj}
\usepackage{graphicx}
\usepackage{subfig}
\usepackage{float}
\usepackage{caption}
\makeatletter
\newcommand{\reffnmark}[1]{%
    \begingroup
        \unrestored@protected@xdef\@thefnmark{\ref{#1}}%
    \endgroup
    \@footnotemark}
\makeatother
\usepackage{natbib}
\usepackage{booktabs}
\newcommand{\Hersc}{{\it Herschel}}
\newcommand{\hevics}{HeViCS}
\newcommand{\SO}{S0+S0a~}
\newcommand{\plus}{$\pm$}

\shorttitle{Dusty Early-types and the Hubble Sequence}
\shortauthors{Smith, Gomez, Eales et al.}

\begin{document}

\title{The \textit{Herschel} Reference Survey: Dust in Early-Type Galaxies and Across the Hubble Sequence\footnotemark[*]}
\submitted{Submitted to ApJ December 2011; accepted January 2012}
\author{M. W. L. Smith\altaffilmark{1},  H. L. Gomez\altaffilmark{1}, S. A. Eales\altaffilmark{1},  L. Ciesla\altaffilmark{2}, A. Boselli\altaffilmark{2},  L. Cortese\altaffilmark{3}, G. J. Bendo\altaffilmark{4}, M. Baes\altaffilmark{5}, S. Bianchi\altaffilmark{6}, M. Clemens\altaffilmark{7}, D. L. Clements\altaffilmark{8}, A. R. Cooray\altaffilmark{9}, J. I. Davies\altaffilmark{1}, I. de Looze\altaffilmark{5}, S. di Serego Alighieri\altaffilmark{6}, J. Fritz\altaffilmark{5}, G. Gavazzi\altaffilmark{10}, W. K. Gear\altaffilmark{1}, S. Madden\altaffilmark{11}, E. Mentuch\altaffilmark{12}, P. Panuzzo\altaffilmark{11}, M. Pohlen\altaffilmark{1}, L. Spinoglio\altaffilmark{13}, J. Verstappen\altaffilmark{5}, C. Vlahakis\altaffilmark{14,15}, C. D. Wilson\altaffilmark{12}, E. M. Xilouris\altaffilmark{16}}

\altaffiltext{1}{School of Physics \& Astronomy, Cardiff University,
  The Parade, Cardiff CF24 3AA, UK \email{matthew.smith@astro.cf.ac.uk}}
  \altaffiltext{2}{Laboratoire d'Astrophysique de Marseille, UMR6110 CNRS, 38 rue F. Joliot-Curie,
  F-13388 Marseille, France}
\altaffiltext{3}{European Southern Observatory, Karl Schwarschild Str. 2, 85748,
  Garching bei Muenchen, Germany}
  \altaffiltext{4}{UK ALMA Regional Centre Node, Jodrell Bank Centre for Astrophysics, School of Physics and Astronomy, University of Manchester, Oxford Road, Manchester M13 9PL, UK}
\altaffiltext{5}{Sterrenkundig Observatorium, Universiteit Gent, Krijgslaan 281 S9, B-9000 Gent, Belgium}
\altaffiltext{6}{INAF-Osservatorio Astrofisico di Arcetri, Largo E. Fermi 5, 50125 Firenze, Italy}
\altaffiltext{7}{INAF-Osservatorio Astronomico di Padova, Vicolo dell'Osservatorio 5, 35122 Padova, Italy} 
\altaffiltext{8}{Astrophysics Group, Imperial College, Blackett Laboratory, Prince Consort Road, London SW7 2AZ, UK}
\altaffiltext{9}{Center for Cosmology and the Department of Physics \& Astronomy, University of California, Irvine, CA 92697, USA}
\altaffiltext{10}{Universita' di Milano-Bicocca, Piazza della Scienza 3,  20126 Milano, Italy}
\altaffiltext{11}{CEA, Laboratoire AIM, Irfu/SAp, Orme des Merisiers, F-91191 Gif-sur-Yvette, France}
\altaffiltext{12}{Department of Physics \& Astronomy, McMaster University, Hamilton, Ontario L8S 4M1, Canada}
\altaffiltext{13}{Istituto di Fisica dello Spazio Interplanetario, INAF, via del Fosso del Cavaliere 100, 00133 Roma, Italy }
\altaffiltext{14}{Joint ALMA Office, Alonso de Cordova 3107, Vitacura, Santiago, Chile}
\altaffiltext{15}{Departamento de Astronomia, Universidad de Chile, Casilla 36-D, Santiago, Chile}
\altaffiltext{16}{Institute of Astronomy and Astrophysics, National
  Observatory of Athens, I. Metaxa and Vas. Pavlou, P. Penteli, 15236
  Athens, Greece}

\begin{abstract}
  We present \Hersc~observations of 62 early-type galaxies (ETGs),
  including 39 galaxies morphologically classified as \SO and 23
  galaxies classified as ellipticals using SPIRE at 250, 350 and 500\,$\mu$m
  as part of the volume-limited \Hersc~Reference Survey (HRS).  We detect dust emission in 24\% of
  the ellipticals and 62\% of the S0s. The mean
  temperature of the dust is $\langle T_d \rangle =23.9 \pm 0.8\rm \, K$, warmer than
  that found for late-type galaxies in the Virgo Cluster. The mean
  dust mass for the entire detected early-type sample is ${\rm log}M_d = 6.1 \pm 0.1\,
  M_{\odot}$ with mean dust-to-stellar mass
  ratio of ${\rm log}(M_d/M_*)= -4.3 \pm 0.1$.  Including the non-detections, 
  these parameters are ${\rm log}M_d = 5.6 \pm 0.1$ and ${\rm log}(M_d/M_*)= -5.1 \pm 0.1$ respectively.   
  The average dust-to-stellar mass ratio for the early-type sample is fifty times 
  lower, with larger dispersion, than the spiral galaxies observed as part of the HRS, and there is an
  order of magnitude decline in $M_d/M_*$ between the S0s and ellipticals.   
  We use UV and optical photometry to show that virtually all the
  galaxies lie close to the red sequence yet the large number of
  detections of cool dust, the gas-to-dust ratios and the ratios of
  far-infrared to radio emission all suggest that many ETGs contain a
  cool interstellar medium similar to that in late-type galaxies.  We
  show that the sizes of the dust
  sources in S0s are much smaller than those in early-type spirals and the 
  decrease in the dust-to-stellar
  mass ratio from early-type spirals to S0s cannot simply be explained
  by an increase in the bulge-to-disk ratio.
  These results suggest that the disks in S0s contain much less dust
  (and presumably gas) than the disks of early-type spirals and this
  cannot be explained simply by current
  environmental effects, such as ram-pressure stripping.  The wide
  range in the dust-to-stellar mass ratio for ETGs and the lack of a
  correlation between dust mass and optical luminosity suggest that
  much of the dust in the ETGs detected by \Hersc~has been acquired as
  the result of interactions, although we show these are
  unlikely to have had a major effect on the stellar masses of the
  ETGs.  The \Hersc~observations tentatively suggest that in the most
  massive systems, the mass of interstellar medium is unconnected to
  the evolution of the stellar populations in these galaxies.
\end{abstract}


\footnotetext[*]{\textit{H\lowercase{erschel}} \lowercase{is an} ESA \lowercase{space observatory with science instruments provided by} E\lowercase{uropean-led} P\lowercase{rincipal} I\lowercase{nvestigator consortia and with important participation from} NASA.}

\section{Introduction}
\label{sec:intro}

In the standard theoretical view of galactic evolution, based on the
hierarchical paradigm, galaxies may move both ways along the Hubble
sequence during their evolution: ellipticals may form as the result of the merging of
late-type galaxies (e.g., Cole et al. 2000; De Lucia et al. 2006);
disks form as the result of the accretion of gas on to bulges (Cole et
al. 2000). There are also plenty of environmental processes that might
transform the morphology of a galaxy, just one example among many
being ram-pressure stripping in clusters, which would move a galaxy
toward the early-type sequence by removing the interstellar medium
(ISM; e.g., Corbelli et al. 2011) and thus quenching star formation (Boselli \& Gavazzi 2006).

The observational evidence that galaxy transformation is occurring is
mixed, especially for galaxies at the current epoch.  The fact that
the morphology--density relation is not just the result of early-type
galaxies (ETGs)---ellipticals and S0s---being more common in clusters, but
extends over a wide range of environmental density, is evidence that galaxy
transformation is not a significant process at the current epoch
(Dressler 1980).  However, there is also plenty of observational
evidence that merging is still important today.  In particular, there
is evidence of recent merging or accretion in nearby ellipticals,
primarily from obscuration in the optical (Goudfrooij \& de Jong 1995),
where fossils of mergers in the form of dust ripples and dust lanes
have been detected (Schweizer \& Seitzer 1992). Furthermore, the
discovery with the {\it Infrared Astronomical Satellite} ({\it IRAS}) of the
ultraluminous infrared galaxies (ULIRGs), which are often late-stage
mergers with optical profiles similar to the profiles of ellipticals
(Wright et al. 1990), was persuasive evidence that ellipticals are
still being formed by mergers at the current epoch.  Nevertheless, it
is not clear whether all the properties of the eventual merged systems
would be the same as present-day ellipticals (Naab \& Ostriker 2009),
and many of the distinctive properties of ellipticals seem more
naturally explained if ellipticals form in a relatively short early-period in cosmic history (Peebles \& Nusser 2010).

The properties of the ISM in ETGs potentially
have something to tell us about the evolution of these systems. Since
late-type galaxies are known to contain much more gas and dust than
ETGs, any process that moves a galaxy along the Hubble sequence has to
change simultaneously the morphology and the ISM
of the galaxy.  In this paper, we present \Hersc~observations of
continuum emission from the dust in a sample of the most massive ETGs
in the nearby universe, which we show is the most sensitive
method currently available for estimating the mass of the cool ISM in
these galaxies.

Although once thought to be devoid of a cool ISM, optical absorption
studies suggested that more than 50\% of ellipticals contain some dust
(Goudfrooij et al.\ 1994; van Dokkum \& Franx 1995; Tran et al.\ 2001;
Ferrarese et al. 2006) and therefore molecular gas, though the
mass of dust is uncertain from these works. In the last decade,
previous far-infrared observations with {\it IRAS}, the {\it Infrared Space
  Observatory} ({\it  ISO}), {\it Spitzer}, and {\it AKARI}\ have
shown that dust in ETGs is rather common, with dust masses estimated at
some 10--100 times larger than those derived from the optical
extinction measurements (e.g., Bregman et al.\ 1998; Ferrari et al.\
2002; Kennicutt et al.\ 2003; Pahre et al.\ 2004; Xilouris et al.\ 2004; Temi et al.\
2004, 2007a; Panuzzo et al.\ 2007; Kaneda et al.\ 2008; Young et al. 2009). Although these observations have probed the peak of
the spectral energy distributions (SEDs) of the brightest giant ellipticals
and S0s, they are hampered by poor
resolution and/or lack of long wavelength coverage.

Our current ideas about the life cycle of dust in ellipticals suggest that
it may be possible for us to test the origin of the dust, particularly
whether it is provided by internal (via stellar mass loss) or external
processes (e.g., fueled by mergers). As long as dust is not introduced
from elsewhere, we essentially have a stellar system in equilibrium:
dust is produced in the atmospheres of evolved stars (Athey et al.\ 2002) and destroyed via sputtering in the hot gas (indeed Bressan et
al. 2006 detect silicate features in elliptical galaxies known to be
produced in circumstellar envelopes). The dust life cycle in
ellipticals may therefore be simpler than that in late-types for the
internal origin scenario (Tsai \& Mathews 1996), and the mass of dust predicted to exist in the
steady state through production in evolved stars and destruction
through sputtering is approximately $10^5\,M_\odot$ (Forbes 1991;
Goudfrooij \& de Jong 1995).

An alternative origin for dust in ETGs is accreted material either due
to a merger or from a tidally-interacting companion: Centaurus A has
long been known to harbor a dusty disk thought to be formed by a
merger (recently revealed in the submillimeter (submm)---Leeuw et al.
2002; Auld et al.\ 2011; T. Parkin et al.\ 2011, submitted). Forbes (1991)
and Temi et al. (2004, 2007a) found no correlation between the dust
emission and optical starlight using {\it ISO} observations of massive ETGs,
suggesting that the dust has an external rather than an internal
origin. In 2010, Gomez et al.\ used \Hersc~observations of the
elliptical galaxy M86 to reveal $10^6\,M_{\sun}$ of cold dust 
coincident with material stripped from the nearby spiral NGC4438
(Cortese et al.\ 2010a). A similar process has been seen with the
``displaced ISM'' of the elliptical NGC3077 (Walter et al.\ 2011).

The unprecedented resolution and sensitivity of the recently launched
\Hersc~{\it Space Observatory} (Pilbratt et al.\ 2010), combined with the
wavelength coverage of the instruments PACS (Poglitsch et al.\ 2010)
and SPIRE (Griffin et al.\ 2010) from 70 to 500\,$\mu$m, allow us to
address long-standing issues such as the origin and quantity of dust
in ETGs. \Hersc~provides us with an unbiased view of the interstellar
dust (and therefore the interstellar medium); we are now sensitive to
the {\it total} dust mass in galaxies rather than only to those
galaxies which had enough warm dust to be detected by {\it IRAS} and {\it
  Spitzer} (e.g., Dunne et al.\ 2011).  Here, we use observations from
the guaranteed time project, the \Hersc~Reference Survey (HRS;
Boselli et al.\ 2010b) of 62 ETGs including 23 ellipticals. The HRS
is a study of dust in nearby galaxies, crucial
for calibrating blind surveys of galaxies with \Hersc~at high
redshifts, e.g., HerMES (Oliver et al.\ 2010) and the
\Hersc-ATLAS (H-ATLAS; Eales et al.\ 2010).

Cortese et al.\ (2011b, hereafter C12) recently used the HRS sample
of $\sim$300 nearby galaxies to obtain dust scaling relations, finding
that the dust is tightly coupled to the atomic mass and therefore to the cold ISM.
C12 measured the dust-to-stellar-mass ratio of galaxies with
different morphological types and environment, finding that the ETGs
contain less dust mass per unit of stellar mass than the late-type
galaxies.  In this work, we further explore the dust properties of the
ETGs in the HRS and measure more precisely the variation of dust
content with Hubble type and the origin of this variation.  In
Section \ref{sec:red} we introduce the sample and describe the data
reduction techniques.  The results are given in Section \ref{sec:res} and
are used to discuss our understanding of the dust content of early-
and late-types in Section\,\ref{sec:morp} and the link between dust and the
hot interstellar medium in Section \ref{sec:hot}. We discuss the
implications for the evolutionary history of ETGs in
Section \ref{sec:disc}.  The conclusions are presented in Section \ref{sec:conc}.

\begin{deluxetable*}{lcllllrrrc}
\tabletypesize{\scriptsize}
\tablecaption{The Sample \label{tab:sample}}
\tablewidth{0pt}
\tablehead{ \colhead{HRS} & \colhead{Other Name} & \colhead{R.A.} & \colhead{Decl.} & \colhead{Type} & \colhead{$D$} 
& \colhead{$D(25)$} & \colhead{$L_B$} & \colhead{$L_K$} & \colhead{Membership} \\
\colhead{} & \colhead{}  & \colhead{($\rm h~~m~~s$)} & \colhead{($^{\circ} ~~\arcmin~~\arcsec$)} & \colhead{} &\colhead{(Mpc)} 
& \colhead{($\arcmin$)} & \colhead{($L_{\odot}$)}& \colhead{($L_{\odot}$)}&\colhead{}
}
\startdata
3 & NGC 3226 & 10 23 27.01$^r$ & +19 53 54.7$^r$ & E2:pec;LINER;Sy3 & 16.7 &  3.16& 10.12 & 10.59 & Leo Cl. \\         
7 & NGC 3245 & 10 27 18.40$^r$ & +28 30 26.3$^r$ & SA(r)0:?;H\sc{ii};LINER & 18.8 &  3.24 &10.08 & 10.47&  Leo Cl. \\
14 & NGC 3301 & 10 36 56.04& +21 52 55.7 & (R')SB(rs)0/a & 19.2 & 3.55& 9.84 & 10.47  &Leo Cl. \\ 
22 & NGC 3414 &10 51 16.19$^r$& +27 58 30.2$^r$  & S0 pec;LINER & 20.2 & 3.55 & 9.98 & 10.73&Leo Cl. \\ 
43 & NGC 3608 &11 16 58.96& +18 08 54.9  & E2;LINER: & 15.8 & 3.16 &10.11 &10.70 & Leo Cl. \\  
45 & NGC 3619 & 11 19 21.51$^r$ & +57 45 28.3$^r$ &  (R)SA(s)0+: & 22.1 &  2.69 & 9.87 & 10.57 &Ursa Major Cl.\\
46 & NGC 3626 &11 20 03.80$^r$ & +18 21 24.3$^r$  & (R)SA(rs)0+ & 21.3 &  2.69 &10.12 & 10.70& Leo Cl. \\ 
49 & NGC 3640 & 11 21 06.85 & +03 14 05.4  & E3  & 17.9 & 3.98 &10.43 &11.06 & Leo Cl. \\  
71 & NGC 3945 & 11 53 13.61$^r$ & +60 40 32.3$^r$ & SB(rs)0+;LINER & 18.0 &  5.25 & 10.01 & 11.06 &Ursa Major Cl. \\ 
87 & NGC 4124 & 12 08 09.64 & +10 22 43.4 & SA(r)0+ & 17.0 & 4.10 &9.70 & 10.38 &Virgo Out. \\  
90 & NGC 4179 & 12 12 52.11 &	+01 17 58.9 & S0\tablenotemark{a} & 17.0 & 3.80 & 9.89 & 10.60 & Virgo Out.\\ 
93 & NGC 4203 & 12 15 05.06$^r$ & +33 11 50.2$^r$ & SAB0--:;LINER;Sy3 & 15.6 & 3.39 & 9.89 & 10.73& Coma I Cl. \\ 
101 & NGC 4251 & 12 18 08.31 & +28 10 31.1 &  SB0? sp & 15.3 & 3.63 & 13.86 & 10.54&Coma I Cl. \\
105 & NGC 4262 & 12 19 30.58 & +14 52 39.8 & SB(s)0--? & 17.0 & 1.87 & 9.69 & 10.43& Virgo A \\ 
123 & NGC 4324 & 12 23 06.18 & +05 15 01.5 &  SA(r)0+ & 17.0 &  3.52 &9.65 & 10.38& Virgo S Cl.\\ 
125 & NGC 4339 & 12 23 34.94 & +06 04 54.2 & E0;Sy2 & 23.0 &  2.31 & 9.71 & 10.34&Virgo B \\ 
126 & NGC 4340 & 12 23 35.31 & +16 43 19.9 &SB(r)0+ & 17.0 &  3.60 & 9.81 &10.44 &Virgo A \\ 
129 & NGC 4350 & 12 23 57.81 & +16 41 36.1 & SA0;Abs. line & 17.0 &  3.20 &  9.91 & 10.64&Virgo A \\
135 & NGC 4365 & 12 24 28.23 & +07 19 03.1 & E3 & 23.0 & 8.73 & 10.34 & 11.26 &Virgo B \\  
137 & NGC 4371 & 12 24 55.43 & +11 42 15.4 & SB(r)0+ & 17.0 & 5.10 &9.92 & 10.68& Virgo A \\
138 & NGC 4374, M84 & 12 25 03.78 & +12 53 13.1 & E1;LERG;Sy2 & 17.0 & 10.07 &10.57 & 11.26 & Virgo A \\
150 & NGC 4406, M86 & 12 26 11.74 & +12 56 46.4 & S0(3)/E3 & 17.0 &  11.37 &10.66 & 11.31& Virgo A \\
155 & NGC 4417 & 12 26 50.62 & +09 35 03.0 & SB0: s & 23.0 & 3.60 & 10.08 & 10.77 & Virgo B \\ 
161 & NGC 4429 & 12 27 26.56 & +11 06 27.1 & SA(r)0+;LINER;H\sc{ii} & 17.0 &  8.12 & 10.23 & 11.06& Virgo A \\
162 & NGC 4435 & 12 27 40.50$^r$ & +13 04 44.5$^r$ & SB(s)0;LINER;H\sc{ii} & 17.0  & 2.92 & 10.05 &10.83 &  Virgo A \\
166 & NGC 4442 & 12 28 03.89 & +09 48 13.0 & SB(s)0 & 23.0 & 5.05 &10.36 & 11.12&  Virgo B \\
174 & NGC 4459 & 12 29 00.04$^r$ & +13 58 42.2$^r$ & SA(r)0+;H\sc{ii};LINER & 17.0 & 3.36   &10.07 & 10.91 &Virgo A \\
175 & NGC 4461 & 12 29 03.01 & +13 11 01.5 & SB(s)0+: & 17.0 &  3.52 & 9.90 & 10.57&Virgo A \\
176 & NGC 4469 & 12 29 28.03 & +08 44 59.7 & SB(s)0/a? sp & 23.0 & 4.33  &9.97 & 10.82 &Virgo B \\
178 & NGC 4472, M49 & 12 29 46.76 & +08 00 01.7 & E2/S0;Sy2 & 17.0 &  10.25  & 10.90 & 11.59 &Virgo S Cl. \\ 
179 & NGC 4473 & 12 29 48.87 & +13 25 45.7 & E5 & 17.0 &  4.04 & 10.15 & 10.90& Virgo A \\
180 & NGC 4477 & 12 30 02.17 & +13 38 11.2 & SB(s)0:?;Sy2 & 17.0 & 3.60 &10.12 & 13.77& Virgo A \\
181 & NGC 4478 & 12 30 17.42 & +12 19 42.8 & E2 & 17.0 &  1.89 &9.89 & 10.41 & Virgo A \\
183 & NGC 4486, M87 & 12 30 49.42 & +12 23 28.0 & E+0-1 pec;NLRG;Sy & 17.0 & 11.00 & 10.85 & 11.43 &Virgo A  \\
186 & NGC 4494 & 12 31 24.03 & +25 46 29.9 & E1-2;Sy & 18.7 & 4.79 &10.62 & 11.20 & Coma I Cl. \\ 
200 & NGC 4526 & 12 34 03.03$^r$ & +07 41 57.3$^r$ & SAB(s)0: & 17.0 &  7.00 & 10.41 & 11.18&Virgo S Cl. \\
202 & IC 3510 & 12 34 19.33& +11 04 17.7 & dE\tablenotemark{a} & 17.0 &1.10
&8.70 & 9.20 & Virgo A  \\
209 & NGC 4546 & 12 35 29.51 & --03 47 35.5 & SB(s)0--: & 15.0 & 3.31 &10.05 & 10.82& Virgo Out. \\     
210 & NGC 4550 & 12 35 30.61 & +12 13 15.4 & SB0: Sy,LINER & 17.0 & 3.95 &9.66 & 10.30& Virgo A \\
211 & NGC 4552, M89 & 12 35 39.88 & +12 33 21.7 & E;LINER;HII;Sy2 & 17.0 &  7.23  &10.29 & 11.06 &Virgo A \\
214 & NGC 4564 & 12 36 26.99 & +11 26 21.5 & E6 & 17.0 &  4.33 & 9.86 & 10.58& Virgo A \\
218 & NGC 4570 & 12 36 53.40 & +07 14 48.0 & S0(7)/E7 & 17.0 & 3.52 & 9.96	&10.70 &Virgo S Cl.\\
219 & NGC 4578 & 12 37 30.55 & +09 33 18.4 & SA(r)0: & 17.0 & 3.77 &9.71 & 10.41& Virgo E Cl.\\
231 & NGC 4596 & 12 39 55.94 & +10 10 33.9 & SB(r)0+;LINER: & 17.0 & 4.76 &10.08 & 10.79& Virgo E Cl.\\
234 & NGC 4608 & 12 41 13.29 & +10 09 20.9 & SB(r)0 & 17.0 & 4.30 & 9.83 & 10.51& Virgo E Cl.\\
235 & NGC 4612 & 12 41 32.76 & +07 18 53.2 & (R)SAB0 & 17.0 & 2.16 & 9.82 & 10.35& Virgo S Cl.\\
236 & NGC 4621, M59 & 12 42 02.32 & +11 38 48.9 & E5 & 17.0 & 7.67  &10.32& 11.05& Virgo E Cl.\\
240 & NGC 4638 & 12 42 47.43 & +11 26 32.9 & S0- & 17.0 & 2.01 &9.82 & 10.49 &Virgo E Cl. \\
241 & NGC 4636 & 12 42 49.87 & +02 41 16.0 & E/S0/1;LINER;Sy3 & 17.0 & 9.63 &10.51 & 11.18& Virgo S Cl.\\ 
243 & NGC 4643 & 12 43 20.14 & +01 58 42.1 & SB(rs)0/a;LINER/H\sc{ii} & 17.0 & 3.00 & 9.98 &10.81& Virgo Out.\\
245 & NGC 4649, M60 & 12 43 40.01 & +11 33 09.4 &  E2 & 17.0 & 5.10 & 10.73 & 11.46 &Virgo E Cl.\\
248 & NGC 4660 & 12 44 31.97 & +11 11 25.9 & E5  & 17.0 & 1.89 & 9.70 & 10.47& Virgo E Cl.\\
250 & NGC 4665 & 12 45 05.96 & +03 03 20.5 & SB(s)0/a & 17.0 & 4.50 & 10.04 & 10.80 &Virgo Out.\\
253 & NGC 4684 & 12 47 17.52 & --02 43 38.6 & SB(r)0+;H\sc{ii}  & 21.3 & 2.88 &9.82 & 10.42 &Virgo Out.\\
258 & NGC 4697 & 12 48 35.91 & --05 48 03.1 & E6;AGN  & 17.7 & 7.24 & 10.55 & 11.16 &Virgo Out.\\
260 & NGC 4710 & 12 49 38.93$^r$ & +15 09 59.1$^r$ & SA(r)0+? sp;H\sc{ii} & 17.0 & 4.30 & 9.91 & 10.74 &Virgo Out.\\
269 & NGC 4754 & 12 52 17.56 & +11 18 49.2 & SB(r)0--: & 17.0 & 5.03 &10.05 & 10.81& Virgo E Cl.\\
272 & NGC 4762 & 12 52 56.05 & +11 13 50.9 & SB(r)0 sp;LINER & 17.0 & 8.70  &10.21 & 10.85 &Virgo E Cl.\\
286 & NGC 4866 & 12 59 27.14 & +14 10 15.8 & SA(r)0+;LINER & 17.0 & 6.00 &9.84 & 10.60& Virgo Out.\\
296 & NGC 5273 & 13 42 08.36$^r$ & +35 39 15.1$^r$ & SA(s)0;Sy1.5 & 15.2 &  2.75 &9.54 & 10.21 &Canes Ven. Spur \\
312 & NGC 5576 & 14 21 03.68 & +03 16 15.6 & E3  & 21.2 & 3.55 &10.16 & 10.89&Virgo-Libra Cl.\\
316 & NGC 5638 & 14 29 40.39 & +03 14 00.2 & E1 & 23.9 & 2.69 &10.09 & 10.72& Virgo-Libra Cl.
\enddata
\tablecomments{The columns are as follows. Column 1: 
the number of the source in the \Hersc~Reference Survey (Boselli et al.\
2010b). Column 2: other common names for the galaxy. Columns 3 and 4: the right ascension and declination of the
galaxy (J2000). This is taken from the NASA Extragalactic Database (NED), except positions with 
a superscript $r$ are radio positions taken from Wrobel (1991) and Filho et al.
(2006). Column 5: morphological type
taken from Virgo Cluster Catalogue (Binggeli et al. 1985) or from our own classification. The AGN/Seyfert/LINER classifications were taken from Ho et al. (1997), Schmitt (2001), and  Veron-Cetty \& Veron (2006).  Column 6:
distance in $\rm Mpc$. We assume all objects in Virgo, including those in the outskirts of Virgo, are at
a distance of $17 \,\rm Mpc$, except for those in Cloud B, which we assume are at a distance
of $23 \,\rm Mpc$. For the other galaxies
we have calculated distances from the heliocentric
velocity using a Hubble constant of $70\rm km\ s^{-1}\ Mpc^{-1}$. Column 7: optical isophotal
diameter (25\,mag\,arcsec$^{-2}$). Column 8: total $K$-band luminosity measured using $K_{S_{\rm tot}}$ from 2MASS (Skrutskie et al. 2006).  Column 9: cluster or cloud membership from
Gavazzi \& Boselli (1999) for Virgo, otherwise from Tully (1988) or Nolthenius (1993) wherever available, 
or failing that our own estimate. }  
\tablenotetext{a}{These sources were morphologically classified by hand using optical images.}
\end{deluxetable*}

\section{The Sample}
\label{sec:red}

The HRS is a volume-limited sample ($15\,\rm Mpc$ $<d<25\,\rm Mpc$) with
\Hersc~SPIRE of 322 galaxies selected by the $K$-band magnitude (a proxy
for stellar mass). The ETGs in the HRS have $K\le 8.7$, and so our
sample consists of the ETGs in this volume of space with the highest
stellar masses; contamination from Galactic cirrus is minimized due
to the high galactic latitudes of the sample (see Boselli et al.\
2010b for full details). The sample includes different morphological types with 260
late-type galaxies and, in the original catalog listed in Boselli et al.\ (2010b), 64 early types. However, after
closer inspection of the optical images, we reclassified the galaxies
NGC4438 (HRS163), NGC4457 (HRS173), NGC4691 (HRS256), and NGC5701
(HRS322) as late types (from S0/a to Sb). We changed two
classifications in the other direction, with the Sb galaxies NGC4179
(HRS90) and IC3510 (HRS202) reclassified to S0 and dwarf elliptical,
respectively.  The revised catalog therefore has 62 ETGs, 39 of which are designated \SO
galaxies and 23 are ellipticals (Table~\ref{tab:sample}) though five
of the latter set are morphologically classified as E/S0. The
morphological classifications are taken from the Virgo Cluster
Catalogue (Binggeli et al.\ 1985) or are our own if another was not
available (Boselli et al.\ 2010b).  The HRS contains galaxies from a range of environments,
including isolated field galaxies, pairs and galaxies in the heart of
the Virgo Cluster.

The HRS observations were carried out with
SPIRE in scan-map mode with a scan speed of 30 arcsec s$^{-1}$. We
chose the size of our maps so that they would be at least as large as
the optical disk of each galaxy, defined as the area of the galaxy
within an optical isophote with a $B$-band brightness of 25 mag
arcsec$^{-2}$ ($D({25})$; Boselli et al. 2010b). In practice, our maps were either
4$\times$4, 8$\times$8, or 12$\times$12 arcmin$^2$, and in most cases
are larger than $D({25})$. For each ETG, we made eight
pairs of orthogonal scans, resulting in a 1$\sigma$ instrumental noise in each pixel of 0.35, 0.20 and 0.11 MJy/sr
for the 250, 350, and 500\,$\mu$m, respectively (integration time of 1199s, 3102s,
and 4948s for each map size). We
deliberately observed the ETGs in the HRS for at least twice as long
as for the late-type galaxies (three pairs of orthogonal scans) because the former are known to contain
much less dust on average.

There are a number of sources in the HRS which overlap with galaxies observed as part of the open-time key project the \Hersc~Virgo Cluster Survey (\hevics; Davies et al.\
2010, 2012), thus the two surveys have a data-sharing agreement for these galaxies.   Here we also include data of the duplicate 19 ETGs formally observed as part of \hevics~with both PACS and SPIRE.  The SPIRE
observations for \hevics~were made in parallel-scan-map mode with a scan speed of
60\,arcsec s$^{-1}$. To cover the Virgo Cluster, four fields with
size $4^{\circ} \times 4^{\circ}$ were observed with eight cross-scans (see
Davies et al.\ 2010, 2012 for a complete description).   Note that the dust content of all of the ETGs in the Virgo
Cluster from \hevics~will be presented in a complementary paper (S. di Serego Alighieri
et al.,\ in preparation).  

\subsection{Data Reduction and Flux Extraction}
\label{sec:reduce}
The HRS and \hevics~SPIRE data were reduced using similar pipeline
procedures.  The SPIRE data were processed up to Level-1 with a custom
script adapted from the official pipeline ({\sl POF5\_pipeline.py},
dated 8 Jun 2010) as provided by the SPIRE Instrument Control Centre
(ICC)\footnote{\label{fn:spire}See `The SPIRE Analogue Signal Chain
  and Photometer Detector Data Processing Pipeline' (Griffin et
  al. 2008 or Dowell et al. 2010) for a more detailed description of
  the pipeline and a list of the individual modules.}.  Our custom
Jython script was run in the \Hersc~Interactive Processing Environment
(HIPE; Ott 2010) with the continuous integration build number:
4.0.1367.  For both surveys we use an optimized deglitcher setting
instead of applying the ICC default settings. For the \hevics~data the
{\sc sigmaKappaDeglitcher} module was used, while for the HRS we
applied the {\sc waveletDeglitcher}; this module was adjusted to mask
the sample following a glitch.  For the HRS, after the flux
calibration was applied, an additional pass with the {\sc
  waveletDeglitcher} was run, as this was found to significantly
improve the glitch-removal process. Furthermore, we did not run the
pipeline default temperature drift correction or the median baseline
subtraction. Instead we use a custom method (BriGAdE; M. W. L. Smith et al.,\
in preparation; see also L. Ciesla et al., in preparation) to remove the temperature drift and bring all
bolometers to the same level.

Our final SPIRE maps were created using the na\"{i}ve mapper provided in
the standard pipeline with pixel sizes of 6\arcsec, 8\arcsec, and 12\arcsec\ at 250,
350, and 500\,$\mu$m respectively. The FWHM of the SPIRE beams for
this pixel scale are 18.2\arcsec, 24.5\arcsec, and 36.0\arcsec\ at 250, 350, and
500\,$\mu$m, respectively (Swinyard et al. 2010). In addition, the 350\,$\mu$m data are multiplied by
1.0067 to update our flux densities to the latest v7 calibration
product. The calibration uncertainty is a combination of the 5\% error
due to correlated errors between bands and the 2\% random uncertainty;
these values are added linearly instead of in
quadrature\footnote{\label{fn:som}SPIRE Observer's Manual (2011)}.

The \Hersc~PACS 100 and 160\,$\mu$m data taken as part of \hevics~were
reduced using the standard pipeline (see Davies
et al. 2012). In brief, we used a two step deglitching process, first with the standard deglitcher and second using one based on sigma clipping. We  masked bright sources and a high-pass filter was used to reduce $1/f$ noise.  The orthogonal scans were combined and the na\"{i}ve mapper was used to create the final maps. Unlike Davies et al. (2012) we use images created using the full HeViCS data set with eight parallel-mode scans per
tile. The FWHM beam sizes are approximately 9\arcsec\ and
13\arcsec\ with pixel sizes of 3.2\arcsec\ and 6.4\arcsec\ for the 100
and 160\,$\mu$m bands, respectively\footnote{\label{fn:pacsPSF} PACS Photometer Point Spread Function Document (2010)}. 
Davies et al. (2012) measured differences in global flux densities of up to 20\% between sets of 
cross-scans, we therefore choose this as a conservative error for our flux estimates as it dominates over other sources
of uncertainty, e.g., the calibration\footnote{\label{fn:pacsCal} PACS Photometer Calibration Document (2011)}.


The 250\,$\rm \mu m$ \Hersc~ SPIRE maps for the detected \SO and all of the elliptical
galaxies in our sample are shown in Figures~\ref{fig:spireearly} and
\ref{fig:spireegals}. Most of the objects visible on the images are
background galaxies, creating a potential problem for determining
whether our targets are detected. Fortunately, these background
galaxies are almost always unresolved, whereas virtually all our
detections are extended. The full description of the flux extraction
process and complete photometry information
for all the HRS galaxies including upper limits on the non-detections are provided in L. Ciesla et al. (in preparation).  The upper limits for the non-detections are estimated using circular apertures with size $0.3\times D_{25}$ for ellipticals and
$0.8 \times D_{25}$ for S0s (chosen as a conservative limit
based on the extent of 250\,$\mu$m emission seen in the detected sample).

We tested the effect of using different definitions of the upper limit on the analysis in this paper.  First, we used simply the statistical noise on the map, taking into account the instrumental noise and confusion to give $\sigma_{\rm rms}$.
Second, we included the additional uncertainty in the photometry measurement due to the background
level ($\sigma_{\rm sky}$; L. Ciesla et al., in preparation), which arises mainly from cirrus contamination.   
For this work, we use the photometry upper limit as in L. Ciesla et al. (in preparation), with $\sigma =  3(\sigma_{\rm rms} + \sigma_{\rm sky})$ which is the most conservative limit on the 250\,$\mu$m flux for our sources. Note that all of the results in this paper are also valid if we use 3$\sigma_{\rm rms}$ or 5$\sigma_{\rm rms}$ (as a typical 3$\sigma$ or 5$\sigma$ detection limit).   

For the HeViCS galaxies in our sample the PACS flux densities we used aperture photometry with suitable background 
regions selected around the source (the assumed uncertainty 
is outlined in Section \ref{sec:reduce}). We present the PACS fluxes in Table~\ref{tab:fluxes}.

\subsection{Data at other wavelengths}
\label{sec:other}
In addition to the \Hersc~data, we used 70--160\,$\mu$m data from {\it
  Spitzer} (Kennicutt et al.\ 2003), reprocessed using the techniques
described in Bendo et al.\ (2010b) and presented in Bendo et al.\ (2011, submitted). The calibration uncertainties were assumed to be 5\% at
70\,$\mu$m and 12\% at 160\,$\mu$m (Gordon et al. 2007; Stansberry et al.\ 2007).  We use IRAS
60 and 100\,$\mu$m measurements originally presented in Knapp et al.\
(1989) but modified in a private communication to NASA Extragalactic Database (NED, Table~\ref{tab:fluxes}). The calibration
uncertainties for \textit{IRAS} were assumed to be 13\% at 60\,$\mu$m and 16\%
at 100\,$\mu$m (e.g., Verter \& Rickard 1998). Four of our elliptical galaxies were also detected by 
{\it ISO} and we use the fluxes given in Temi et al. (2004).


Optical photometry for the sample was obtained from the Sloan
  Digital Sky Survey (SDSS) DR7 (Abazajian et al. 2009) database. We
estimate stellar masses for the sample from the \textit{i}-band luminosities
using the relationship between stellar mass and galaxy color for a
Chabrier initial mass function (Zibetti et al.\ 2009). The full
description of this method applied to the entire HRS sample can be
found in Cortese et al.\ (2011) and C12. NUV photometry is available
from the {\it Galaxy Evolution
  Explorer} (\textit{GALEX}; Martin et al.\ 2005) from the GR6 data release
(see Boselli et al.\ 2011; Cortese et al.\ 2011).

X-ray luminosities for 38 of our sources were obtained from the
catalogs of O'Sullivan et al. (2001) and Pellegrini
(2010). A further 57 of our sources are included in the ATLAS$^{\rm
  3D}$ sample of ETGs (Cappellari et al. 2011a). ATLAS$^{\rm 3D}$
includes molecular hydrogen masses estimated from the CO(1--0) intensity ($\rm
I(CO)$; Young et al.\ 2011) with a conversion factor from $\rm H_2$
to CO of $N({\rm H_2})/{I\rm (CO)} = 3 \times 10^{20} \,\rm
cm^{−2}\,(K\,km\,s^{-1})^{-1}$ (e.g., Young \& Scoville 1982). The CO
data only exist for the central region of the galaxies (i.e., within
$30{\arcsec}$) and would be an underestimate of the total molecular
gas if the real gas distribution is extended, though Young et al.\
find that there is no strong evidence for extended emission in any of
the galaxies except for NGC4649. Atomic H{\sc i} masses and upper
limits were obtained for 48 sources from the GOLDMINE database
(Gavazzi et al.\ 2003), the ALFALFA survey (Haynes et al. 2011) or
Noordermeer et al.\ (2005) and Springob et al.\ (2005). The CO and H{\sc i} masses are presented
in Table~\ref{tab:masses}. In all cases, these have been corrected to
the distances assumed for the ETGs in this paper.

\begin{figure*}
\begin{center}
\subfloat[]{\includegraphics[trim=6mm 10mm -3mm
  -4mm,clip=true,width=18cm]{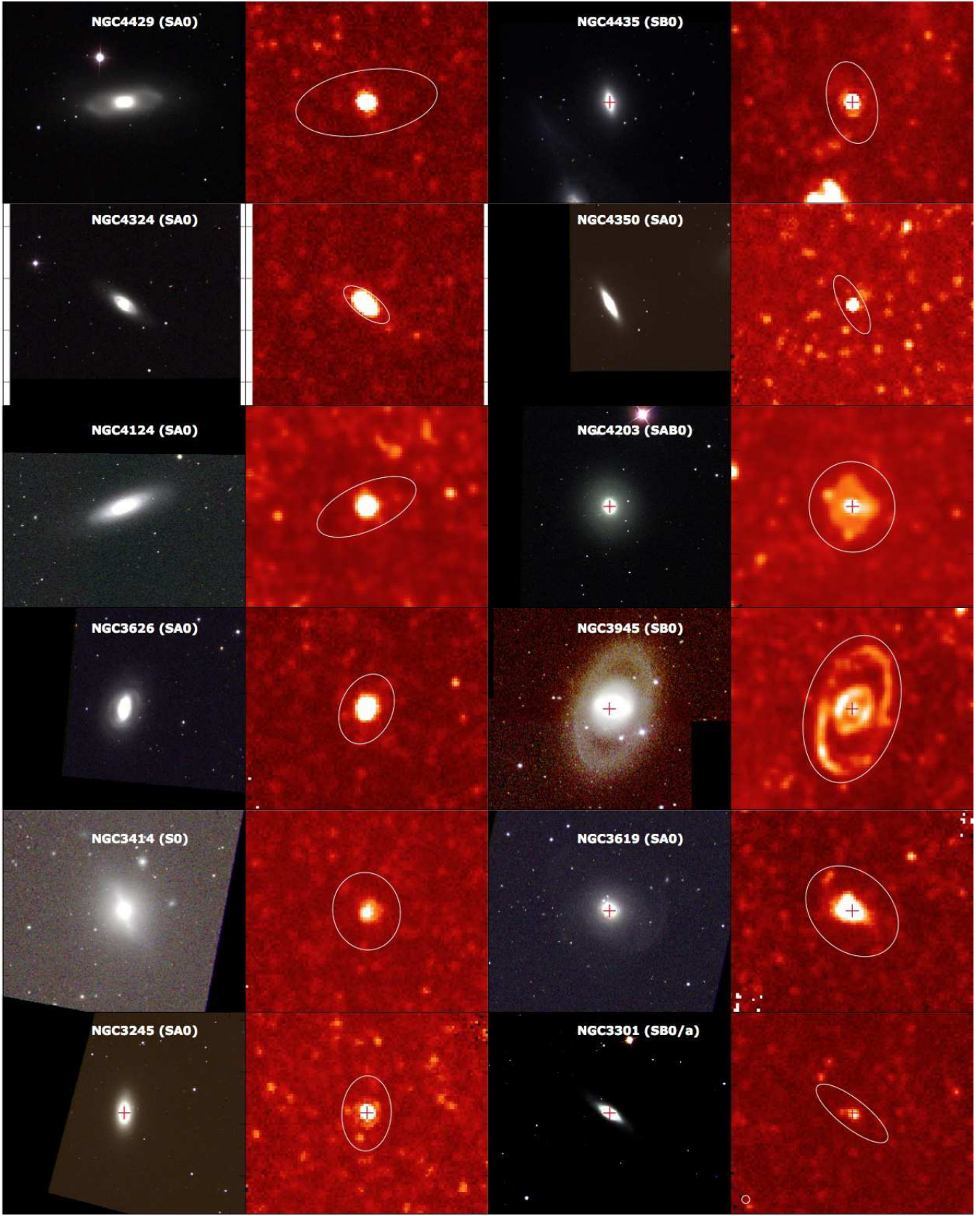}}
\end{center}
\figcaption{SDSS $gri$ three-color maps and 250\,$\mu$m \Hersc~SPIRE maps
  of the 24 detected \SO galaxies from the \Hersc~Reference Survey.
  North is up and east is left.
  The white aperture shows the elliptical optical $B$-band isophotal
  diameter ($D(25)$ at $25\,\rm mag\,arcsec^{-2}$---
  Table~\ref{tab:sample}). The region shown is $8.4\arcmin \times 8.4 \arcmin$.
  Red crosses show the location of the VLA FIRST radio
  detection.   The 250\,$\mu$m beam is shown in the corner of (a) NGC3301 and (b) NGC4469. \label{fig:spireearly}}
\end{figure*}

\begin{figure*}
\ContinuedFloat
\begin{center}
\subfloat[]{\includegraphics[trim=9mm 5mm -3mm
  -4mm,clip=true,width=18cm]{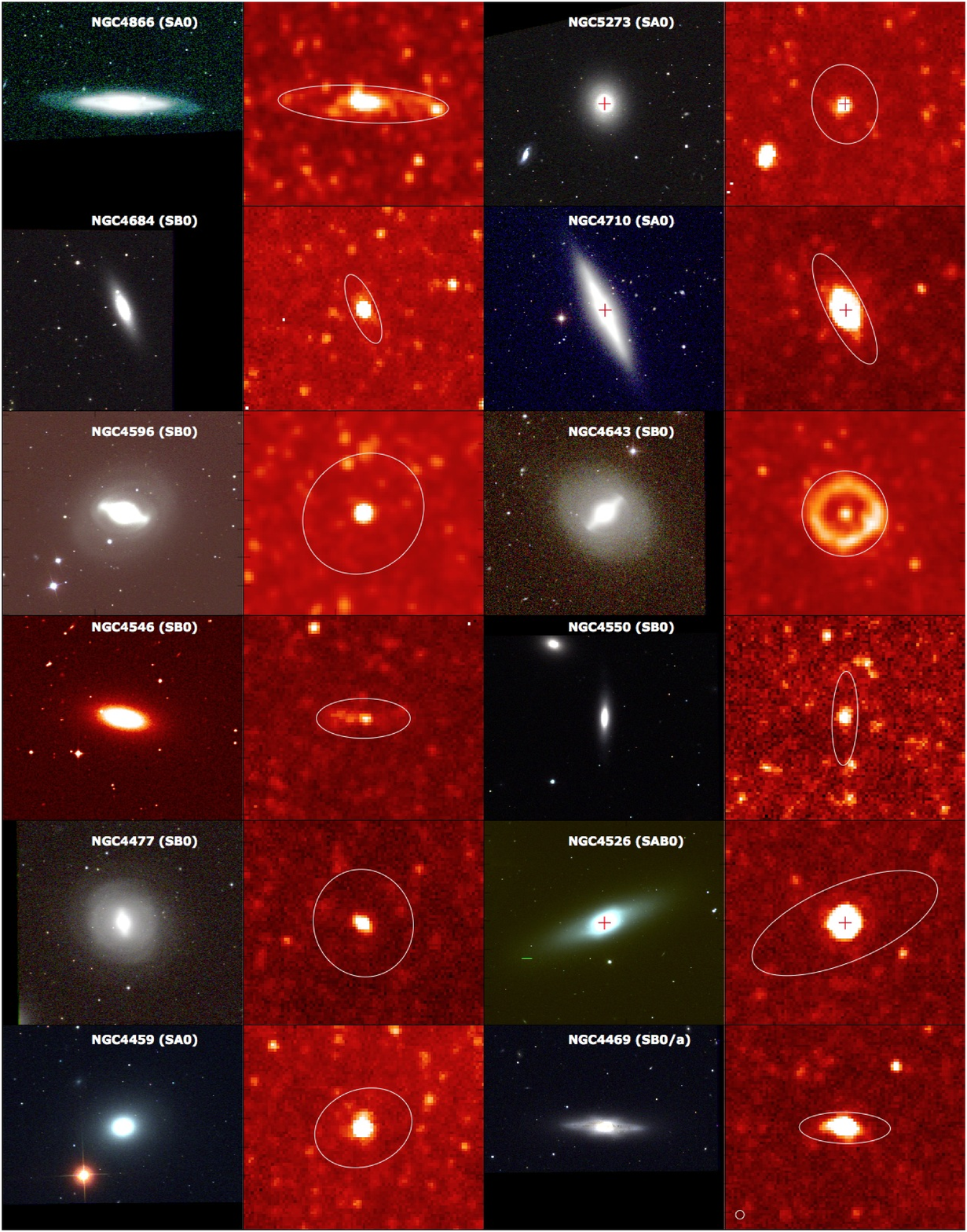}}
\end{center}
\figcaption{{\em (Continued)}}
\end{figure*}

\begin{figure*}
    \begin{center}
\subfloat[][]{\includegraphics[trim=38mm 40mm -1mm
  1mm,clip=true,width=18cm]{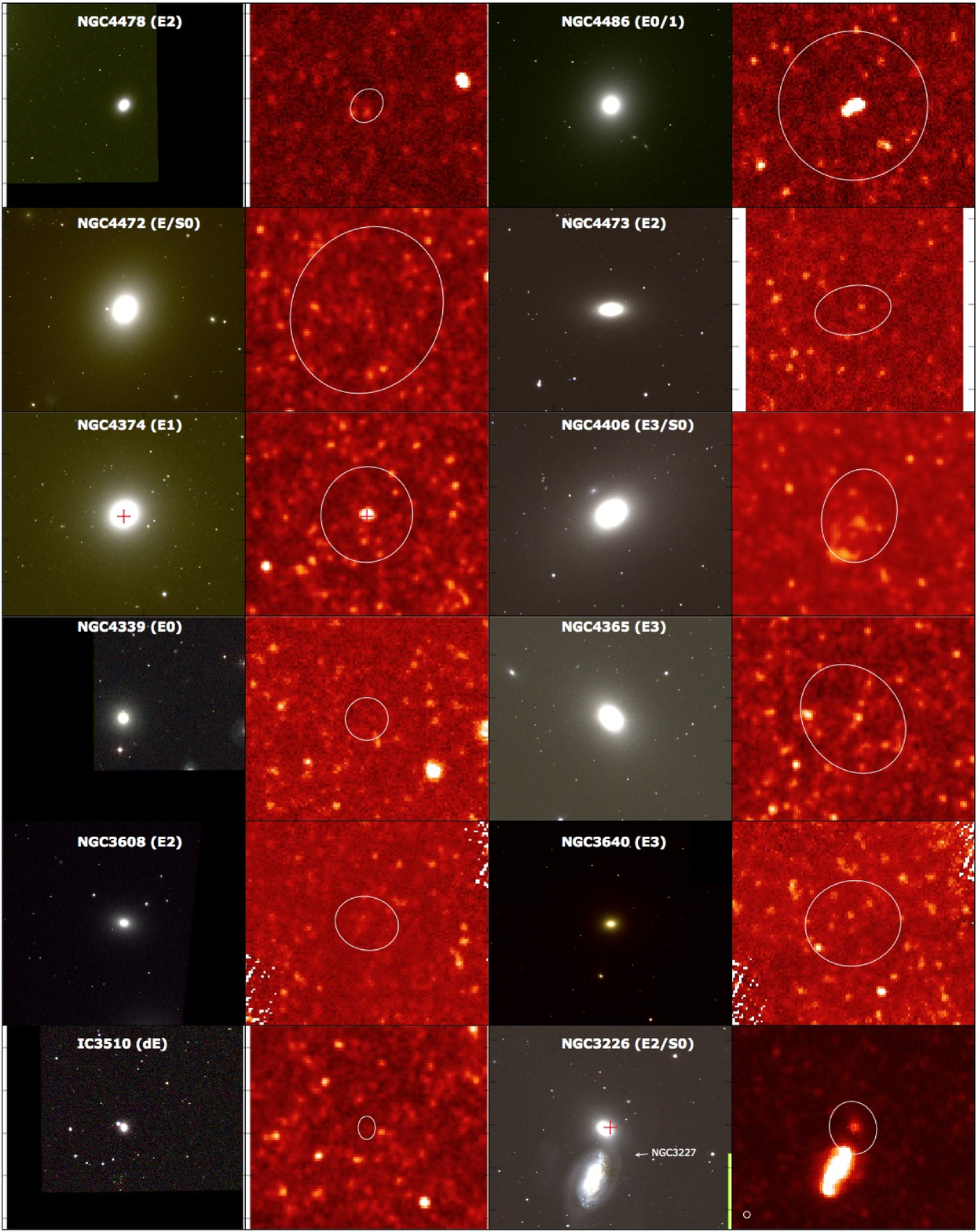}}
\end{center}
\figcaption{SDSS $gri$ three-color maps and 250-$\mu$m \Hersc~SPIRE maps
  of the entire sample of 23 elliptical galaxies from the HRS and
  \hevics~programs (sources denoted with a (*) indicate that it has been detected by \Hersc). 
  North is up and east is left.
  The white aperture shows the optical $B$-band
  isophotal diameter ($D(25)$ at $25\,\rm mag\,arcsec^{-2}$---Table~\ref{tab:sample}).  
  The boxes are $9.6\arcmin \times 9.6 \arcmin$.
  Red crosses show the location of the VLA FIRST radio
  detection. The 250\,$\mu$m beam is shown in the corner of (a) NGC3226 and (b) NGC4552. \label{fig:spireegals}}
\end{figure*}

\begin{figure*}
\ContinuedFloat
    \begin{center}
\subfloat[][]{\includegraphics[trim=38mm 40mm -1mm 1mm,clip=true,width=18cm]{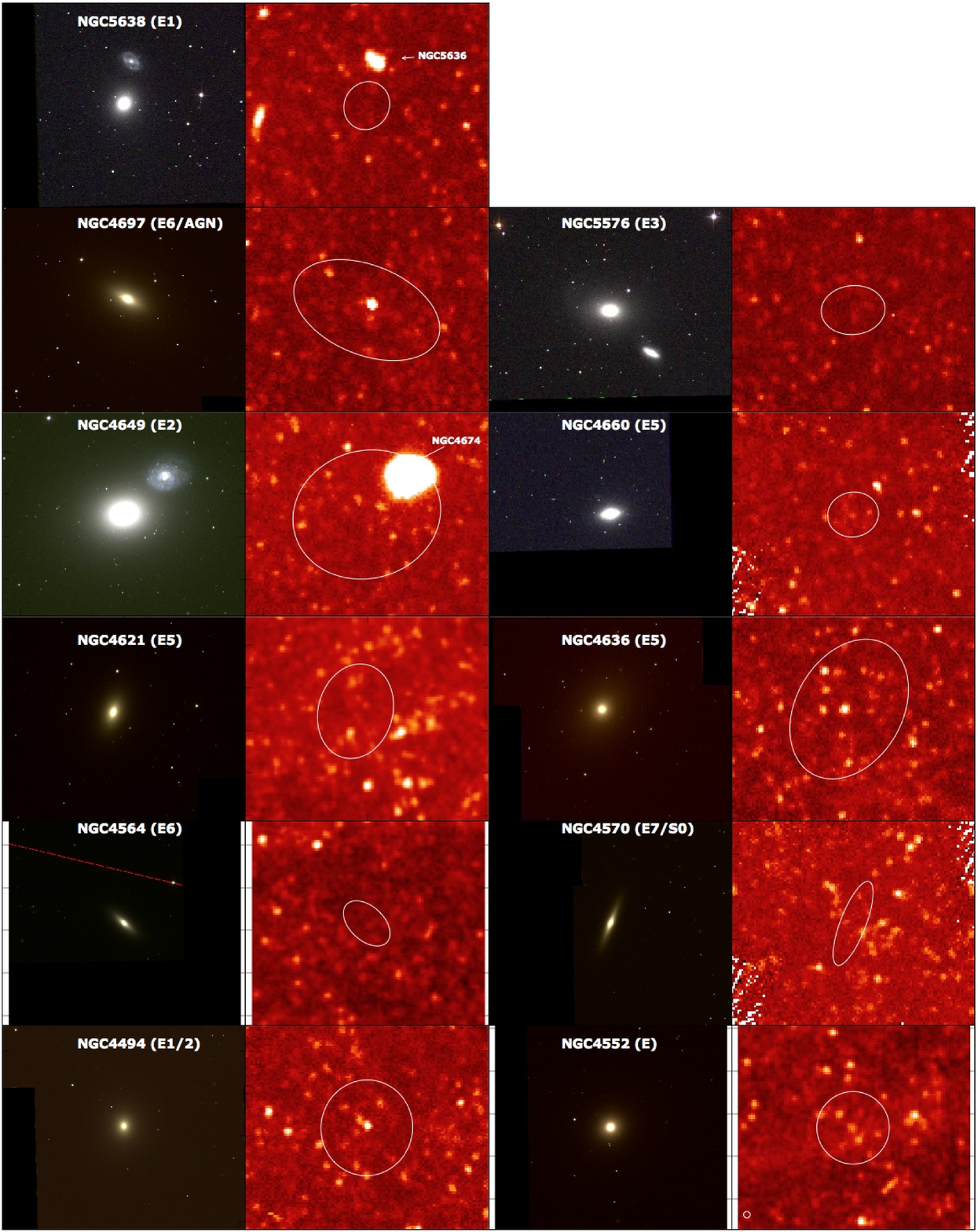}}
\end{center}
\figcaption{{\em (Continued)} \label{fig:spireegals}}
\end{figure*}

\section{Results}
\label{sec:res}

In Figure~\ref{fig:spireearly} we show the detected \SO galaxies and all of the ellipticals in
Figure~\ref{fig:spireegals}; the sources show a variety of morphologies, from small blobs
to spectacular ring structures. The difference in the appearance of
the \SO and ellipticals at 250\,$\mu$m is very clear in these figures. The
detected \SO sources are brighter than their elliptical counterparts
and the FIR emission is more extended within the $D(25)$ isophotes. In
general, this appears to be a trend as we move along the Hubble
sequence: an aperture of size $0.8 \times D(25)$ is sufficient to
include the submm emission from the ETGs, yet L. Ciesla et al. (in preparation)
require an aperture size of $1.4 \times D(25)$ to incorporate all the
emission from the late-type galaxies. This effect was also noticed by
Bendo et al.\ (2007) and Mu\~{n}oz-Mateos et al.\ (2009).  

The difference between late-types and early-types when viewed in the
submm and optical wavebands is quite startling. For example, at
optical wavelengths, the giant elliptical NGC4649 (HRS245) is
extremely bright compared to the smaller, fainter spiral galaxy to the
northwest (NGC4674). At 250\,$\mu$m (Figure~\ref{fig:spireegals})
however, the situation is reversed, with the spiral galaxy extremely
bright in the FIR, but no sign of the elliptical. In
Appendix~\ref{sec:desc}, we provide brief notes on the elliptical
galaxies that are detected by \Hersc.

\begin{deluxetable}{lrrrr}
\tabletypesize{\scriptsize}
\tablecaption{{\it Herschel} PACS and {\it IRAS} Fluxes for Detected Galaxies. \label{tab:fluxes}}
\tablewidth{0pt}
\tablehead{ \colhead{HRS Name} 
 & \colhead{F100} &
\colhead{F160} & \colhead{F100} & \colhead{F60} \\ 
\colhead{} & \colhead{(Jy)} &
\colhead{(Jy)} & \colhead{(Jy)} & \colhead{(Jy)} }
\startdata
7 &  & & 3.97$\pm$0.64 & 2.09$\pm$0.28 \\
14 &  & & 0.92$\pm$0.16 & 0.48$\pm$0.07 \\
22 & && 0.56$\pm$0.21 & 0.25$\pm$0.04 \\
45 & & & 1.83$\pm$0.37 & 0.38$\pm$0.06 \\
71 &  & & 1.36$\pm$0.24 & 0.26$\pm$0.04 \\
87 &&& 1.75$\pm$0.29 & 0.43$\pm$0.07 \\
93 & && 2.16$\pm$0.36 & 0.59$\pm$0.08 \\
101 & & &\multicolumn{1}{c}{.\,.\,.}& 0.12$\pm$0.04\\
105  &\multicolumn{1}{c}{\,.\,.\,.} &  2.71$\pm$0.54 & 0.39$\pm$0.14 & 0.18$\pm$0.04 \\
123 & 1.80$\pm$0.6~ &2.42$\pm$0.48 & 1.99$\pm$0.32 & 0.41$\pm$0.07 \\
126 & & & 0.37$\pm$0.09 & 0.09$\pm$0.03 \\
129 & & & 1.09$\pm$0.19 & 0.36$\pm$0.06 \\
138 & 0.78$\pm$0.16 &  0.46$\pm$0.09 & 1.16$\pm$0.22 & 0.50$\pm$0.07 \\
150 &  & & 0.33$\pm$0.09 & 0.11$\pm$0.04 \\
161 & 4.74$\pm$0.95&  4.54$\pm$0.91 & 5.15$\pm$0.83 & 1.56$\pm$0.21 \\
162 & 4.48$\pm$0.90&  3.75$\pm$0.75&\multicolumn{1}{c}{.\,.\,.} &\multicolumn{1}{c}{.\,.\,.} \\
174 &4.59$\pm$0.92 &3.78$\pm$0.75 & 4.82$\pm$0.78 & 1.87$\pm$0.25 \\
176 & 3.33$\pm$0.67& 3.89$\pm$0.78 & 3.40$\pm$0.57 & 1.02$\pm$0.14 \\
180 &  0.94$\pm$0.19&0.97$\pm$0.19  & 1.41$\pm$0.25 & 0.57$\pm$0.09 \\
183 & 0.38$\pm$0.08 & 0.60$\pm$0.12& 0.41$\pm$0.12 & 0.39$\pm$0.07 \\
186 & & &\multicolumn{1}{c}{.\,.\,.} & 0.19$\pm$0.06 \\
200 &15.81$\pm$3.16 &15.53$\pm$3.11& 17.10$\pm$2.74 & 5.56$\pm$0.72 \\
209 &  &  & 0.89$\pm$0.26 & 0.26$\pm$0.06 \\
210 & 0.28$\pm$0.06 & 0.27$\pm$0.05 & 0.25$\pm$0.10 & 0.14$\pm$0.04 \\
211 & && 0.53$\pm$0.10 & 0.16$\pm$0.05 \\
231 &  & & 0.75$\pm$0.13 & 0.40$\pm$0.06 \\
243 & && 2.06$\pm$0.34 & 0.62$\pm$0.09 \\
253 &  &&  2.15$\pm$0.36 & 1.27$\pm$0.17 \\
258 & & & 1.24$\pm$0.21 & 0.46$\pm$0.06 \\
260 & & & 14.79$\pm$2.37 & 5.73$\pm$0.75 \\
286 & & & 1.02$\pm$0.23 & 0.15$\pm$0.06 \\
296 & & & 1.56$\pm$0.28 & 0.90$\pm$0.12 \\
312 & && 0.21$\pm$0.28 & 0.09$\pm$0.03 \\
316 &&& 0.45$\pm$0.14 &\multicolumn{1}{c}{.\,.\,.}
\enddata
\tablecomments{The {\it Spitzer}, and {\it Herschel} photometry is presented in 
Temi et al. (2004), G. Bendo et al. (2011, submitted). and L. Ciesla et al. (in preparation), respectively.  The columns are as follows.
Column 1: 
the number of the source in the \Hersc~Reference Survey (Boselli et al.
2010b). 
Columns 2 and 3: {\it Herschel} PACS fluxes at 100 and 160\,$\mu$m. Columns 4 and 5: \textit{IRAS} fluxes at 100 and 60\,$\mu$m.}
\end{deluxetable}

\subsection{Detection Rates}

We detect 31 out of the 62 ETGs, including 24 \SO galaxies (62\%) and
7 ellipticals (30\%, Table~\ref{tab:percents}).  We detect all three
of the ETGs classified as `peculiar' (two ellipticals, M87 and
NGC3226, and one S0, NGC3414).  
Our detection rates are higher than
published in an \textit{IRAS} study of ETGs (Bregman et al. 1998), which
detected 12\%--17\%.  Although the \textit{IRAS} study excluded peculiar galaxies
and active galactic nuclei (AGNs; with approximately half their sample at distances larger than the HRS), 
the small number of these sources in our sample suggests that the
most likely explanation of the large difference in detection rates is
that \Hersc~is more sensitive to cold dust than \textit{IRAS}. This is supported
by the higher mean dust temperature reported by Bregman et al. (1998). 
Temi et al. (2004) detected a higher fraction of ETGs (41\% of ellipticals and 79\% of S0s)
with {\it ISO}, but their sample contains peculiar galaxies and giant
ellipticals. 
The detection rates for the SPIRE and multiwavelength
data for our sample are listed in Table~\ref{tab:percents}; these
rates are significantly\footnote{\label{fn:fis}Using the two-sided
  Fisher's exact test, the equivalent of the $\chi$-squared test but
  for small samples.} different for S0s and ellipticals at the 96\%
level.  
Our overall detection rate of 50\% for the ETGs is higher than
the 22\% obtained by the ATLAS$^{\rm 3D}$ CO study of 260 ETGs (Young et al. 2011),
though the ATLAS$^{\rm 3D}$ sample has a larger volume out to distances 46\,Mpc (compared to 25\,Mpc for the HRS).
Comparing those HRS galaxies which are in both the ATLAS$^{\rm 3D}$ and HRS (i.e., observed in both CO
and dust continuum), the detection fraction in CO is a factor of two lower
than at 250\,$\mu$m. Similarly for those galaxies with \Hersc\ and H{\sc i} observations, the detection
fraction in H{\sc i} is $\sim$1.5 times lower (Table\,\ref{tab:percents}).
The \Hersc~observations of the continuum
emission from the dust are therefore currently the most sensitive way of
detecting the ISM in the HRS sample of ETGs.

\begin{deluxetable*}{lcccccccccccccccc}[ht]
\tabletypesize{\scriptsize}
\tablecaption{The Number of Early-type and Elliptical
  Galaxies Detected with \Hersc  \label{tab:percents}}
\tablewidth{0pt}
\tablehead{
 \colhead{Type} &\multicolumn{4}{c}{SPIRE Sample} &
    \multicolumn{6}{c}{X-Ray Sample} &  \multicolumn{6}{c}{CO Sample}\\
    \cmidrule(rl){2-5} \cmidrule(rl){6-11} \cmidrule(rl){12-17}
    \colhead{} &\multicolumn{2}{c}{Observed} &
    \multicolumn{2}{c}{Detected 250} & \multicolumn{2}{c}{Observed} & \multicolumn{2}{c}{Detected X-ray} &
    \multicolumn{2}{c}{Detected 250} & \multicolumn{2}{c}{Observed} & \multicolumn{2}{c}{Detected CO} &
    \multicolumn{2}{c}{Detected 250} \\
    \colhead{} &\colhead{$N$} &\colhead{\%} 
   &\colhead{$N$} & \colhead{\%} & \colhead{$N$} &\colhead{\%}
    &\colhead{$N$} & \colhead{\%} &\colhead{$N$} & \colhead{\%} &
    \colhead{$N$} & \colhead{\%} &\colhead{$N$} & \colhead{\%}
    &\colhead{$N$} & \colhead{\%}
}
\startdata
Total & 62 & 100\% & 31 & 50\%  & 38 & 61\% & 29 & 76\% & 15 & 52\%  & 56
&90\%  & 16 & 28\% & 16 & 100\% \\ 
&&&&&&&&&&&&&&&&\\
E & 21 &34\% & 5 & 24\%  & 20 &53\% & 16 & 80\% & 4 & 25\% & 20  &
34\%& 0  & 0\% & 0& 0\% 
\\
S0 &38 & 61\%& 23 & 63\% &  15 & 39\%&  11  & 77\% & 9 & 82\%  &35 &
63\%& 16 & 55\% & 16 & 100\% 
\\
pec &3  & 5\%& 3 & 100\%& 3 & 8\%& 2 & 67\% & 2  & 100\%  & 1& 2\%
& 0  & 0\% & 0 & 0\%  \\ 
\cmidrule{1-17}
& \multicolumn{6}{c}{H{\sc i} sample} &
\multicolumn{8}{c}{Environment} & &\\
\cmidrule(rl){2-7} \cmidrule(rl){8-15}
& \multicolumn{2}{c}{Observed} & \multicolumn{2}{c}{Detected H{\sc i}} &
\multicolumn{2}{c}{Detected 250} &
\multicolumn{2}{c}{Obs. Virgo} & \multicolumn{2}{c}{Det. Virgo} &
 \multicolumn{2}{c}{Obs. Not Virgo} &  \multicolumn{2}{c}{Det. Not
   Virgo}  & & \\
& \colhead{$N$} &\colhead{\%}  &\colhead{$N$} & \colhead{\%} &\colhead{$N$}
& \colhead{\%} & \colhead{$N$} &\colhead{\%}  &\colhead{$N$} & \colhead{\%} &\colhead{$N$}
& \colhead{\%} & \colhead{$N$} &\colhead{\%}  & & \\
\cmidrule(rl){1-15}
Total  & 49 & 79\% & 17 & 35\% & 16 & 94\% & 47 & 76\% & 21 & 45\%& 15
& 24\% & 10 & 66\% & &  \\ 
E & 15 &31\% & 4 & 27\% & 4 & 100\% & 17 & 74\% & 5 & 29\% & 6 & 26\% & 2 & 33\%
 & & \\
S0 &  34 & 69\% & 13 & 38\% & 12 & 92\% & 30 & 77\%& 16 & 53\% & 9 & 23\% & 8 & 88\% 
& & \\
pec & 0  & 0\% & 0 & 0\% & 0 & 0\%&.\,.\,.&\,.\,.\,.&.\,.\,.&.\,.\,. & .\,.\,. &.\,.\,.& .\,.\,.&
.\,.\,.& & \\
\cmidrule(rl){1-17}
& \multicolumn{8}{c}{($b/a \le 0.5$)} &
\multicolumn{8}{c}{($b/a> 0.5$)} \\
\cmidrule(rl){2-9} \cmidrule(rl){10-17}
& \multicolumn{2}{c}{Observed} & \multicolumn{2}{c}{Detected}
&\multicolumn{2}{c}{Det. Virgo} & 
 \multicolumn{2}{c}{Det. Not Virgo}  &
\multicolumn{2}{c}{Observed} &
\multicolumn{2}{c}{Detected} &\multicolumn{2}{c}{Det. Virgo} & 
 \multicolumn{2}{c}{Det. Not Virgo}  \\
& \colhead{$N$} &\colhead{\%}  &\colhead{$N$} & \colhead{\%} &\colhead{$N$}
& \colhead{\%} & \colhead{$N$} &\colhead{\%}  &\colhead{$N$} & \colhead{\%} &\colhead{$N$}
& \colhead{\%} & \colhead{$N$} &\colhead{\%}  &\colhead{$N$} & \colhead{\%}  \\ \hline 
&&&&&&&&&&&&&&&&\\
S0 & 18 & 46\% & 12 & 67\%  & 13/17 & 76\%  & 1/1 & 100\%  & 21 &54\% & 12 &
57\% & 2/10 & 20\% & 9/11 & 82\%\\
Barred & 11 & 28\% & 6 & 55\%  &  5/10 & 50\% & 1/1 & 100\% &  13 & 33\% & 6 & 46\% &
4/10 & 40\% & 2/3 & 67\%
\enddata
\tablecomments{Top Panel: the detection numbers for the
  250\,$\mu$m,
  X-ray, and CO samples in this work. Columns are as follows.
  Column 1: morphological type for the sample.  
  Columns 2--5:{\it SPIRE}-detected sources including---Column 2: number of sources observed; Column 3: percentage of the observed sample which falls into the different types; Column 4: number of sources detected with SPIRE; Column 5: percentage detected with SPIRE compared to the observed sources in each morphological class. 
  Columns 6--11:{\it X-ray sources} including---Columns 6 and 7: number of sources in our sample observed in X-ray from the literature and the percentage which falls into each morphological class (O' Sullivan \& Ponman 2004; Pellegrini 2010); Columns 8 and 9: number and percentage of the observed X-ray sources which are detected in X-ray; Columns 10 and 11: number of sources detected in X-rays {\em and} detected with SPIRE.  
  Columns 12--17:{\it CO sources} including---Columns 12 and 13: number of
  sources in our sample observed in CO from the literature and the
  percentage which falls into each morphological class (Young et al.\
  2011); Columns 14 and 15: number and percentage of the observed CO
  sources which are detected in CO; Columns 16 and 17: number of sources
  detected in CO {\em and} detected with SPIRE. 
  Middle Panel: the detection numbers for the H{\sc i} and
  250\,$\mu$m sample, and for the entire 250\,$\mu$m sample split by environment.  Columns are as follows.
  Columns 2--7:{\it H{\sc i} sources} including---Columns 2 and 3: number of
  sources in our sample observed in H{\sc i} from the literature and the
  percentage which falls into each morphological class; Columns 4 and 5:
  number and percentage of the observed H{\sc i} sources which are detected
  in H{\sc i}; Columns 6 and 7: number of sources
  detected in H{\sc i} {\em and} detected with SPIRE. 
  Columns 8--15:{\it Environment}: the number and percentage in each
  subset defined as within the Virgo Cluster (Columns 8--11) and outside
  (Columns 12--15).  Given the small numbers in the Virgo/non-Virgo
  samples, we do not separate the peculiar galaxies in this analysis.
  Bottom Panel: the detection numbers for \SO and barred S0
  galaxies split into regimes depending on their axial ratios, where
  $b/a <0.5$ and $b/a>0.5$, indicating edge-on or face-on galaxies,
  respectively.
  Columns are as follows.
  Columns 2--9:{\it Edge-on galaxies} including---Columns 2 and 3:
  number and percentage of edge-on sources ($b/a<0.5$); Columns 4 and 5: number and percentage of these
  sources which are detected with SPIRE. Columns 6 and 7: number of
  sources detected with SPIRE inside the Virgo Cluster and Columns 8 and 9
  for those outside Virgo.  
  Columns 10--17:{\it Face-on galaxies}---Columns 10 and 11 - number and percentage of face-on
  sources ($b/a>0.5$); Columns 12 and 13:
  number and percentage of those sources which are detected with
  SPIRE. Columns 14 and 15: number of sources detected with SPIRE inside
  the Virgo Cluster and Column 16 and 17 for those outside Virgo.  }
\end{deluxetable*}

Given the small number of ellipticals that have been detected by
\Hersc, it is interesting to ask the question of what makes these
special. Twenty of the ETGs in our sample are classified as having
low-ionization narrow emission line regions (LINERs). The origin of
LINER activity in ETGs can be ionizing photons from the old stellar
population (e.g., di Serego Alighieri et al. 1990;
Sarzi et al.\ 2010), heating by the hot ISM (Sparks et al.
1989) or AGN heating (e.g., Gonz\'{a}lez-Mart\'{i}n et al.\ 2009).
At least 12 of the HRS sample have bright radio cores
(Figure~\ref{fig:spireearly}) suggesting AGN photoionization is
important for at least 20\% of the sample.  For the most massive
ellipticals, the hot ISM could be a significant photoionizing source;
nine of the detected LINER sources in our sample have X-ray
luminosities well above that expected from the discrete stellar
population i.e., $L_X/L_B \sim 10^{29.5}$ (O'Sullivan et al.\ 2001;
Section \ref{sec:hot}) suggesting that a significant fraction of the LINER
emission may originate from the hot ISM rather than an AGN source. Of
the seven detected ellipticals, two are classified as LINERs, two as Seyferts and a further two are unambiguously identified as containing a
bright AGN (Ho et al. 1997; Schmitt 2001; Veron-Cetty \& Veron 2010). Only one of the seven, NGC 4406 (HRS150) is a `pure'
elliptical with no sign of an active nucleus, but this galaxy has
acquired dust and atomic gas through a recent interaction (Gomez et
al. 2010).  Nine other `pure' ellipticals in our sample are not
detected at 250\,$\mu$m.

\begin{figure*}[ht]
  \centering
  \includegraphics[trim=23mm 11mm 19mm
    20mm,clip=true,width=8.4cm]{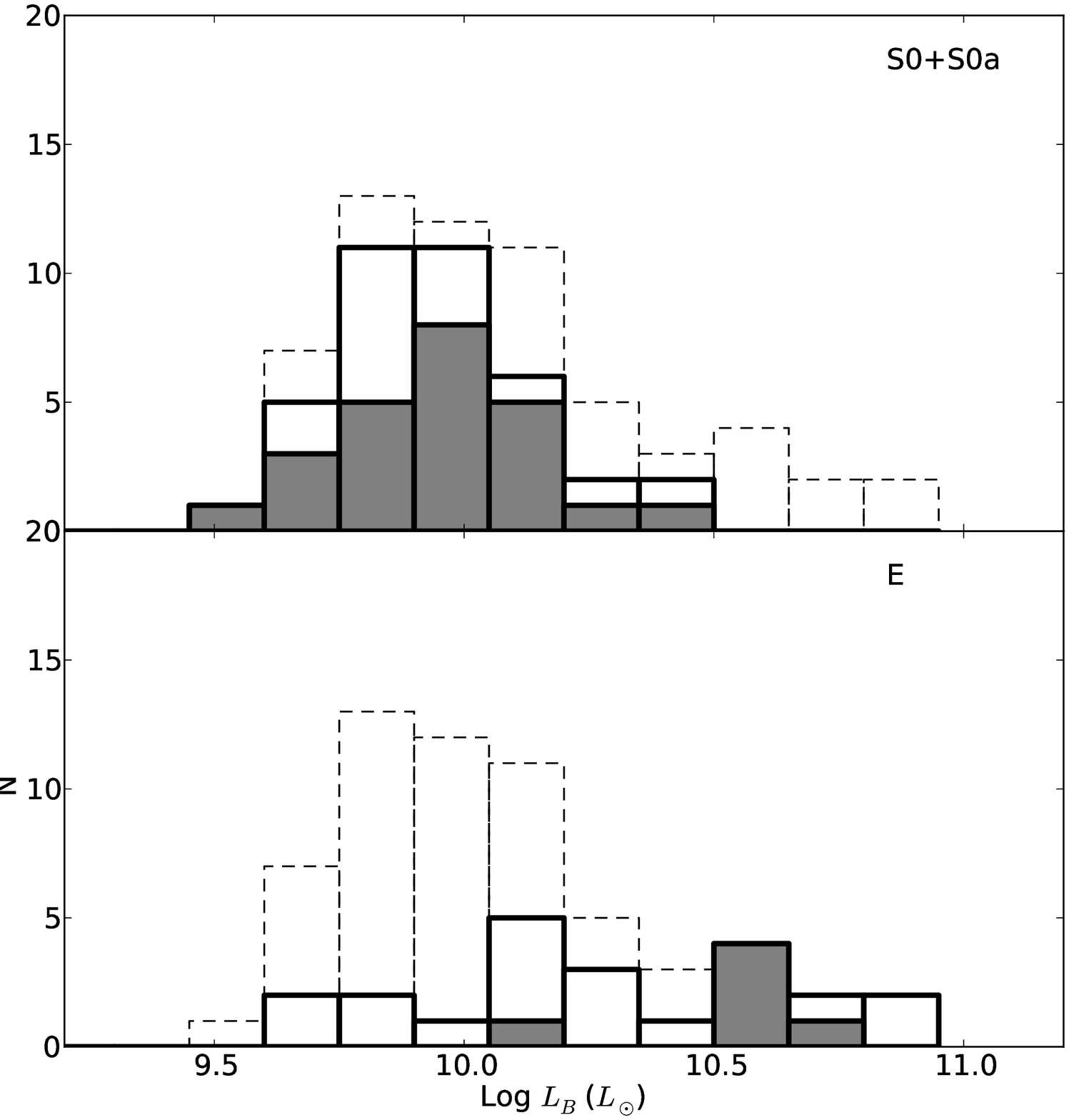}
  \includegraphics[trim=23mm 11mm 19mm
    20mm,clip=true,width=8.4cm]{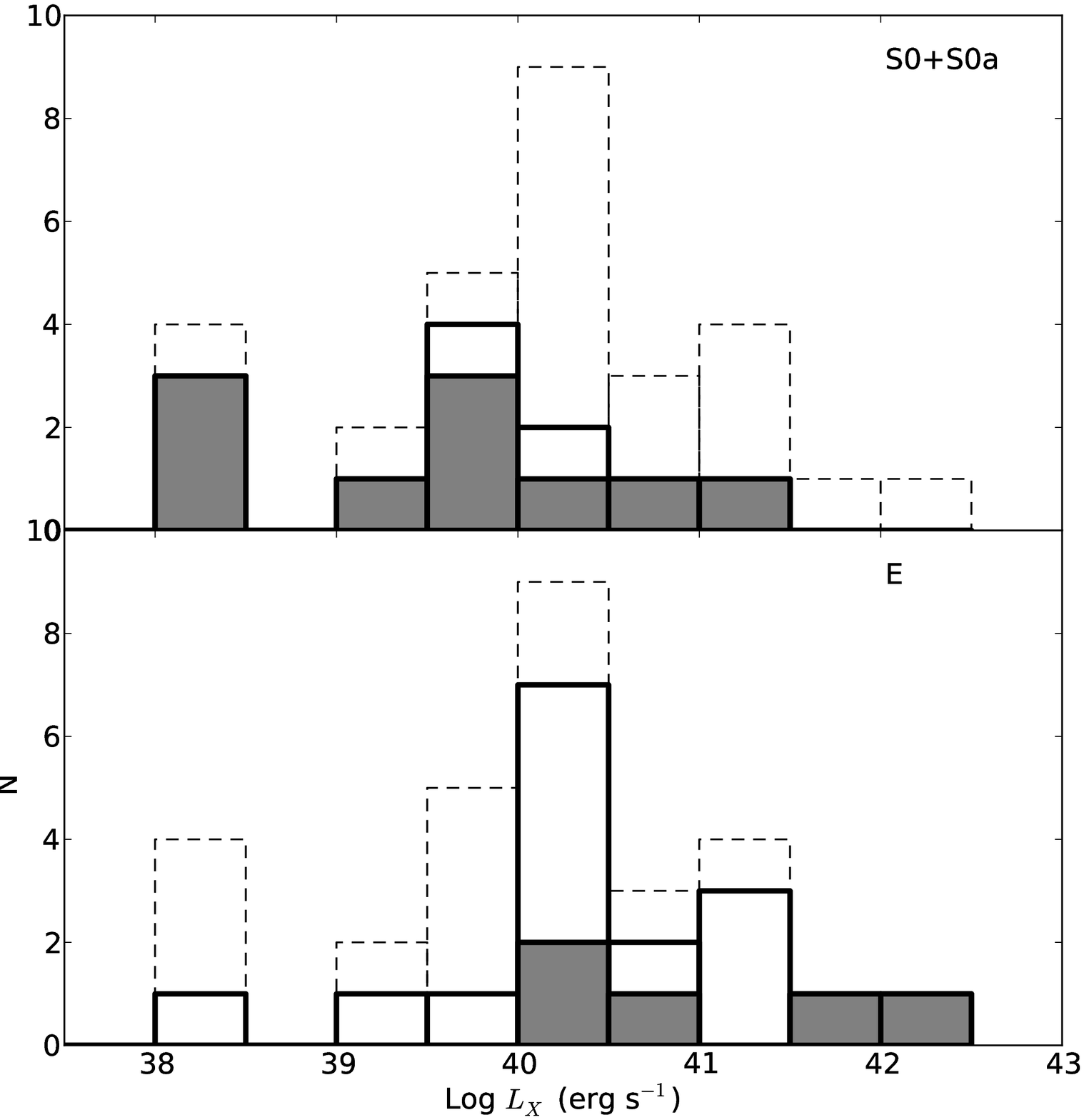}
  \includegraphics[trim=23mm 11mm 19mm
    20mm,clip=true,width=8.4cm]{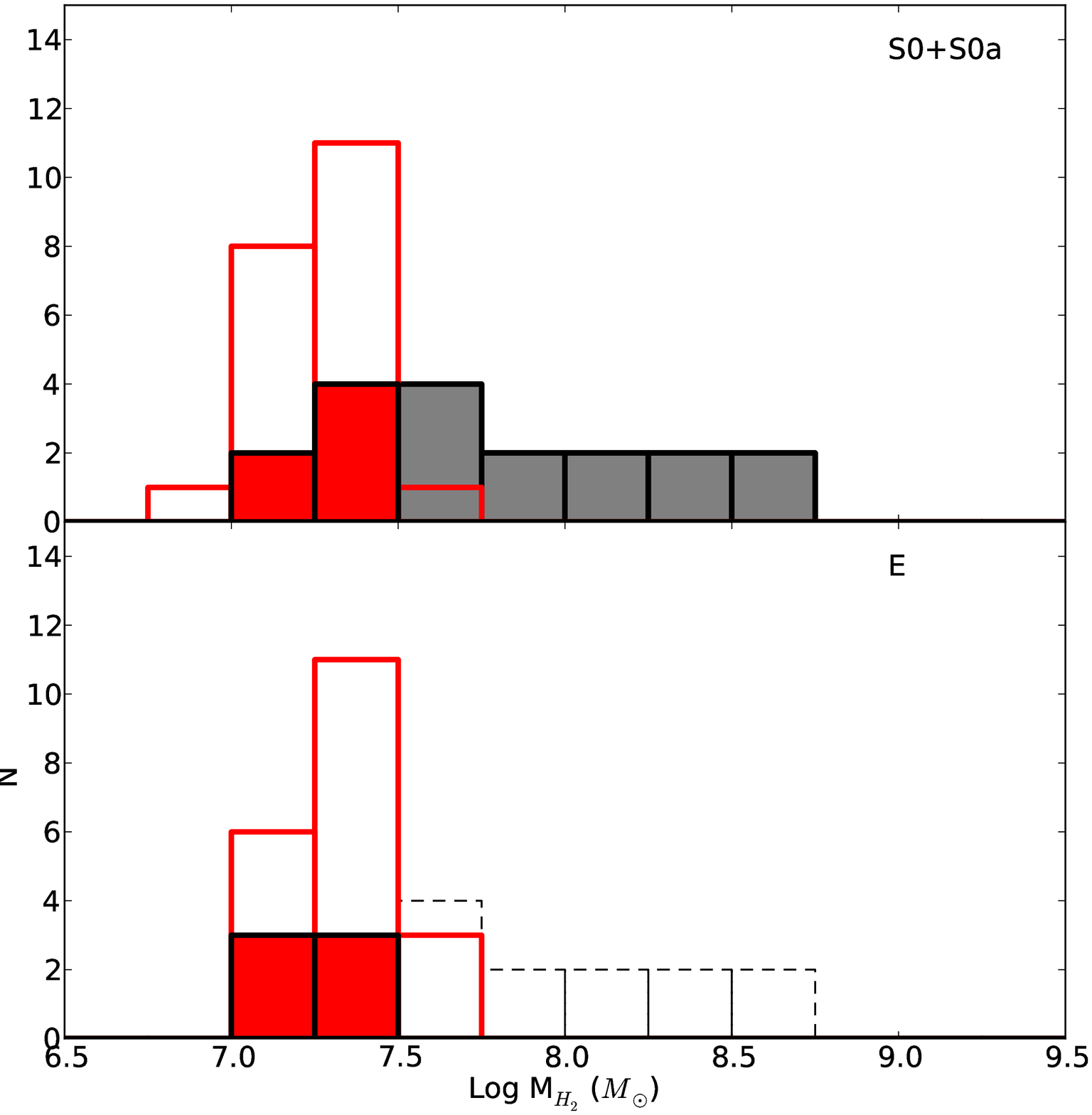}
 \includegraphics[trim=23mm 11mm 19mm
    20mm,clip=true,width=8.4cm]{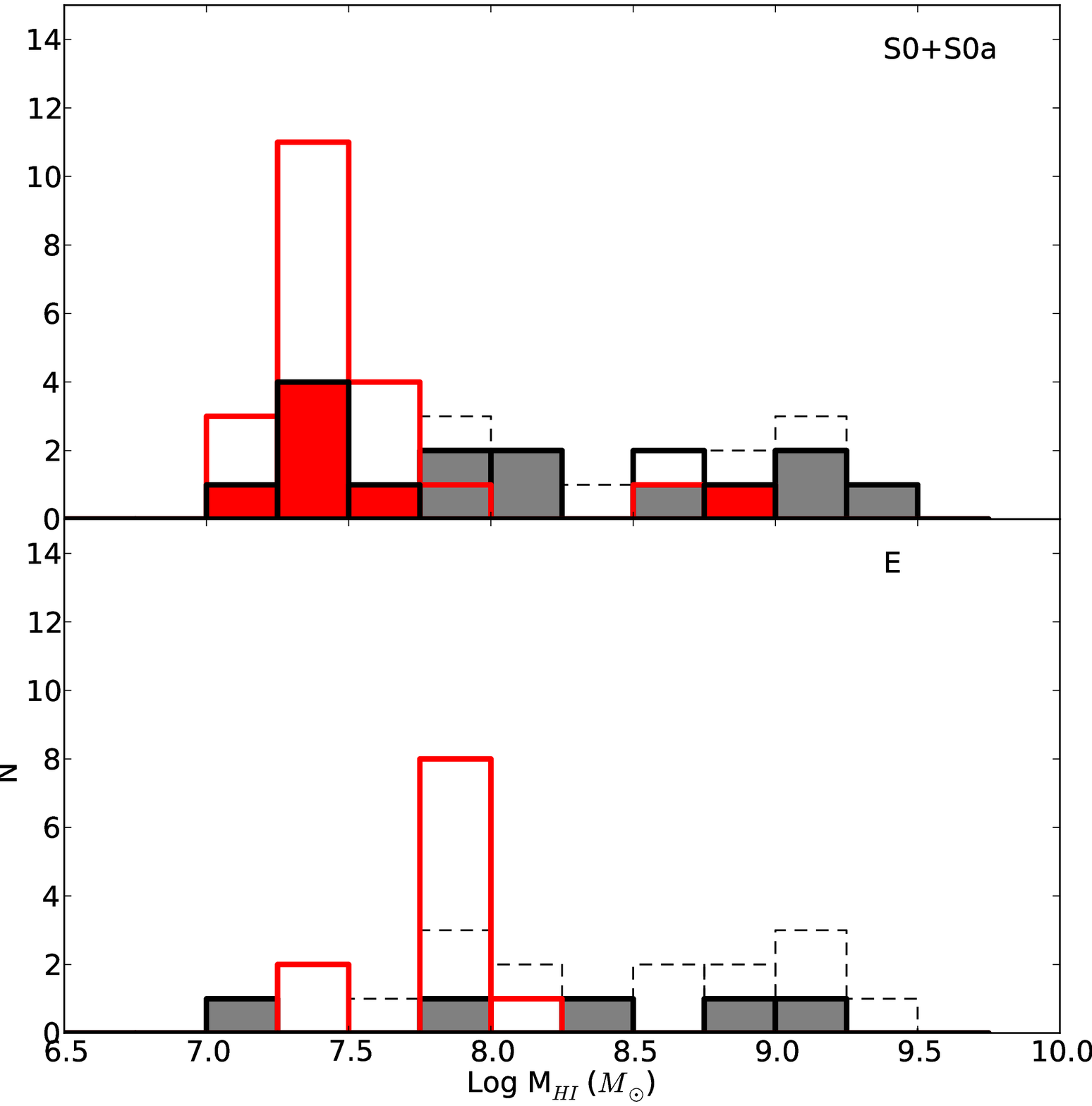}
    \figcaption{Top left: the range of
      $B$-band luminosities; top right: X-ray luminosities
      for the \SO (top) and elliptical (bottom) galaxies. The outer
      dashed histogram shows the range of the whole sample (\SO or E
      combined), the thick black unfilled histograms are the
      subsamples (\SO and E separated out), and the detected sources
      are shaded gray. Bottom left: the molecular gas mass
      estimated from CO observations; bottom right: the atomic gas
      mass from H{\sc i}. The red histograms show the upper limits on the
      gas mass obtained from the CO (H{\sc i}), i.e., these are sources in our
      sample not detected either at 250\,$\mu$m or CO (H{\sc i}).  The red shaded
      histogram shows the sources detected at 250\,$\mu$m but {\em
        not} detected in CO (H{\sc i}).  \label{fig:kbandhist}}
\end{figure*}

Another way to look at this question is to compare the properties of
the detected and undetected sources in other wavebands.  The
distribution of the $B$-band luminosities for the sample is shown in
Figure~\ref{fig:kbandhist} along with the distributions of the detected
sources.  The figure shows that while there is no tendency for
  the detected S0 galaxies to be optically luminous, the detected
  elliptical galaxies are optically the most luminous ones.

  Figure~\ref{fig:kbandhist} shows a similar comparison for the X-ray
  luminosities.  In this case, one might suspect that the galaxies
  with the highest X-ray luminosities from the hot halo are the least likely to be
  detected by \Hersc~because of sputtering.
  The figure shows that the elliptical galaxies detected at
  250\,$\mu$m tend to
  have higher X-ray luminosities so that elliptical
  galaxies more luminous in X-rays and optical (i.e., the most
  massive stellar systems) tend to be detected by \Hersc.  The detected S0 galaxies span a wide range of $L_B$ and $L_K$,
  indicating that the stellar mass of these systems is not the
  critical factor for detection by \Hersc.  

  In contrast to the X-rays, we find that galaxies containing large
  masses of cool gas are more likely to be detected by \Hersc~(Figure~\ref{fig:kbandhist}).
  CO observations of our sources taken from the ATLAS$^{\rm 3D}$
  survey (Young et al.\ 2011) suggest that 28\% of our ETGs have
  molecular gas (Table~\ref{tab:percents}), similar to the detection
  rate of 22\% for the entire ATLAS$^{\rm 3D}$ sample of 260 ETGs.
  Atomic hydrogen observations are available in the literature for 79\% of
  our sample, although only 35\% of these galaxies are detected.  

  We can also consider whether certain morphological types
are more likely to be detected. Eighty percent of the undetected
  galaxies are morphologically classified as barred systems (NED, Table~\ref{tab:sample}), which at
  first glance suggests that barred galaxies have less dust content
  than galaxies without a bar.  However, we actually detect 50\% of the barred
  sources from the HRS sample (Table~\ref{tab:percents}), suggesting
  that the presence of a bar does not determine whether it is detected
  at 250\,$\mu$m.    Finally, we investigated whether the inclination angle might affect
  whether or not the galaxy was detected by \Hersc~using the measured
  axial ratios of the galaxies (minor axis/major axis---$b/a$).  The
  detection rate for edge-on galaxies ($b/a < 0.5$) is 67\% and 
  for face-on galaxies ($b/a > 0.5$) is 57\% though this is not statistically significant.  Whether a galaxy
  has rings, bars or high inclination angles therefore does not affect
  whether it is detected by \Hersc.

  One important effect on whether we detect ETGs with \Hersc~is
  environment.  It has long been known that galaxies in dense
  environments are depleted in atomic gas, and recent
  \Hersc~observations have shown that this is also true of dust
  (Cortese et al. 2010b; C12). Gravitational interactions might also
  either remove or add gas and dust to a galaxy. We can do a rough
  comparison of the detection rates in different environments using
  the cluster or cloud membership listed in Table~\ref{tab:sample}.
  Following C12, we split our sample into two sets: one set includes
  galaxies inside the Virgo Cluster and its outskirts ($N=47$
  galaxies) and the other set includes galaxies outside Virgo
  ($N=15$).  We detect 53\% (16/30) of the S0s and 29\% (5/17) of the
  ellipticals within the Virgo Cluster.  Outside Virgo, we detect 89\%
  (8/9) and 33\% (2/6) respectively (Table~\ref{tab:percents}).  The
  detection rates are therefore lower for galaxies within the Virgo
  Cluster, but this is not at a significant
  level.\reffnmark{fn:fis} 

\subsection{Residual Star Formation?}
\label{sec:nuvr}

We now discuss the NUV--$r$ color for our sample; for star-forming
galaxies, this is a useful measure of the specific star formation rate
(SFR$/M_*$) since the NUV traces recent star formation and the $r$
band is a proxy for stellar mass (e.g., Wyder et al.\ 2007). For Milky Way like extinction curves, the 
NUV is less affected by internal dust obscuration compared to the FUV,
but the NUV can also be affected by the presence of evolved hot
stars (the UV-upturn, e.g.. Greggio \& Renzini 1990; Yi et al. 2011).
Blue, star-forming galaxies occupy the region NUV--$r<3.5$ (Kaviraj et
al.\ 2007), the so-called `blue cloud', with the red, passive
ETGs found at NUV--$r >4.5$ in the ``red sequence''.
Although the color cut of NUV--$r >4.5$ is often used to select
galaxies on the red sequence, the most passive galaxies are
significantly redder than this. Saintonge et al. (2011) have found
that galaxies with NUV--$r>5$ have very little molecular gas and
Kaviraj et al. (2007), Schawinski et al.\ (2007), and Yi et al.\ (2005) have
showed that quiescent non-star-forming ETGs occupy the region above
the boundary NUV--$r=5.4$.

\begin{figure*}
\centering
\includegraphics[trim=30mm 0mm 30mm
  10mm,clip=true,width=14.5cm]{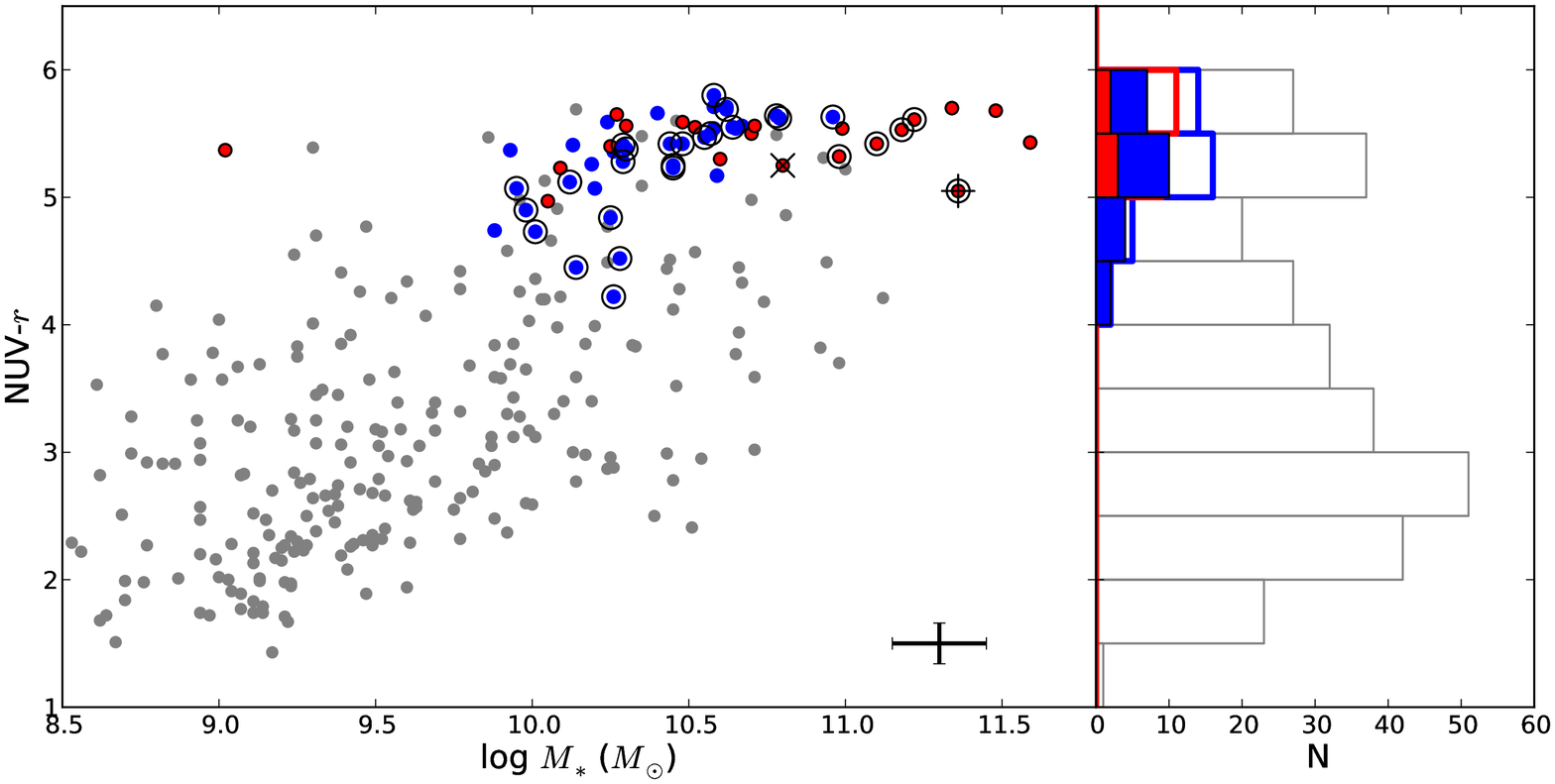}
  \figcaption{Left: the NUV--$r$ color vs. the stellar mass of
    the HRS sample (L. Cortese et al. in preparation) including our morphologically classified spirals
    (gray, C12), \SO (blue), and elliptical galaxies
    (red).  The \SO and elliptical galaxies detected with SPIRE are
    highlighted with the outer black circles.  Note that the galaxy
    NGC4486 is shown as detected in this plot but we do not see
    evidence for dust (above the synchrotron emission), NGC4486 is
    highlighted with ``+''. The well-known
    galaxy with no evidence for star formation (NGC4552) but with somewhat bluer NUV--$r$ color than
    the average elliptical in our sample is
    highlighted with a ``$\times$''.  UV photometry exists for 87\% of the
    ellipticals and 95\% of the S0 sample. Right: the
    distribution of NUV--$r$ colors for the observed samples split into
    spirals (gray), \SO (blue), and ellipticals (red; unfilled
    histograms) including detections for the early -types (shaded
    histograms).
    \label{fig:nuvr}}
\end{figure*}

Figure~\ref{fig:nuvr} shows the NUV--$r$ color plotted against stellar
mass for all the HRS galaxies taken from C12 and
L. Cortese et al.\ (in preparation). Sources detected at 250\,$\mu$m are
highlighted with the outlined black circles. All but two of our
early-type sample have NUV--$r >4.5$ and most cluster around the line
NUV--$r=5.5$. If NUV--$r<5.4$, the blue colors may be due to recent
star formation but they may also be caused by the UV upturn. NGC4552
is a famous example of this, with NUV--$r \sim 5.2$, even though there
is no residual star formation (Figure~\ref{fig:nuvr}). Excluding
NGC4552, 12 out of 18 S0s which lie below this boundary are detected
with \Hersc, and 2 out of 7 ellipticals. The two ellipticals are the
AGN NGC4486 (HRS138) and the (unusual) X-ray source NGC4636 (HRS241);
the NUV emission from these galaxies may be related to their central
activity.

As we detect dust emission in almost half of the galaxies, the UV
fluxes should be corrected for the dust attenuation since this will
cause the galaxies to be redder in NUV--$r$ than they truly are. Given
the complexity of disentangling the old and young stellar populations,
we leave this for future work and, given the sensitivity of
the NUV emission to internal dust extinction, we make no attempt to estimate
the SFRs from the NUV--$r$ colors.  Following
Schawinski et al. (2007), we simply classify galaxies as quiescent
where NUV--$r>5.4$ and possibly undergoing recent star formation (RSF)
where NUV--$r<5.4$. In this case (including sources not detected with
\textit{GALEX} but excluding NGC4486 and NGC4552), $\sim 40\%$ of our early-type
sample is in the RSF group, including 30\% of our ellipticals. This
agrees with the fraction of SDSS ETGs found with RSF (Schawinski et al. 2007).  Of course, our estimate may
be biased in a number of ways, since we have not corrected for dust
extinction and the colors of some of the galaxies in the RSF sample
may be due to LINER-type activity or to the UV upturn.

The most interesting result, however, that can be seen in
Figure~\ref{fig:nuvr} is the very large number of ETGs that lie on the
red sequence but are still detected by \Hersc.  It is often assumed
that the red sequence contains galaxies with virtually no current star
formation and little interstellar gas, but the detection with
\Hersc~of $\simeq 50\%$ of these galaxies clearly shows that they
contain significant reservoirs of interstellar material agreeing with
recent observations of CO in ETGs (Young et al. 2011).

\subsection{Dust Masses and Temperatures}
\label{sec:seds}

\begin{figure*}[ht]
\centering
\includegraphics[trim=5mm 6mm 5mm
  25mm,clip=true,width=13cm]{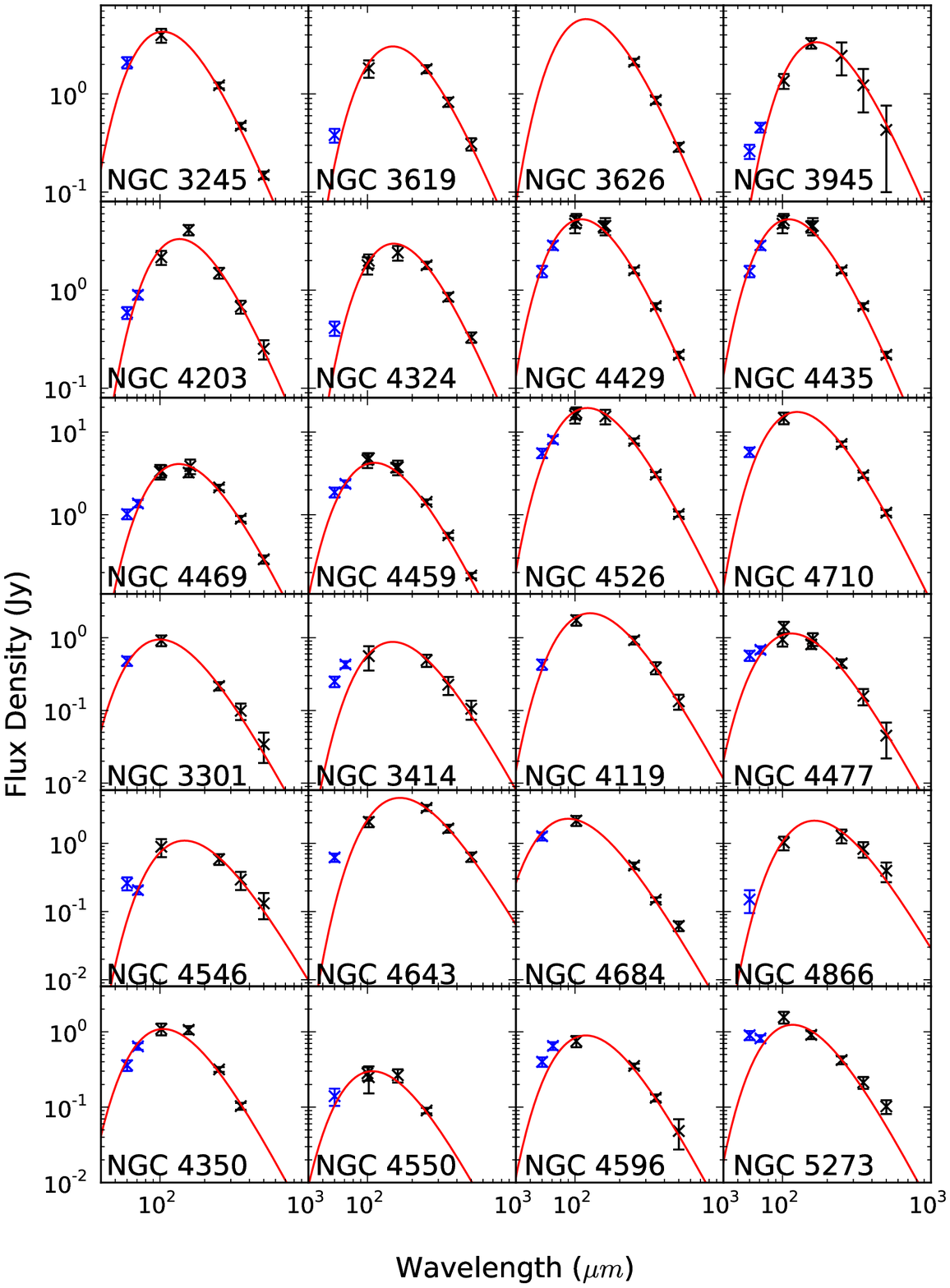}
  \figcaption{Spectral energy distributions of the 24 detected \SO
    galaxies in our sample.  The red solid line is the best-fit
    modified blackbody, blue points are fluxes used as upper limits,
    and black crosses are the photometry used for the cold dust component, including \Hersc~SPIRE fluxes and PACS, {\it IRAS},
    and {\it Spitzer} where available.  Errors
    shown also include calibration errors (L. Ciesla et al., in preparation).   The best-fit model parameters are
    provided in Table~\ref{tab:masses}.  \label{fig:sedearly}}
\end{figure*}

\begin{figure*}[ht]
\centering
\includegraphics[trim=5mm 3mm 5mm
  10mm,clip=true,width=13cm]{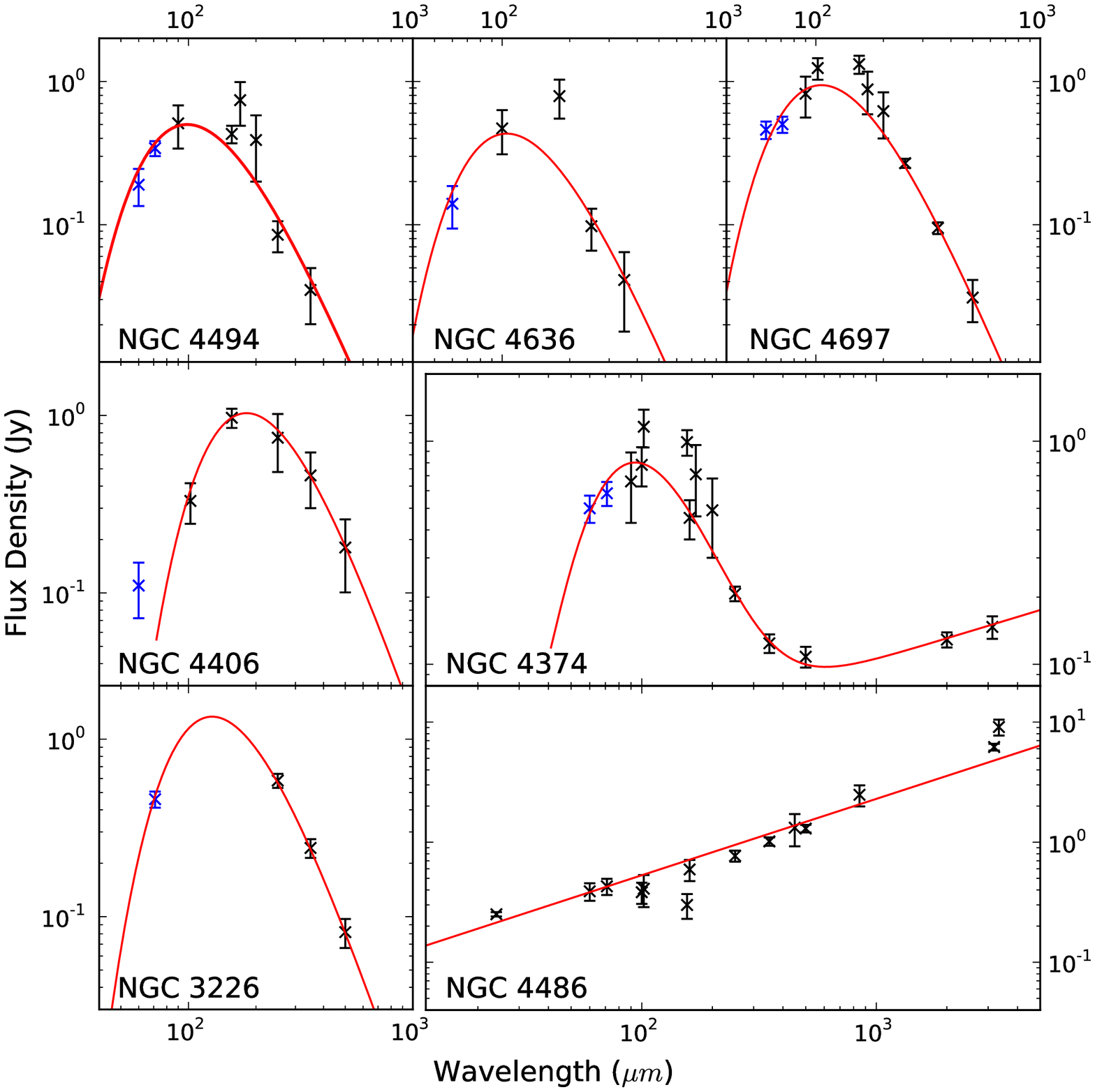}
  \figcaption{Spectral energy distributions of the seven detected
    ellipticals in our sample.  Blue points are those fluxes used as upper limits
    and black crosses are the photometry used for the cold dust component, including \Hersc~SPIRE and PACS, {\it IRAS}, {\it ISO},
    and {\it Spitzer} where available.   Radio fluxes are also shown
    by black points for NGC4374 and NGC4486 (M87). The
    red solid line is the best fit to the data and includes a modified
    (single temperature component) blackbody with $\beta=2$ and a
    synchrotron component.  Errors shown also include calibration
    errors (L. Ciesla et al. in preparation).  These galaxies
    are discussed in more detail in Appendix~\ref{sec:desc} and the best-fit
    model parameters are listed in
    Table~\ref{tab:masses}. \label{fig:sedegals}}
\end{figure*}

The global IR-submm SEDs of the
detected \SO and elliptical galaxies with available data from
60 to 500\,$\mu$m are presented in Figures~\ref{fig:sedearly} and
\ref{fig:sedegals}, respectively.   NGC4486 (M87) and NGC4374 are both bright
radio sources and their SEDs are well fitted by a power law at the
longer FIR wavelengths.  For the former, we find no evidence for dust
emission above the power-law synchrotron component (in agreement with
Baes et al. 2010).  In
NGC4374, we do see a strong excess from dust emission in the FIR.   We
fit the SEDs from 100 to 500\,$\mu$m with a modified blackbody model,
where
\begin{equation}
S_{\nu} = {{\kappa_{\nu}M_dB({\nu},T_d)}\over{D^2}},
\end{equation}
$M_d$ is the dust mass, $T_d$ is the dust temperature,
$B(\nu,T_d)$ is the Planck function and $D$ is the distance to the
galaxy. $\kappa_{\nu}$ is the dust absorption coefficient described by
a power law with dust emissivity index $\beta$, such that
$\kappa_{\nu} \propto \nu^{\beta}$. For $\beta = 2$ (typical of
Galactic interstellar dust grains), we use $\kappa_{350\mu \rm m} =
0.19\,\rm m^2\,kg^{-1}$ (Draine 2003). Although $\kappa$ is
notoriously uncertain, virtually all of our analysis relies on
$\kappa$ being constant between galaxies rather than on its absolute
value.

We assumed $D=17\,\rm Mpc$ for the Virgo Cluster (Gavazzi et al.\
1999) and 23\,Mpc for the Virgo B cloud; for all other galaxies, we
estimated $D$ from the recessional velocities using a Hubble constant
of $H_o = 70\,\rm\,km\,s^{-1}\,Mpc^{-1}$.

We found the best-fit solution by minimizing the chi-squared
($\chi^2$) difference between the model and the measured fluxes,
involving the model with the filter transmission functions. To
account for the uncertainties in the fluxes that are correlated
between the SPIRE bands (5\%), we used the full covariance matrix in
the $\chi^2$ calculation. The covariance matrix has diagonal elements
with the total variance of each band and non-diagonal elements
consisting of the covaraince between bands from the correlated calibration uncertainties.
We treated flux measurements at wavelengths
$\leq70\,\mu$m as upper limits for the cold dust emission as previous
works (Smith et al. 2010; Bendo et al. 2010a) found that there is a
significant contribution at these wavelengths from a warmer component
of dust. We estimated uncertainties in the dust temperatures and
masses using a bootstrap technique, in which we generated artificial
sets of fluxes from the measured fluxes and errors, and then applied
our fitting technique to each set of fluxes. The dust temperatures,
and therefore dust masses, depend on the choice of $\beta$;
$\beta=1.5$ would yield slightly higher dust temperatures for the
whole sample (Bendo et al.\ 2003) and therefore lower dust masses. 

The SED for NGC5273 (Figure~\ref{fig:sedearly}) is the only source which
appears to have an additional component present. Using the $\beta=2$
one-temperature component described above provides an adequate fit to
the photometry data if an additional synchrotron component is present.  The SED is also adequately fit with a one temperature dust component model with $\beta$ as a free parameter, $\beta=0.9$, or with a
two-temperature dust component model. To fit the data with the latter model would require an additional
component of $\sim 9\,\rm K$ dust which would increase the dust mass by an order
of magnitude. Both the $\beta$ = 2 + synchrotron model and $\beta=1$
model give a dust mass of $10^{5.45}\,M_{\odot}$ which is
somewhat lower than the dust mass estimated using the $\beta=2$ one-temperature dust component without synchrotron ($10^{5.65}\,M_{\odot}$) but within the errors (Table~\ref{tab:masses}).   As no millimeter data are
available for this source, we cannot rule out synchrotron
contamination, so in this analysis we have
chosen to stick with the $\beta=2$ one-component model to be
consistent with the other sources.

We have estimated upper limits to the dust masses for the galaxies
undetected by \Hersc~using the flux upper limit at 250\,$\mu$m
(including the noise on the image, an estimate of confusion noise and
background variance in the map as described in L. Ciesla et al., in preparation)
and the average temperature for the detected galaxies.

The dust masses and temperatures are presented in
Table~\ref{tab:masses}, and we show the range of dust temperatures,
masses, and stellar masses for the \SO and elliptical detected sample
in Figure~\ref{fig:temphist}. These range from $M_d = 10^{5.0-7.1}\,
M_{\odot}$ and $\rm 16-32\,K$ with mean values of $10^{6.14}\,
M_{\odot}$ and $24\,\rm K$, respectively (Table~\ref{tab:average}). The
stellar masses of these galaxies (C12) range from $10^{9.9-11.2}
\,M_{\odot}$ with an average of $\langle {\rm log}   M_* (M_{\sun})\rangle = 10.89$;
the average dust-to-stellar-mass ratio is ${\rm log} (M_d/M_*)=-4.33
\pm 0.14$.  The $M_d/M_*$ ratios are plotted in
Figure~\ref{fig:duststellar} and clearly shows that, at the same stellar
mass, there is a sharp fall in the dust-to-stellar-mass as we move
from spirals to S0s (by roughly a factor of 10), and that this fall
continues as we move from S0s to ellipticals (again, by a factor of
10).  The two ellipticals with anomalously high values given their
stellar mass are NGC3226 (HRS3) and NGC4406 (HRS150).  Both galaxies
show signs of a tidal interaction (see Appendix~\ref{sec:desc}),
suggesting that the dust may have been acquired as the result of an
interaction with a dust-rich galaxy.

\section{Dust, Stars and Gas}

In this section we compare the dust temperatures, masses, and
dust-to-stellar ratios with other \Hersc~results of late-type and
early-type galaxies.  First, the temperatures for the HRS ETGs are
generally higher than temperatures for the galaxies detected in blind
submm surveys or in other samples of nearby galaxies (see
Planck collaboration 2011; Dunne et al.\
2011) when corrected for different choices of $\beta$. We have tested whether the temperatures of ETGs are
systematically higher than those of late types by comparing
our dust temperatures with those of 71 Virgo galaxies in Davies et
al. (2012) who used an identical method.  We find that the temperatures of the early types are
systematically higher (with $U=1638$, $n_1=70$, $n_2=30$ and $P >99.99\%$\footnote{\label{fn:mann}Using
  the Mann--Whitney \textit{U} statistic, appropriate when comparing two groups
  where the underlying distribution of the data is not necessarily
  normal. The statistic assumes that the observations are independent
  and are continuous (i.e., able to be ranked) and is more robust than
  the Student's $t$-test.}), in agreement with the result that the
ETGs in the HRS have warmer IR colors than late types (Boselli et al.
2010a; see also Bendo et al.\ 2003; Engelbracht et al.\ 2010).

In Figure~\ref{fig:duststellar}, we compare the ratio of dust-to-stellar-mass for the HRS ETGs with the \Hersc~KINGFISH (Kennicutt et al. 2011) results based on 10
nearby early types (Skibba et al.\ 2011). They find warmer dust temperatures,
with a mean of $\langle T_d \rangle = 30\, \rm K$ (mostly due to the use of a lower
emissivity index $\beta= 1.5$ compared to this work), and their sample
has a lower average stellar
mass compared to the HRS. Though their dust masses are similar, the
dust-to-stellar-mass ratios are higher in KINGFISH than
for the HRS. This may be a result of the different
environments (a large fraction of the HRS galaxies are in a rich
cluster) and/or the selection of the sample (KINGFISH ETGs may include
unusual galaxies rather than the flux-selected ETGs in the HRS).

Rowlands et al.\ (2012) detected only $\simeq 5\%$ of ETGs in a blind submm
survey with \Hersc~at 250\,$\mu$m, finding a median dust mass for
their detected galaxies (in their lowest redshift bin) of
$10^{7.8}\,M_{\sun}$ with average dust temperature consistent with the HRS ETG sample.  Their dust mass is significantly higher than the dust masses of
the HRS ETGs and their mean value of their dust-to-stellar mass ratio
(${\rm log}(M_d/M_*)=-2.95$, Figure~\ref{fig:duststellar}) is also larger. Many of their sources have bluer
UV--optical colors than the HRS (with a significant fraction lying below NUV--$r<4$: their
Figure~15), suggesting that the larger amount of dust in the H-ATLAS
galaxies is associated with increased star formation (Rowlands et al.
2012). The simplest explanation of the differences is that the H-ATLAS
study is picking up the rare, very dusty ETGs, which the HRS misses,
partly because it is not large enough, and possibly also because it is
dominated by the ETGs in the Virgo Cluster. A possible local example
of these rare dusty ETGs is the local elliptical Cen A (T. Parkin et al.
2011, submitted). Since the depth of the SDSS images used for the morphological
classification in the H-ATLAS study made it impossible to distinguish
between ellipticals and S0s, it is also possible that some of the ETGs
detected in H-ATLAS are S0s or early-type spirals rather than ellipticals.

Rowlands et al. (2012) were also able to estimate the dust
masses of the ETG population {\em as a whole} using a stacking
analysis. They estimated a mean dust mass of $10^{6.3}\,
M_{\odot}$, with a mean dust temperature of 25\,K and a mean
dust-to-stellar-mass ratio of ${\rm log}(M_d/M_*) = -4.87$. This is in
reasonable agreement with the average parameters of the early-type
sample from this work and in particular, with the \SO sample and our
most dusty ellipticals (Figure~\ref{fig:duststellar}).

\subsection{Dust and stellar mass along the Hubble Sequence} 
\label{sec:morp}

\begin{figure}
\centering
\includegraphics[trim=5mm 10mm 5mm 15mm,clip=true,width=0.49\textwidth]{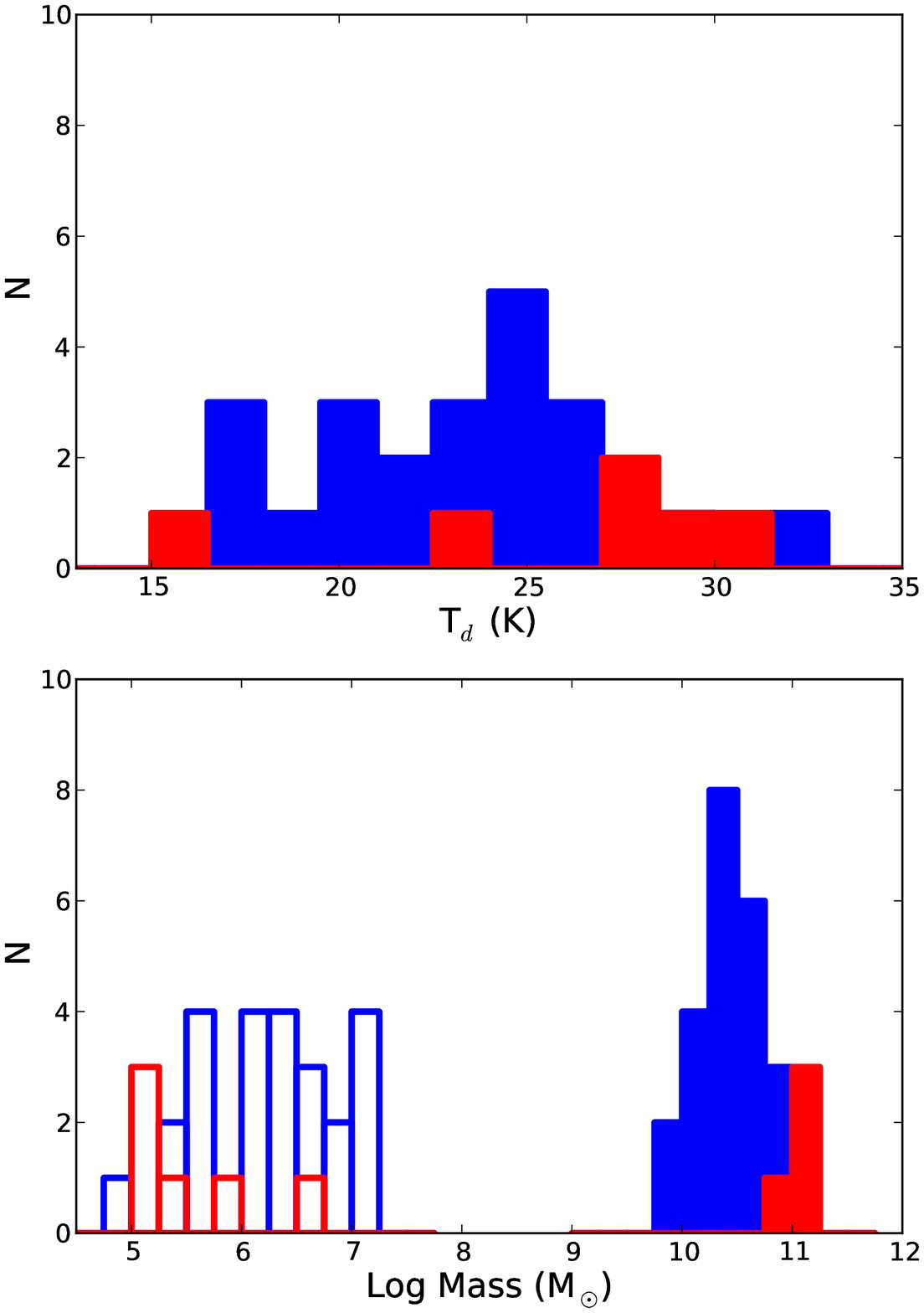}
  \figcaption{Top: histogram of the dust temperatures based on single modified blackbody fits;
      bottom: dust (left) and stellar masses (right shaded; L. Cortese et al.,
    in preparation) for the detected ellipticals
    (red) and \SO galaxies (blue)---see Table~\ref{tab:masses}.   The
    populations of the dust and stellar masses of the detected ellipticals and S0s are different with
    significance level $P>97.9\%$ ($M_d$) and $P>99.9\%$ ($M_*$) using a Mann--Whitney U-test. \label{fig:temphist}}
\end{figure}

The average dust mass for the \SO and elliptical galaxies detected by
\Hersc~is $10^{6.3}\,M_{\sun}$ and $10^{5.5}\,M_{\sun}$
(Figure~\ref{fig:temphist}, Table~\ref{tab:average}) with average dust
temperatures of 23.5 and 25.7\,K, respectively.  The dust masses for
the two morphological groups are significantly different (
Mann--Whitney $U$=117 with $n_1=6$, $n_2=24$ at $P>99.7\%$ for two-sided
test).\reffnmark{fn:mann} This does
not take into account the large number of upper limits so we have used
the Astronomical Survival Analysis programs (Feigelson \& Nelson
1985), implemented through the IRAF {\sc STSDAS statistics} package,
to compare the two samples. Using three different tests we find that
the probabilities that the elliptical and \SO dust masses are drawn from
the same population are 0.0005, 0.0006, and 0.004, and for $M_d/M_*$,
the probabilities are zero (Table~\ref{tab:twosampt});
these parameters for \SO and ellipticals in our sample almost certainly arise from
different distributions. We have used the Kaplan--Meier
estimator\footnote{\label{fn:km}It is likely that the censoring in
  this data set is not random: the censored data are a multiple of the
  noise in the data and we have observed to a flux limit in $K$ with shorter integration times
  for the \SO sample compared to the
  ellipticals. The censored data may have some randomization
  introduced by the SED-fitting technique since the SED can have
  different distributions but still give the same $M_d$. We note that
  the censored data could be biased.} to estimate the
mean\footnote{\label{fn:kmmean}The median of the distribution function
  returned by Kaplan--Meier estimator is well defined if the censored
  data are restricted to less than half of the sample.  Given the small
  sample size of the ellipticals which is dominated by upper limits
  (70\% of the data) we use the mean in this work.} for the two
samples, making the necessary approximation that the lowest upper
limit is actually a detection. With this approximation, we find that
the mean dust mass for the S0s is $10^{5.87}\,M_{\odot}$ and
the mean dust mass for the ellipticals is 
$10^{5.21}\,M_{\odot}$ (Table~\ref{tab:average}).

\begin{deluxetable*}{lcrlcrrc}[h]
\tabletypesize{\scriptsize}
\tablecaption{SED Parameters and Gas Masses\label{tab:masses}}
\tablewidth{0pt} 
\tablehead{ \colhead{HRS Name} & \colhead{Type} &
  \colhead{${\rm log}L_{\rm
      FIR}$} & \colhead{$T_d$} & \colhead{${\rm log} M_d$} & \colhead{{\rm log}$M_{\rm H_2}$} & \colhead{{\rm log}$M_{\rm HI}$} &\colhead{Rotator}  \\
  \colhead{} & \colhead{} &\colhead{($L_{\odot}$)} & \colhead{($\rm
    K$)} & \colhead{($M_{\odot}$)} & \colhead{($M_{\odot}$)} &
  \colhead{($M_{\odot}$)} & \colhead{} }
  \startdata
  3 & E &8.50&22.7 \plus 1.29&5.96 \plus 0.09 &   $<$7.13 &\multicolumn{1}{c}{\,.\,.\,.}& F\\
  7 & S0 & 9.20&27.7 \plus 1.5&6.14 \plus 0.05&  7.20 & \multicolumn{1}{c}{\,.\,.\,.}& F\\
  14 & S0& 8.57&29.0 \plus 1.9&5.41 \plus 0.12 &  $<$7.31 & 7.80\tablenotemark{3} & F\\
  22 & S0&8.42&19.6 \plus 2.0&6.26 \plus 0.21&  $<$7.02 &\multicolumn{1}{c}{\,.\,.\,.}& S\\
  43 & E& $<$7.56&\multicolumn{1}{c}{\,.\,.\,.} &$<$4.88& $<$7.28 & \multicolumn{1}{c}{\,.\,.\,.}&S \\
  45 & S0&9.04&19.6 \plus 1.0&6.89 \plus 0.07&  8.11 & 8.89\tablenotemark{3} & F\\
  46 & S0&9.38&24.4 \plus 4.0&6.65 \plus 0.17&  8.29 & 9.02\tablenotemark{3} & F\\
  49 & E & $<$7.88&\multicolumn{1}{c}{\,.\,.\,.}&$<$5.20 &  $<$7.25 &\multicolumn{1}{c}{\,.\,.\,.}& F \\
  71 & S0&8.84&17.1 \plus 1.1&7.05 \plus 0.18&  $<$7.28 & \multicolumn{1}{c}{\,.\,.\,.}& F\\
  87 & S0& 8.73&22.9 \plus 1.0&6.17 \plus 0.09& 7.91 & \multicolumn{1}{c}{\,.\,.\,.}& F \\
  90 & S0&$<$8.38&\multicolumn{1}{c}{\,.\,.\,.}&$<$5.70 & $<$7.31 & $<$7.68\tablenotemark{1}& F \\
  93 & S0&8.81&21.4 \plus 0.6&6.43 \plus 0.07& 7.63 & 9.44\tablenotemark{4}& F \\
  101 & S0& $<$8.41&\multicolumn{1}{c}{\,.\,.\,.}&$<$5.73& $<$6.92 & $<$7.39\tablenotemark{1} & F\\
  105 & S0 &$<$8.12&\multicolumn{1}{c}{\,.\,.\,.}&$<$5.44 & $<$7.16 & 8.71\tablenotemark{1} & F\\
  123 & S0&8.79&19.3 \plus 0.6&6.68 \plus 0.07& 7.72 & 8.73\tablenotemark{1}&  F\\
  125 & E & $<$7.85&\multicolumn{1}{c}{\,.\,.\,.}&$<$5.17& $<$7.2 & $<$7.84\tablenotemark{1} & F\\
  126 & S0& $<$8.48&\multicolumn{1}{c}{\,.\,.\,.}&$<$5.80&\multicolumn{1}{c}{\ \ \ \ \,.\,.\,.}& $<$7.31\tablenotemark{1} & F\\
  129 & S0& 8.51&27.6 \plus 0.6&5.47 \plus 0.04& $<$7.27 & $<$7.38\tablenotemark{1}& F \\
  135 & E & $<$8.85&\multicolumn{1}{c}{\,.\,.\,.}&$<$6.17& $<$7.62 & $<$8.18\tablenotemark{1} &S\\
  137 & S0& $<$8.91&\multicolumn{1}{c}{\,.\,.\,.}&$<$6.23& $<$7.29 & $<$7.38\tablenotemark{1}&  F\\
  138 & E &8.4&31.1 \plus 1.0&5.05 \plus 0.06& $<$7.16 & 8.96\tablenotemark{1} &S\\
  150 & E & 8.25&16.0 \plus 1.1&6.63 \plus 0.16& $<$7.4 & 7.95\tablenotemark{2} &S\\
  155 & S0& $<$8.71&\multicolumn{1}{c}{\,.\,.\,.}&$<$6.03 &  $<$7.54 & $<$7.64\tablenotemark{1}&  F\\
  161 & S0&9.18&26.3 \plus 0.6&6.26 \plus 0.04& 8.05 & $<$7.44\tablenotemark{1} & F\\
  162 & S0&9.13&25.0 \plus 1.2&6.34 \plus 0.10& 7.87 & $<$7.31\tablenotemark{1}& F \\
166 & S0&$<$9.06&\multicolumn{1}{c}{\,.\,.\,.}&$<$6.38 & $<$7.48 & $<$7.64\tablenotemark{1} & F\\
174 & S0&9.11&26.4 \plus 0.6&6.18 \plus 0.03&8.29 & $<$7.61\tablenotemark{1}& F \\
175 & S0 &$<$8.29&\multicolumn{1}{c}{\,.\,.\,.}&$<$5.61& $<$7.23 & $<$7.13\tablenotemark{1}& F \\
176 & S0&9.25&21.6 \plus 0.7&6.83 \plus 0.05&\multicolumn{1}{c}{\ \ \ \ \,.\,.\,.}& $<$7.64\tablenotemark{1}&  F\\
178 & E & $<$8.45&\multicolumn{1}{c}{\,.\,.\,.}&$<$5.77& $<$7.25 & $<$7.92\tablenotemark{1} &S\\
179 & E & $<$7.87&\multicolumn{1}{c}{\,.\,.\,.}&$<$5.19 & $<$7.16 & $<$7.92\tablenotemark{1} & F\\
180 & S0 & 8.49&25.2 \plus 1.6&5.68 \plus 0.11& 7.54 & $<$7.31\tablenotemark{1}&  F\\
181 & E &$<$7.53&\multicolumn{1}{c}{\,.\,.\,.}&$<$4.85& $<$7.31 &\multicolumn{1}{c}{\,.\,.\,.}& F\\
183 &E & ...&\multicolumn{1}{c}{\,.\,.\,.}&\multicolumn{1}{c}{\ .\,.\,.}&\multicolumn{1}{c}{\ \ \ \ \,.\,.\,.}&\multicolumn{1}{c}{\,.\,.\,.}& S \\
186&E & 8.28&29.3 \plus 1.3&5.08 \plus 0.11 & $<$7.35 & 8.26\tablenotemark{1} & F\\
200& S0 &9.7&23.9 \plus 0.6&7.03 \plus 0.04 & 8.62 & 7.13\tablenotemark{1}& F \\
202 & E &$<$7.61&\multicolumn{1}{c}{\,.\,.\,.}&$<$4.93 &\multicolumn{1}{c}{\ \ \ \ \,.\,.\,.}& $<$7.37\tablenotemark{1} & F\\
209 & S&8.26&19.8 \plus 0.8&6.08 \plus 0.11& $<$7.05 &\multicolumn{1}{c}{\,.\,.\,.}& F \\
210 & S0&7.94&26.8 \plus 1.8&4.97 \plus 0.08& $<$7.32 & 8.79\tablenotemark{1}&S \\
211 & E & $<$8.35&\multicolumn{1}{c}{\,.\,.\,.}&$<$5.67& $<$7.36 & $<$7.92\tablenotemark{1}& S\\
214 & E & $<$7.98&\multicolumn{1}{c}{\,.\,.\,.}&$<$5.30 & $<$7.33 & $<$7.79\tablenotemark{1} & F\\
218 & E& $<$8.35&\multicolumn{1}{c}{\,.\,.\,.}&$<$5.67 & $<$7.47 & $<$7.31\tablenotemark{1} & F\\
219 & S0& $<$8.41&\multicolumn{1}{c}{\,.\,.\,.}&$<$5.73 & $<$7.24 & $<$7.24\tablenotemark{1} & F\\
231& S0 & 8.37&24.5 \plus 1.3&5.64 \plus 0.08& 7.34 & $<$7.19\tablenotemark{1} & F\\
234 & S0& $<$8.55&\multicolumn{1}{c}{\,.\,.\,.}&$<$5.87& $<$7.33 & $<$7.44\tablenotemark{1} & F\\
235 & S0 & $<$7.95&\multicolumn{1}{c}{\,.\,.\,.}&$<$5.27 &$<$7.22 & $<$7.38\tablenotemark{1}  & F\\
236 & E &$<$ 8.44&\multicolumn{1}{c}{\,.\,.\,.}&$<$5.76 & $<$7.24 & $<$7.92\tablenotemark{1}& F \\
240 & S0 & $<$8.20&\multicolumn{1}{c}{\,.\,.\,.}&$<$5.52  & $<$7.30 & $<$7.44\tablenotemark{1} & F \\
241 & E&8.11&27.6 \plus 2.2&5.06 \plus 0.19& $<$7.02 & 9.0\tablenotemark{1}& S \\
243& S0 &8.94&17.5 \plus 0.7&7.08 \plus 0.07& 7.30 & 8.06\tablenotemark{1}& F  \\
245 & E& $<$8.08&\multicolumn{1}{c}{\,.\,.\,.}&$<$5.40 & $<$7.59 & $<$7.92\tablenotemark{1}&  F\\
248 & E& $<$7.53&\multicolumn{1}{c}{\,.\,.\,.}&$<$4.85& $<$7.30 & $<$7.92\tablenotemark{1} & F\\
250& S0 & $<$9.14&\multicolumn{1}{c}{\,.\,.\,.}&$<$6.46&\multicolumn{1}{c}{\ \ \ \ \,.\,.\,.}& $<$8.66\tablenotemark{1}& F\\
253& S0 &9.1&32.2 \plus 1.4&5.65 \plus 0.06& 7.63 & 8.22\tablenotemark{3}& F \\
258 & E & 8.48&27.4 \plus 0.7&5.46 \plus 0.04& $<$7.25 &\multicolumn{1}{c}{\,.\,.\,.}&F\\
260 & S0 &9.64&23.1 \plus 1.2&7.06 \plus 0.05& 8.72 & 7.76\tablenotemark{2}&F\\
269 & S0 & $<$8.95&\multicolumn{1}{c}{\,.\,.\,.}&$<$6.27& $<$7.23 & $<$7.44\tablenotemark{1}&F\\
272 & S0 &$<$9.16&\multicolumn{1}{c}{\,.\,.\,.}&$<$6.48& $<$7.23 & $<$7.92\tablenotemark{1} &F\\
286 & S0&8.61&17.7 \plus 1.2&6.73 \plus 0.17&\multicolumn{1}{c}{\ \ \ \ \,.\,.\,.}& 9.13\tablenotemark{1}&F \\
296 & S0 &8.43&24.8 \plus 2.6&5.65 \plus 0.20 & 7.26 &\multicolumn{1}{c}{\,.\,.\,.}& F\\
312 &E & $<$8.19&\multicolumn{1}{c}{\,.\,.\,.}&$<$5.51 & $<$7.46 &\multicolumn{1}{c}{\,.\,.\,.}& S \\
316 & E & $<$7.91&\multicolumn{1}{c}{\,.\,.\,.} & $<$5.23& $<$7.54 &\multicolumn{1}{c}{\,.\,.\,.}&F
\enddata
\tablecomments{Column 1: HRS ID;  Column 2: morphological classification;  Columns 3 and 4:  the FIR luminosity (of the cold dust component) and dust temperature; Columns 5: dust masses estimated using $\beta = 2$ and $\kappa_{350} = 0.19\,\rm m^2\,kg^{-1}$,  3\,$\sigma$ upper limits (L. Ciesla et al., in preparation) are quoted where the source is not detected using the mean dust temperature;  Columns 6 and 7: molecular masses estimated from CO (Young et al.\ 2011) and H{\sc I}  masses (references below); Columns 8: slow (S) or fast (F) rotator as defined by ATLAS$^{\rm 3D}$ (Emsellem et al.\ 2011). {\bf References}. \tablenotetext{1}{Gavazzi et al.\  (2003)} \tablenotetext{2}{Haynes et al.\ (2011).}
  \tablenotetext{3}{Springob et al.\ (2005).}
  \tablenotetext{4}{Noordermeer et al.\ (2005).}
\tablenotetext{a}{No SDSS data are available for HRS209 or HRS186.}}
\end{deluxetable*}

C12 has recently shown that the dust-to-stellar-mass ratio decreases
when moving from late- to early-type galaxies, but used a conservative approach by treating non-detections as upper limits.  Here we
incorporate the information in the upper limits on the non-detected sources and quantify the change as we move along
the Hubble sequence.  We estimate that the mean value\reffnmark{fn:kmmean}
of the dust-to-stellar-mass for the ellipticals is ${\rm log}
(M_d/M_*)= -5.83 \pm 0.1$ and for the S0s is $-4.42 \pm 0.1$ (Table~\ref{tab:average}).  

Following C12 (his Figure 5), we plot the dust-to-stellar-mass ratio against
morphological type in Figure~\ref{fig:hubble}. We split the late-type sample into the subgroups: Sa+Sab,
Sb+Sbc, Sc+Scd and Sd+Sdm. The figure also shows the strong drop in
mean dust-to-stellar mass ratio from early-type spirals to S0s (and
the further decline as we move to ellipticals).  Figure~\ref{fig:hubble}
suggests that the variation in this ratio is larger for early types than late types. This is particularly true for early-type spirals and \SO
galaxies with $M_d/M_*$ ranging by a factor of approx. 50.  Although
ellipticals probably have a similarly wide range in dust-to-stellar-mass, 
the high number of non-detections make it difficult to be sure
of this.

\subsection{The origin of dust in ETGs}
\label{sec:hot}

One of the big questions about dust in ETGs is its origin.  There are three main possible sources:
(1) the dust is formed in the old stellar population in the
atmospheres of AGB stars, (2) the dust is acquired externally as the
result of a galaxy merger or a tidal interaction, and (3) the dust has
the same rather uncertain sources as in late-type galaxies
(e.g., Morgan \& Edmunds 2003). In (3), the dust might be
continuously created by grain growth in the ISM or current star
formation via dust production in supernovae (e.g., Barlow et al.\ 2010;
Matsuura et al.\ 2011; Gomez et al.\ 2012), or it might be left over
from a more vigorous star-forming epoch.  The strongest prediction is
made by the first hypothesis, since this makes the clear prediction
that there should be a strong correlation between dust mass and
stellar mass.

To test point (1) we compare the FIR emission from the dust with the
starlight and the emission from the hot ISM.  If the source of the hot
gas is mass loss from stars, one would expect the X-ray luminosity
($L_X$) to correlate with the optical luminosity ($L_B$) which is true
for our sample (Appendix~\ref{sec:laws}, Equation~(\ref{eq:optlx})).
Similarly, if the source of the dust is stellar mass loss, we would
expect to see a correlation between dust mass and optical luminosity;
alternatively, a lack of correlation may indicate that mergers are
important for delivering dust.

The FIR luminosity, $L_{\rm FIR}$ (Table~\ref{tab:masses}) was
determined by integrating over the SED (Figures~\ref{fig:sedearly} and
\ref{fig:sedegals}). (This is an underestimate of the total IR
luminosity as the 70\,$\mu$m fluxes are treated as upper limits in our
SED fitting; we estimate that including a warm temperature component
to fit the MIR--70\,$\mu$m emission would contribute (on average) 3\% to
the FIR luminosity, though for some sources this can be up to 17\%---
see also Mu\~{n}oz-Mateos et al.\ 2009.) The mean ratio of FIR to
optical luminosity ${\rm
  log}(L_{\rm FIR}/L_B)$ for S0s is $-1.1\pm 0.1$ and $-2.2\pm 0.1$ for
ellipticals, therefore ellipticals have less FIR luminosity per unit
blue luminosity than S0s (also seen in an {\it IRAS} sample of isolated
galaxies - Lisenfeld et al.\ 2007). In Figure~\ref{fig:firtrends}, we
compare the $L_{\rm FIR}$ with the optical luminosity which does
indeed increase as $L_B$ increases, but the large scatter suggests
that dust may have been acquired from tidal interactions or mergers as
well as stellar mass loss. There is no evidence of a correlation
between $L_{\rm
  FIR}$ and $L_B$ (Figure~\ref{fig:firtrends},
Appendix~\ref{sec:laws}). We also find no correlation with $L_{\rm
  FIR}$ and the X-ray luminosity $L_X$.

\begin{figure*}
\centering
\includegraphics[trim=0mm 3mm 0mm 5mm,clip=true,width=0.62\textwidth]{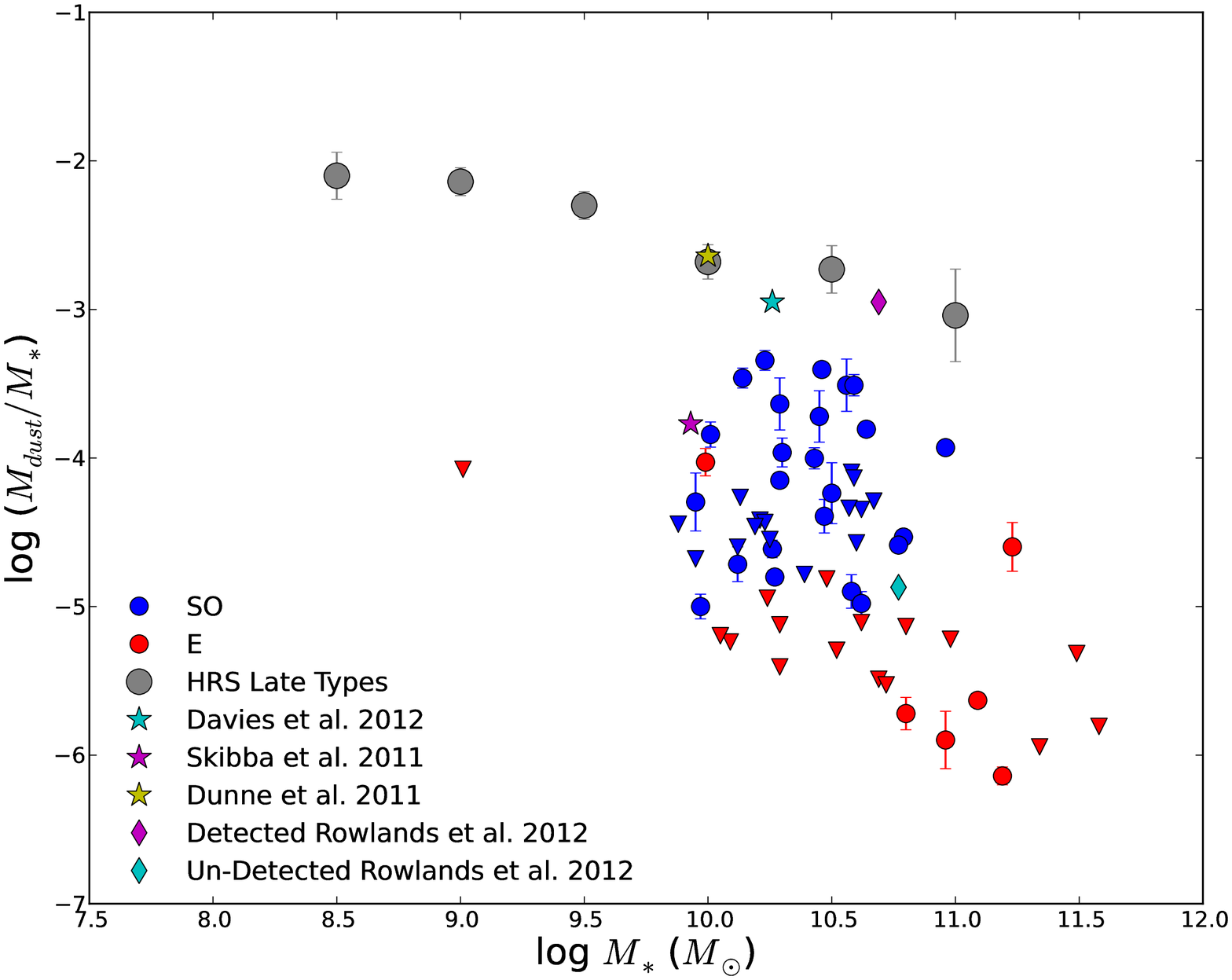}
\figcaption{Dust-to-stellar-mass ratio vs. stellar mass.  Blue points
  are the detected \SO galaxies, with red points for the ellipticals.
  Upper limits on the dust mass are shown with triangles.  The mean
  $M_d/M_*$ for the HRS spirals in each stellar mass bin is
  shown with gray circles (see C12, his
  Figure 5) with error bars indicating the error on the mean.  We also
  compare the mean $M_d/M_*$ results of late types in \hevics~(blue star; Davies et al.\ 2011)
  and H-ATLAS (yellow star; Dunne et al.\ 2011), early types with H-ATLAS (Rowlands
  et al.\ 2012 for detected (purple diamond) and undetected (cyan diamond) early-type populations), and early types with KINGFISH (purple star; Skibba et al.\
  2011).  The
  stellar masses are from L. Cortese et al.\ in preparation.  \label{fig:duststellar}}
\end{figure*}

\begin{deluxetable*}{lccclclcl}
\tabletypesize{\scriptsize}
\tablecaption{Mean Parameters for the Sample\label{tab:average}}
\tablewidth{0pt}
\tablehead{
\colhead{Type} & \colhead{$N$} &\colhead{$T_d$} & \colhead{${\rm log} M_d$} & \colhead{{\rm log}$M_*$} &\colhead{{\rm log}$M_d/M_*$}  & \colhead{{\rm log}$L_{X}$}   & \colhead{{\rm log}$M_{\rm H_2}$}  & \colhead{{\rm log}$M_{\rm HI}$}  \\
\colhead{} &\colhead{} & \colhead{$\rm (K)$} & \colhead{($M_{\odot}$)}
& \colhead{($M_{\odot}$)} & \colhead{} &\colhead{($\rm erg\,s^{-1}$)}
& \colhead{($M_{\odot}$)}& \colhead{($M_{\odot}$)}
}
\startdata
Detected sample: & & & & & & & &\\
Total & 31 & 23.9 \plus\ 0.8  & 6.12 \plus\ 0.12  & 10.49 \plus\ 0.07 &
--4.33 \plus\ 0.13 & 40.02 \plus\ 1.17 (15) &\multicolumn{1}{c}{.\,.\,.}& 8.40 \plus\ 0.17 (15)\\
E (+E/S0+pec) & 7  & 25.7 \plus\ 2.5  & 5.54 \plus\ 0.28 & 10.89 \plus\ 0.19
& --5.26 \plus\ 0.24 & 40.96 \plus\ 3.11 (5) & $<$ 7.21 (0) & 8.26
\plus\ 0.14 (5)\\ 
S0 (+pec) & 24 & 23.5 \plus\ 0.8 & 6.26 \plus\ 0.13 &  10.40 \plus\ 0.06
& --4.13 \plus\ 0.12 & 39.34 \plus\ 1.90 (10) & 7.84 \plus\ 0.25 (16)  &
8.47 \plus\ 0.19 (10)\\
\cmidrule{1-9}
Entire sample:& & & & & & & \\
 Total & 62  &.\,.\,. & 5.59 \plus\ 0.09 & 10.09 \plus\ 0.12 & --5.12 \plus\ 0.14 & 39.89 \plus\
 0.18 & 7.13 \plus\ 0.67 (57) & 7.57 \plus\ 0.11 (48)\\
 E (+E/S0+pec) & 22 &.\,.\,.& 5.21 \plus\ 0.09 & 10.66 \plus\ 0.13  & --5.83
 \plus\ 0.11 & 40.13
 \plus\ 0.24 & 7.02 \plus\ 1.29 (21) & 7.72 \plus\ 0.17 (15)\\
 S0 (+pec) &  39 &.\,.\,. &  5.87 \plus\ 0.13 & 10.38 \plus\ 0.04 & --4.42
 \plus\ 0.10  & 39.25 \plus\ 0.24 &  7.25 \plus\ 0.75 (36) & 7.58 \plus\
 0.13 (33)
\enddata
\tablecomments{Top: the mean parameters for the detected sample
  of early types. Bottom: the mean parameters estimated using
  survival analysis, including the 
  upper limits. The error quoted is the
  standard error of the mean.  The columns are as follows.\\
 Column 1: morphological type. 
 Column 2: the number of sources. 
Column 3: the average dust temperature. 
Column 4: the average dust mass (dust masses are estimated using a single temperature modified blackbody with $\beta = 2$ and $\kappa_{350} = 0.19\,\rm m^2\,kg^{-1}$).
Column 5: the average stellar masses using the optical colors (L. Cortese
et al., in preparation). 
Column 6: the average dust-to-stellar-mass
  ratio.
Column 7: average X-ray luminosity---the numbers in brackets
  give the number of sources with X-ray {\it and} 250\,$\mu$m
  detections. 
Column 8: the average molecular gas masses estimated from
  the CO data. 
Column 9: the average atomic gas masses estimated from H{\sc i} data.  
} \end{deluxetable*}

Since $L_{\rm FIR}$ depends strongly on the temperature of the dust,
it is instructive to compare the dust mass, $M_d$, with the optical
and X-ray luminosities (Figure~\ref{fig:firtrends}, Equations~(\ref{eq:mdso})
and (\ref{eq:mde})). We find that the dust mass increases weakly with
$L_B$ but there is no statistical correlation between the two.  The
same is true for $M_d$ and $L_X$. We compare the predicted
relationships for a model in which dust is replenished by stellar
mass loss from evolved stars and destroyed by dust sputtering
(Goudfrooij \& de Jong 1995; Figure~\ref{fig:firtrends}). The mass of
gas lost in these winds is estimated by $\dot{M}_{\rm gas} = 1.5\times
10^{-11}(L_B/{L_{\odot}})\,M_{\odot}\rm \,yr^{-1}$ (Faber \&
Gallagher 1976), which ranges from 0.005 to 0.75\,$M_{\odot}\rm \,yr^{-1}$
for the HRS ellipticals\footnote{We assume that the gas-to-dust ratio
  in the stellar winds of evolved stars is $\sim 150$.}. The estimates
of the final dust masses depend critically on the assumption of the
lifetime of the dust: dust destruction in a hot plasma probably occurs
on timescales of $\tau_d \sim 10^6$--$10^7\,\rm yr$ (Draine \& Salpeter
1979). Goudfrooij \& de Jong note that $\tau_d \sim 10^{7.5}\rm \,
yr$ is the maximum plausible destruction timescale since this
corresponds to electron densities in the hot X-ray plasma
$n_{\rm H}<10^{-3}\,\rm cm^{-3}$. In the figure we plot the predicted
relationships for values of ${\rm log}_{10} (\tau_d)$ from 6.5 to 7.5. 
The ETG sample detected by \Hersc~have dust masses well above the
theoretical curves, suggesting an alternative source of dust is needed
for most of the sources. Longer grain lifetimes for some of the galaxies
may also occur if a significant component of cool gas exists (see Section \ref{sec:cool});
in this case dust may reside in conditions similar to the Milky Way, with destruction
timescales of the order of ${\rm log}_{10}(\tau_d) = 8.5$.
Three elliptical galaxies have dust masses
consistent with the boundary marked by the maximum destruction timescale ${\rm log}_{10}
(\tau_d) = 7.5$.  Note that the figure shows that dust
originating from AGB-stellar mass loss (Clemens et al.\ 2010) could be
consistent with the upper limits estimated for 50\% of the ETGs.

\begin{figure*}
\centering
\includegraphics[trim=5mm 0mm 15mm 10mm,clip=true,width=0.6\textwidth]{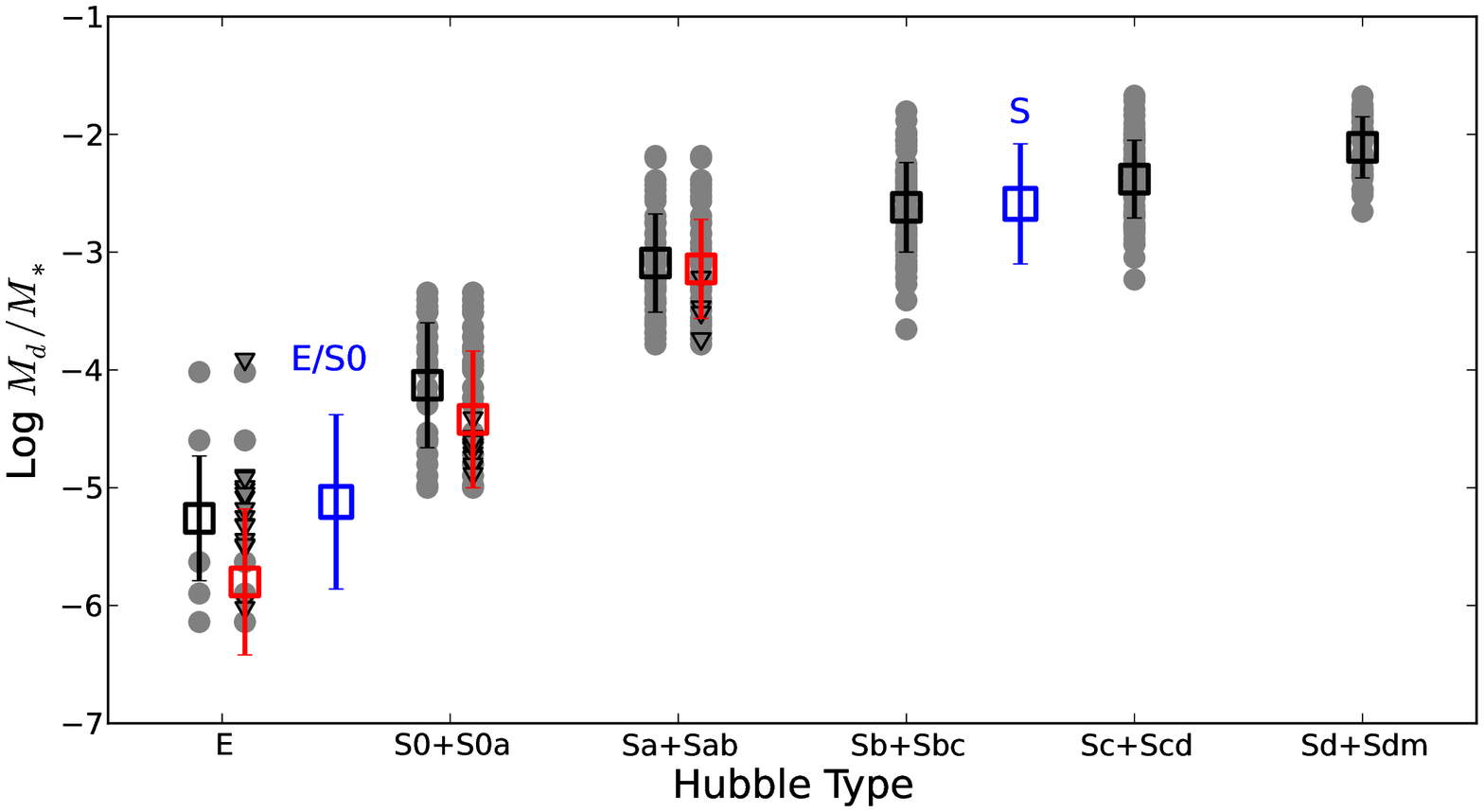}
  \figcaption{Dust-to-stellar-mass ratio vs. morphological type for the detected and undetected sources.
    Gray circles show the $M_d/M_*$ estimated for detected sources, with
    upper limits highlighted with gray triangles. The mean and standard deviation of $M_d/M_*$
    for the {\em detected} samples are highlighted with black boxes and error
    bars.  Offset from the detected sources, we also show the entire HRS sample in each Hubble type including both detected sources and upper limits.  The
    mean and standard deviation for the samples {\em including} the
    upper limits are shown with the red squares and
    error bars (see Table~\ref{tab:average}).  We also compare the
    mean and standard deviation of the complete early-type galaxies (\SO and E) and the complete sample of spiral
    galaxies (C12) in blue.  The stellar masses are from L. Cortese et al. (in preparation). \label{fig:hubble}}
\end{figure*}

\begin{figure*}
\centering
\includegraphics[trim=24mm 0mm 20mm
  10mm,clip=true,width=15cm]{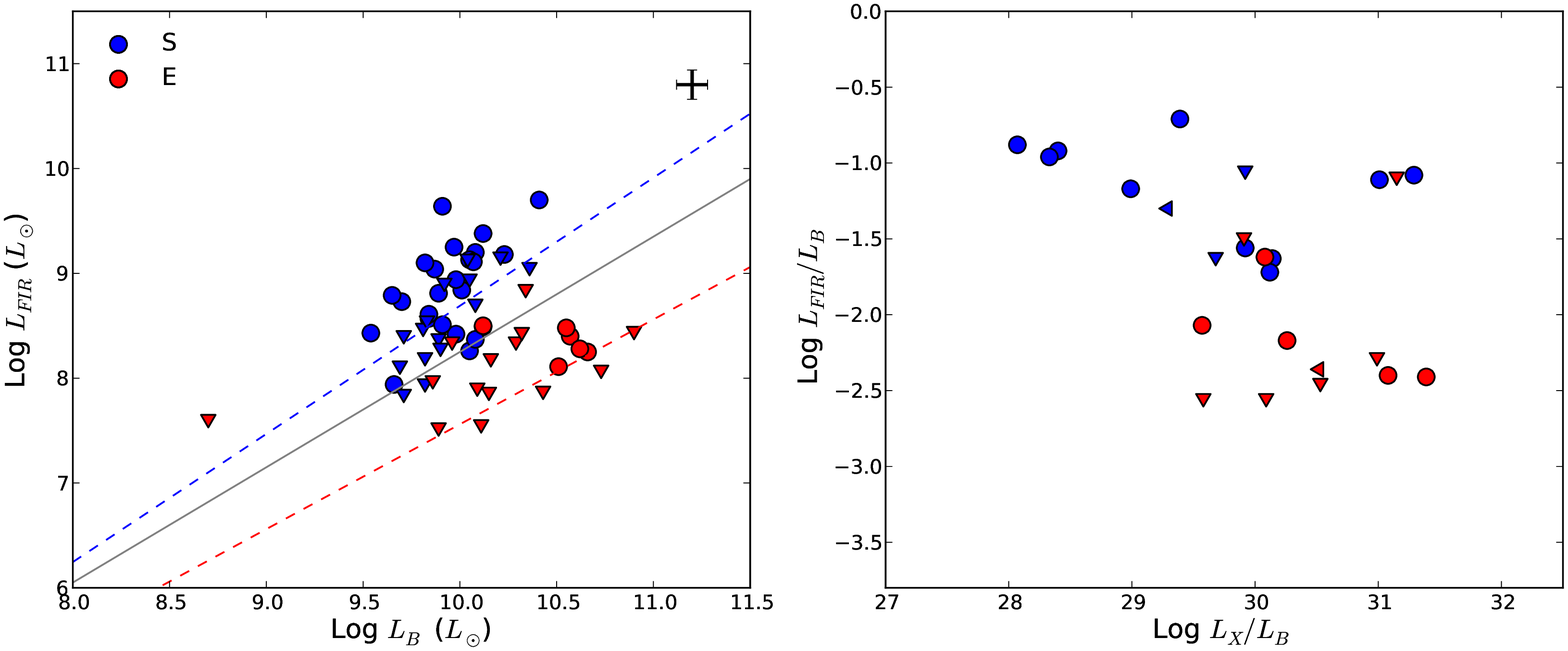}
\includegraphics[trim=24mm -3mm 20mm
  5mm,clip=true,width=15cm]{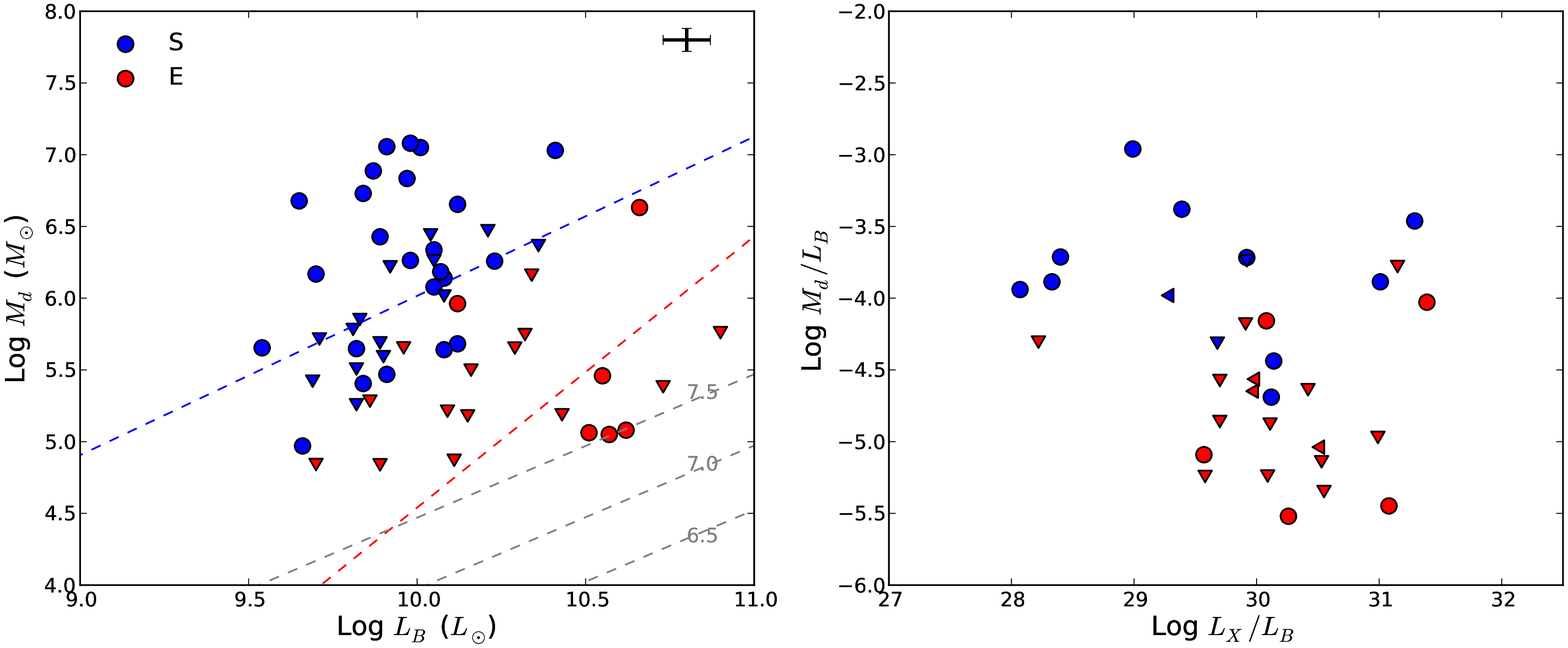}
  \figcaption{Top panel: the FIR luminosity from the cold dust component $L_{\rm FIR}(\rm cold)$
    vs. the $B$-band luminosity (left).  Blue points are \SO
    galaxies and red points are the ellipticals.Upper limits are
    plotted as downward-pointing triangles.  The gray solid line shows
    the expected relationship (arbitrarily normalized) for $L_{\rm FIR} \propto L_B^{1.1}$ if
    the FIR emission is produced by dust from stellar atmospheres.
    The best-fit line to the \SO sample including the upper limits is
    shown by the dashed blue line ($m=1.22,~c=-3.53, \sigma_r = 0.39$) as
    estimated from the Buckley James method.  The correlation
    parameters for ellipticals are shown by the dashed red line
    ($m=1.00,~c=-2.44,~\sigma_r = 0.3$). Right: the FIR emission vs. X-ray
    luminosity ($L_X$) normalized by the blue luminosity ($L_B$).
    Bottom panel: the dust mass $M_d$ vs. the $B$-band
    luminosity (left) and vs. X-ray luminosity (right).
    The ${\rm log}M_d$ vs {\rm
      log}$L_B$ relationship is described with $m=1.11, c=-5.08, \sigma_r = 0.58$ for
    the \SO (blue dashed) and $m=1.89, c=-14.40, \sigma_r =0.97$ for the
    ellipticals (red dashed).  The dashed gray lines represent the
    loci where dust is replenished by stellar mass loss and destroyed
    by dust sputtering in the hot gas (the labels represent the log of
    the dust destruction timescale assumed).  This model assumes a gas-to-dust ratio of 150.  The errors on the dust mass are similar to the symbol size for most galaxies, and a representative error bar is shown.
    \label{fig:firtrends}}
\end{figure*}

We conclude then that there is no strong statistical evidence of a correlation in FIR
luminosity or dust mass with blue- or X-ray luminosity. The dust masses
of the ETGs detected by \Hersc~are also larger than the estimates from
a model in which dust is produced from stellar mass loss from evolved
stars, even with the most generous assumption about how long the dust
grains will survive in the ISM.  Furthermore, if the dust in ETGs is
in equilibrium, with dust formation in old AGB stars balanced by dust
destruction via sputtering, we would also expect a small range in
$M_d/M_*$. Figures~\ref{fig:duststellar} and \ref{fig:hubble} show that
the variation in this ratio is much larger for early types than late
types; this is particularly true for S0 galaxies with $M_d/M_*$
ranging by a factor of $\sim$50.     Finally, another argument against
hypothesis (1) and the supernova origin of dust in hypothesis (3) is that if stellar mass loss was responsible for the
dust, the dust would be distributed in a similar way to the stars.
However, for five of the six ellipticals in which \Hersc~has detected
dust (excluding M87), there is evidence from the appearance of the dust in absorption
against the optical or near-IR continuum emission that the dust is
distributed in a different way from the stars (see Appendix~\ref{sec:desc}).

The evidence therefore points toward a merger origin of dust in this sample, 
this is supported by evidence from other recent studies that the cool ISM in ETGs 
may have been acquired as the result of gravitational interactions.  
For example, Davis et al. (2011) have used the misalignment between the kinematics 
of the stars and the gas in ETGs to conclude that at least one-third of ETGs have acquired much of their ISM by this means.

\begin{deluxetable}{lclll}
\tabletypesize{\scriptsize}
\tablefontsize{\scriptsize}
\tablecaption{Results of the Two-sample Tests\label{tab:twosampt}}
\tablewidth{0pt}
\tablehead{Sample& \colhead{\scriptsize Parameter} &\colhead{\scriptsize $P$(GGW)} & \colhead{\scriptsize $P$(LR)}
    & \colhead{\scriptsize $P$(PPGW)}
}
\startdata
S0 $\times$ E & $M_d$ & 0.0005 &  0.0006 &0.004  \\
& $M_d/M_*$ & 0.000 &  0.000 & 0.000  \\
& $M(\rm H_2)$ & 0.007&  0.009 &0.003  \\
& & & & \\
Virgo $\times$ not Virgo  & $M_d$ & 0.14 & 0.09& 0.14\\
& $M_*$ & 0.189 &  0.151 &0.165  \\
& $M_d/M_*$ & 0.055 &  0.082 &0.078  \\
& $M$(H{\sc i}) & 0.008&  0.011 &0.005  
\enddata
\tablecomments{The \SO versus elliptical (top) and Virgo versus non-Virgo galaxy samples tested for the
  probability they are drawn from the same population. The tests
  include GGW:
  Gehan's Generalized
  Wilcoxon test; LR: log-rank test; PPGW: Peto and Peto Generalized
  Wilcoxon test.  These are determined using the IRAF STSDAS task {\sc
  twosampt}.}
\end{deluxetable}

\subsection{A cool interstellar medium in early-type
  galaxies?}
\label{sec:cool}

In the last decade, it has become clear that a significant fraction of
ETGs do contain a cool ISM similar to that of spirals. Morganti et
al.\ (2006) detected H{\sc i} in 70\% of their sample of nearby ETGs
(see also Oosterloo et al.\ 2010; Serra et al. 2009; di Serego Alighieri et al. 2007) and detections of
molecular gas have been made in many (Sage et al. 2007; Lucero
\& Young 2007; Combes et al. 2007). A
number of studies indicate ongoing star formation in a significant
fraction of ETGs (Kaneda et al. 2005, 2008; Bressan et al. 2006;
Panuzzo et al. 2007, 2011; Temi et al.\ 2007a, 2007b) and the
detection of dust in 50\% of our sample strongly supports the
conclusion that a significant fraction of ETGs contain a cool ISM. 

We can compare the ISM in the ETGs with that in late types in two
other ways. First, using the measurements of the dust, atomic, and
molecular gas that exist for eight galaxies from our sample
(Table~\ref{tab:masses}), we estimate that the mean gas-to-dust ratio
is $\rm 10^{2.08\pm0.08}$, similar to the typical value for late-type
galaxies: $\sim$ 100--200.

Second, in late-type galaxies there is a tight correlation between FIR
and radio emission (Wrobel \& Heeschen 1988; Helou et al. 
1985; Devereux \& Eales 1989).  This is usually explained as the
result of a correlation between the FIR emission and the SFR, with FIR emission originating either from dust heated by young
stars (e.g., Vlahakis et al.
2007 and references therein; G. P. Ford et al. in preparation) or due to dust itself tracing the gas mass, which in turn fuels star formation (Rowan-Robinson et al. 2010; Bendo et al.\ 2011).  The radio emission is then also correlated with the SFR as it originates from relativistic electrons produced by
supernova remnants after the young stars die. Whether or not this correlation is truly caused by the SFR,
we can at least see whether the ETGs in our sample have a similar
ratio of FIR--radio as spiral galaxies. Sixteen ETGs are detected at
1.4\,GHz (see Figures~\ref{fig:spireearly} and \ref{fig:spireegals}) by
the Faint Images of the Radio Sky at Twenty centimeters (FIRST) radio survey (Becker et al. 1995). For these
sources we have estimated the value of the parameter, $q$, introduced
by Helou et al. (1986), defined as
\begin{equation} q= {\rm
    log}\left(
    {L_{\rm FIR}\over{3.75\times 10^{12}}\,\rm W\, m^{-2}} \right) - {\rm
    log} \left( {L_{1.4\,\rm GHz}\over{\rm W\,m^{-2}}}\right),
\label{eq:qir}
\end{equation}

where $L_{\rm FIR}$ and $L_{1.4\rm GHz}$ are the FIR and radio
luminosities, respectively.  Yun et al. (2001) found a median value for
nearby galaxies of $q = 2.64 \pm 0.02$,
with most star-forming galaxies having values of $q$ between 2 and 3.
Wrobel \& Heeschen (1988), and more recently Combes et al.
(2007) and Lucero \& Young (2007), have found similar values of $q$ for
ETGs in which star formation is occurring.

Two of our galaxies (NGC4636, NGC4374) have $q<1.8$, which
suggests they host radio-loud AGNs. Thirteen out of 16 sources with
FIRST detections in our HRS sample have $2.15<q<3.32$, similar to
star-forming late types. Nine of these sources also have CO detections
(Table~\ref{tab:masses}) and therefore 70\% of the galaxies which the
$q$ values suggestive of star formation also have a reservoir of
molecular gas. These galaxies lie below the `quiescent' UV--optical
boundary defined as NUV--$r<5.4$ (Section \ref{sec:nuvr}) and a literature
search reveals evidence of residual star formation in seven out
of the nine galaxies, with signatures including mid-IR emission
(Panuzzo et al. 2007; Shapiro et al.\ 2010), line diagnostics
(Sil'chenko et al. 2010; Crocker et al.\ 2011), and UV
emission (Cortese \& Hughes 2009).

Converting the $q$ ratio into SFRs for ETGs is extremely complex due
to the difficulty in disentangling the AGN component, other processes
that may be heating the dust, and the thermal contribution to the
radio emission. If the $q\sim 2$ ratios for those galaxies with
molecular gas are an indication of star formation, then there are not
only cool gas and dust in these galaxies but also stars forming at a
measurable rate for at least 15\% of the sample. We defer a full
analysis of SFRs for future work, though we note that
a literature search reveals SFRs ranging from 0.03 to 0.40\,$M_{\odot}\rm \,yr^{-1}$ 
in 23\% of the HRS ETG sample (see references listed above).

In summary, although the mass of the ISM in ETGs is less than the mass
of the ISM in late types, observations of the radio continuum, CO, and
H{\sc i} are consistent with the ISM in ETGs being quite similar to that
found in late types. The only difference is that, on average, the dust
in ETGs has a temperature that is a few degrees higher, possibly due to the
dust grains in early types being exposed to a more intense interstellar
radiation field or from the more centrally distributed dust emission
in ETGs (e.g., Sauvage \& Thuan 1992; G. Bendo et al. 2011, submitted).

\section{Discussion}
\label{sec:disc}

\subsection{Evidence for dust-depleted disks in S0s}
\label{sec:dep}
The large fall in the dust-to-stellar-mass ratio between early-type
spirals and \SO galaxies (Figures~\ref{fig:duststellar} and
\ref{fig:hubble}) suggests that S0s contain a smaller mass of ISM per
mass of stars than early-type spirals (i.e., those with Hubble type
ranging from Sa to Sbc).  However, the bulge-to-total mass ratio is known to
be larger for S0s than early-type spirals: is the fall in the
dust-to-stellar-mass ratio we observe here simply a reflection of the
more massive bulges of S0s?  We can test this idea by making the
assumptions that (1) bulges do not contain dust and (2) the
dust-to-stellar-mass ratio is the same for all disks. With these
assumptions, to explain the factor of 10 decrease in the
dust-to-stellar-mass ratio of S0s relative to early-type spirals, the
bulge-to-total stellar-mass ratio must change by a very large
factor. If the bulge-to-total stellar-mass ratio is 0.1 for early-type
spirals, it must be 0.91 for S0s.  For larger values of the
bulge-to-disk stellar-mass ratio for the early-type spirals (Sa--Sc), even
larger ones are required for the S0s.  Weinzirl et al. (2009) have
found that the proportion of the total mass of a galaxy that is in the
bulge increases from $\sim$0.1 for Sbc galaxies to $\sim$0.2 for \SO
galaxies (their Figure~14).  In larger samples, Graham \& Worley (2008) and
Laurikainen et al. (2010) find
values for this ratio of approximately 0.1--0.2 for Sab-Sbc galaxies and
0.3--0.5 for S0 galaxies (their Figure~4), a somewhat larger change but still too
low to explain the different dust-to-stellar-mass ratios.

\begin{figure}
\includegraphics[trim=8mm 5mm 0mm 5mm,clip=true,width=0.49\textwidth]{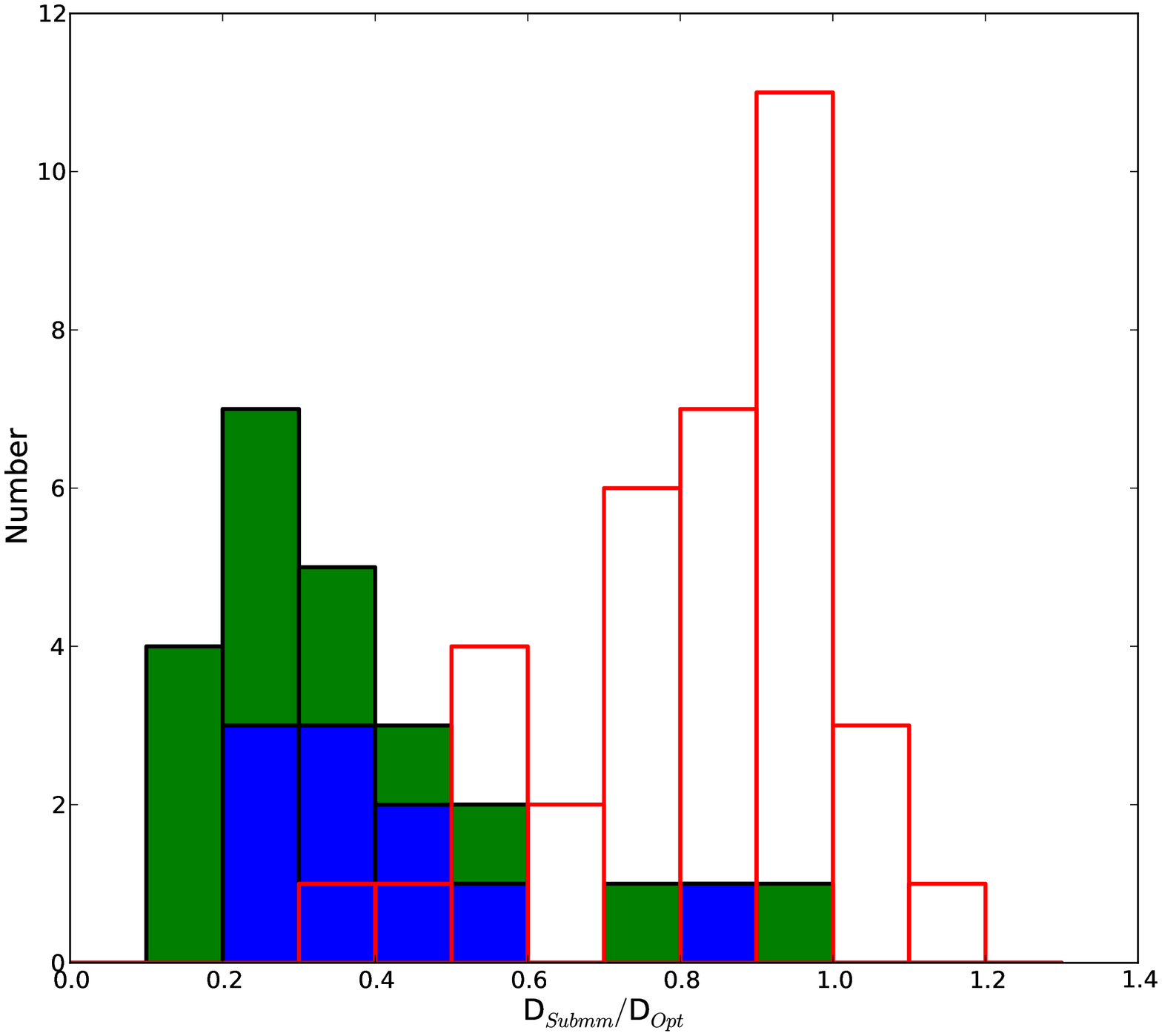} 
\figcaption{Ratio of the
    submm and optical sizes for the HRS sample.  The red histogram
    shows the distribution of $D_{\rm submm}/D_{\rm opt}$ for the
    early-type spiral galaxies (Sa-Sbc) with the same color cut as the S0 and ellipticals, i.e., $K<8.7$ (the original sample of early types is presented in Cortese et al.\ 2010b and C12). The green
    shaded histogram shows the distribution for the S0s in this paper,
    with the blue shaded histogram showing the S0s that are not in one
    of the Virgo clouds or the outskirts of Virgo (Boselli et
    al. 2010b).
    \label{fig:dopt}}
\end{figure}

Another way to investigate this is to compare the ratio of the submm
diameter to the optical diameter (e.g., Cortese et al.  2010b), where
the optical diameter is likely to be a good measure of the size of the
disk.  We measured this ratio for two samples: (1) all of the S0
galaxies and (2) all early-type spiral galaxies in the HRS with $K<8.7$ (the
limit used for ETGs).  For the optical size of the galaxy we used
$\theta_{\rm major}$, the diameter along the major axis measured to
the standard B 25th mag arcsec$^{-2}$ isophote ($D(25)$). For the submm size, we
used the same definition, finding the ellipse that provided the best
fit on the 250\,$\mu$m image to an isophotal brightness of
$6.7\times10^{-5}\rm \,Jy \,arcsec^{-2}$.  
Figure~\ref{fig:dopt} shows
the distributions of $D_{\rm submm}/D_{\rm opt}$ for the \SO galaxies and
the early-type spirals with the same $K$-band magnitude selection;
these populations are significantly different ($P<0.001$
level\footnote{\label{fn:kol}Using the Kolmogorov--Smirnov two-sample
  test.}).  
We do not apply a correction for beam smearing, which increases 
the angular sizes of the objects, an effect which is biggest for the objects with smallest angular size.
If a correction was made for this bias, the difference between \SO and early-type spiral populations seen
in Figure\,{\ref{fig:dopt}} would increase.
We have also compared the submm--optical sizes for the S0 galaxies outside of Virgo in
Figure~\ref{fig:dopt}. The values for these are still clearly different
from the early-type spirals, strongly suggesting that the
dust-depleted disks of S0s are not caused by a current
cluster-environmental effect.

Both approaches imply that the dust-to-stellar-mass ratio is lower for
the disks of S0 galaxies than the disks of early-type spirals,
suggesting that the disk of an S0 contains a smaller mass of ISM than
the disk of an early-type spiral with the same stellar mass.  This is
not an entirely surprising result, since for the last 40 years the
working definition of an S0 galaxy is that it is a galaxy with a disk
but no sign of spiral arms or, if it is an edge-on galaxy, of a dust
lane (Sandage 1961).  Nevertheless, even if the qualitative result is
not surprising, \Hersc~has allowed us to determine how little ISM the
disks of S0 galaxies do contain.

\subsection{Ellipticals and S0s or slow rotators and fast rotators?}
\label{sec:nature}

So far in this paper, we have classified ETGs based on their optical
morphology into ellipticals and S0s. Based on the ATLAS$^{\rm 3D}$
study of the stellar kinematics of 260 ETGs, Cappellari et al. (2011)
have argued that a physically more meaningful way is to divide ETGs
into slow rotators and fast rotators.  Emsellem et al.\ (2011) found
that 66\% of the galaxies in ATLAS$^{\rm 3D}$, which are
traditionally morphologically classified as ellipticals, have disk-like kinematics and are therefore fast rotators. Given the similarity in the kinematics, the suggestion is then that early-type fast rotators are part of the same evolutionary sequence as the late types, but are at different stages of their evolution.  Slow rotators are defined as galaxies with bulge-like properties, likely to be the end-point of systems
that have undergone complex merger histories; these are massive, lie on the red sequence, and contain very little cool interstellar material.

\begin{figure}
\includegraphics[trim=0mm 0mm -29.5mm 5mm,clip=true,width=0.49\textwidth]{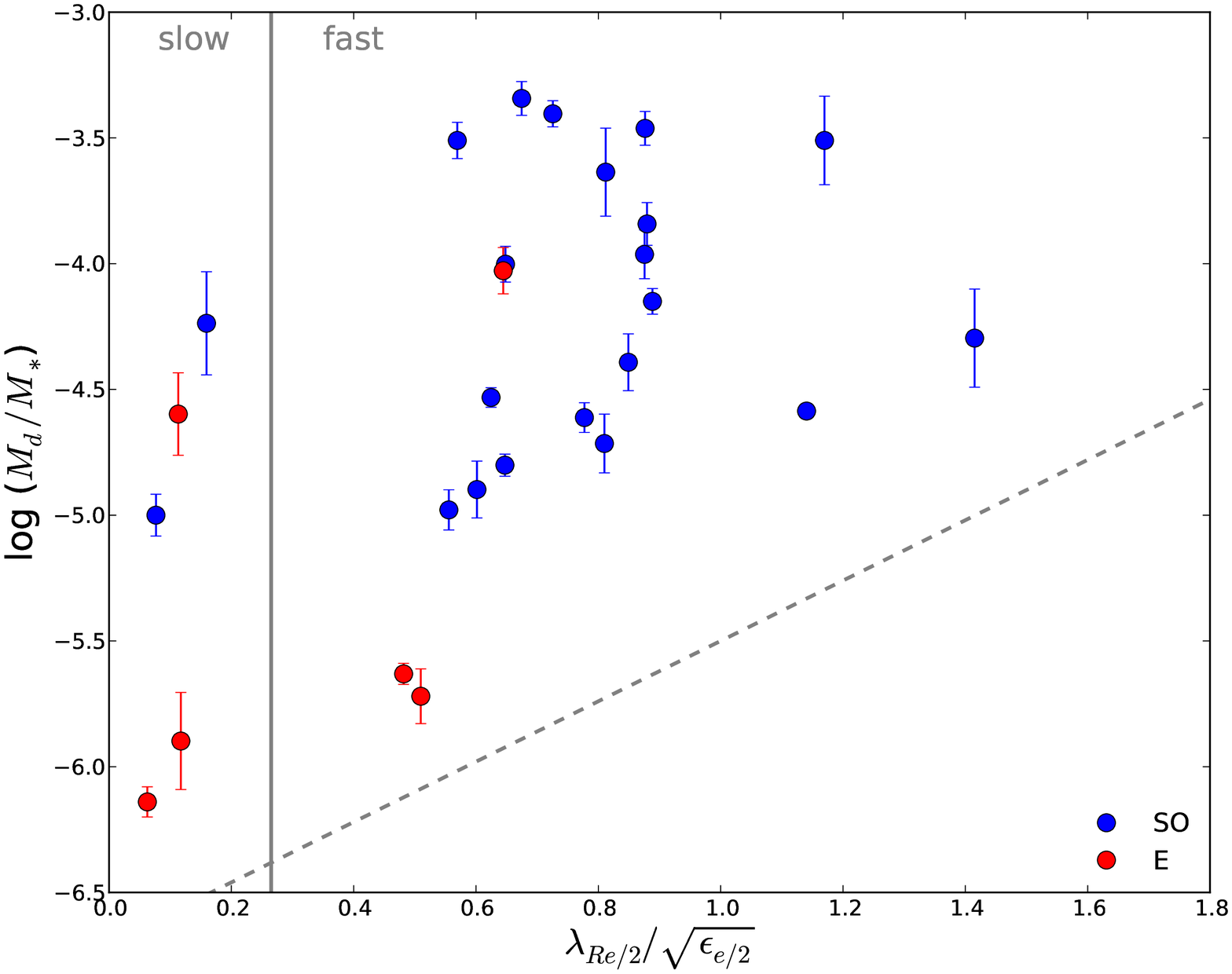} 
\includegraphics[trim=2mm 12mm 0mm 5mm,clip=true,width=0.49\textwidth]{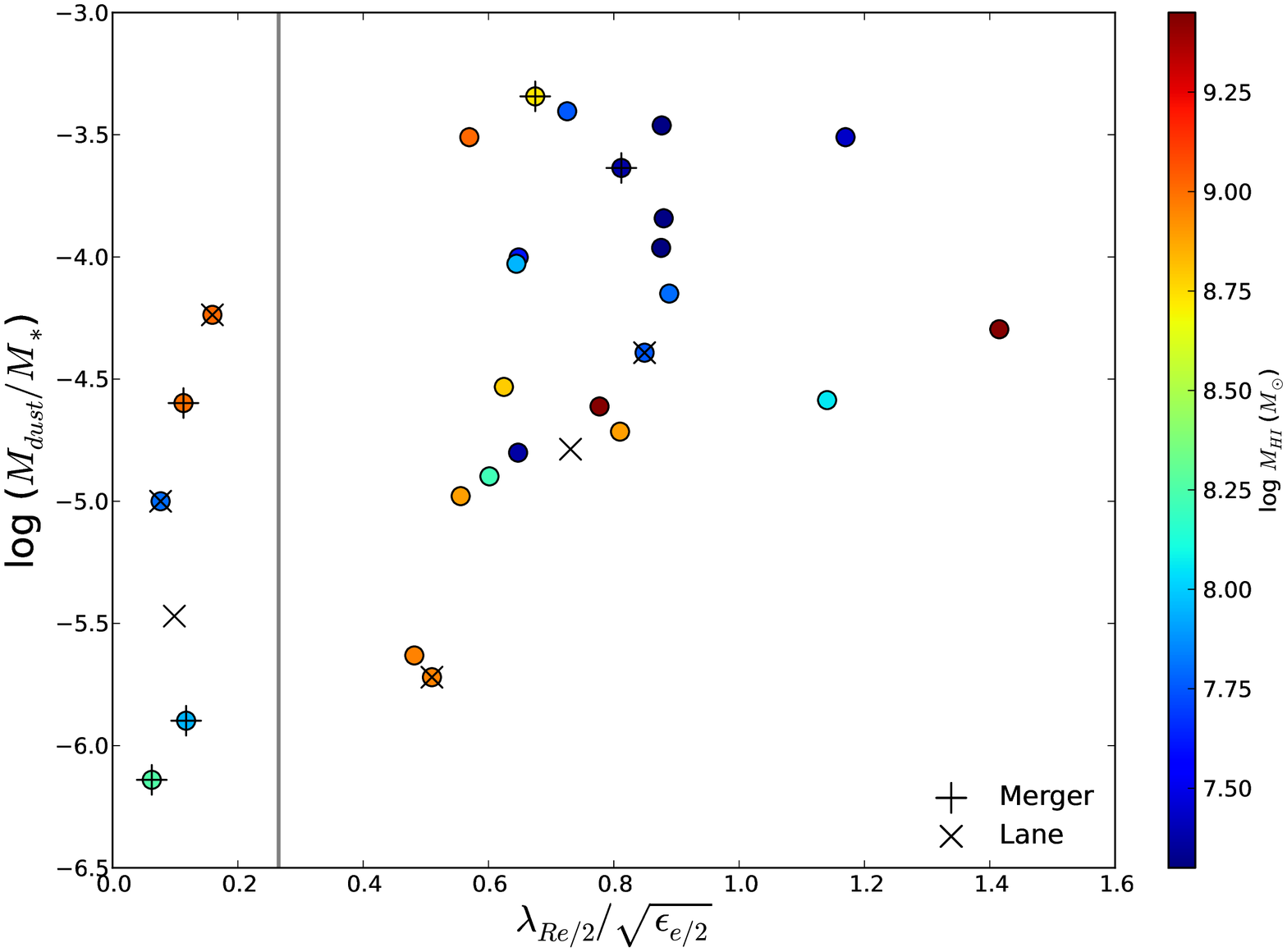} 
  \figcaption{Top: the dust-to-stellar-mass ratio vs. the kinematical
    parameter $\lambda_{Re/2}/\sqrt{\epsilon_{e/2}}$, where
    $\epsilon_{e/2}$ is the ellipticity measured within an aperture
    defined at $0.5R_e$ and $R_e$ is the effective radius of the
    galaxy.  $\lambda_{R}$ is a proxy for the angular momentum per
    unit stellar mass and the $\sqrt{\epsilon_{e/2}}$ term accounts for the shape of a galaxy since a more flattened galaxy would be expected to have a stronger anisotropy.     Blue points are
    \SO galaxies and red points are 
    ellipticals. The gray solid line marks the region defined as the
    boundary between fast and slow rotators (Emsellem et al.\ 2011),
    i.e., between galaxies with disk-like rotation and bulge-like
    kinematics. Bottom: as above but color-coded with respect to the H{\sc i} mass.    Galaxies which are known to be mergers or have obvious dust lanes in the optical are indicated with $+$ (mergers) and $\times$ (dust lanes).     Galaxies without filled circles  are the {\it Herschel}-detected galaxies that have no corresponding H{\sc i} observations.  Many of the galaxies at the lower end of the H{\sc i} mass scale are upper limits (Table~\ref{tab:masses}).    The gray dashed line roughly marks the detection limit of the survey and note that NGC4486 (HRS183) is not shown in these plots.
    \label{fig:rotator}}
\end{figure}

Eleven (18\%) of the ETGs in our sample are slow rotators, including
nine ellipticals (39\% of the ellipticals) and two S0s (5\%). In
Figure~\ref{fig:rotator}(a), we plot dust-to-stellar-mass ratio versus
the quantity $\lambda_{R_{e/2}}/\sqrt{\epsilon_{e/2}}$, where
$\epsilon$ is the ellipticity of the galaxy and $R_e$ is the effective
radius. This parameter is suggested by the ATLAS$^{\rm 3D}$ team as
appropriate to separate out the slow and fast rotators since it is a
good measure of the projected stellar angular momentum per unit mass
(Emsellem et al. 2007). There is no sign of a correlation between
the two parameters for galaxies in our sample. Indeed, with
\Hersc~we have detected dust emission from 50\% of the slow rotators,
including four ellipticals, which is not very different from the
detection fraction for all the ETGs (Table~\ref{tab:percents}).

The fact that there is a clearer distinction in the dust properties of
the ETGs when they are put in morphological classes
(Figures~\ref{fig:duststellar} and \ref{fig:hubble}) than when they are
put into kinematic classes does not mean that the kinematic division is not
physically meaningful. It is possible that the morphological
classification is more connected to the mass of ISM in a galaxy than
the kinematic classification, although the kinematic definition may be
telling us more about the evolutionary state of the galaxy. As
\Hersc~detects a number of slow rotators and the lack of
correlation seen in Figure~\ref{fig:rotator}(a), this may be additional evidence that
the cool ISM in ETGs is acquired by random gravitational encounters.  Again, the amount of dust
may not be connected with the evolutionary state of
the galaxy.  In Figure~\ref{fig:rotator}(b), we show the same diagram
but this time labeling the galaxies by their H{\sc i} mass.
There is a tendency for galaxies with larger atomic gas masses to have
higher dust-to-stellar mass ratios. Since the 21\,cm morphologies have
often been adduced as evidence for an external origin of the ISM
(e.g., Serra et al. 2009), this figure may be additional evidence that
the mass of the ISM is rather unconnected to the evolutionary history of the stellar populations in
the galaxies.

Finally, we note that connections between the dust properties and
stellar kinematics of the ETGs may become evident in a larger sample.
The HRS sample of ETGs is a statistically complete
quasi-volume-limited sample, but the volume contains the rich
environment of the Virgo Cluster and is therefore not a representative
sample of the local universe. The HRS ETGs also cover a limited
range of stellar mass (Section \ref{sec:res}, Figure~\ref{fig:kbandhist}). An
important future project would be to observe the dust in a sample of
ETGs containing more galaxies in low-density environments and with a
wider range of stellar mass.

\subsection{Galaxy evolution}
\label{sec:evolution}

We can use the results from the \Hersc~observations to make an
inference about the evolution of early types.  On the assumption that the dust we observe is delivered to the galaxies externally, we can make a rough estimate of the effect such an
interaction would have on the stellar mass of the ETG. We use the mean
values of $M_d/M_*$ shown in Figure~\ref{fig:hubble}
(Table~\ref{tab:average}) for the different morphological classes,
($\rm 10^{-5.9}$ for ellipticals, $\rm 10^{-4.4}$ for S0s, and $\rm
\sim 10^{-3.0}$ for spirals). On the assumption that all the dust we
detect in an ETG is the result of a past merger with a spiral, the
ratio of the mass of the spiral to the original ETG must be $\sim 0.05$
for an elliptical and $\sim$0.1 for an S0. These estimates suggest
that such interactions are only minor mergers and do not represent
significant galaxy-building events. The estimates are also upper
limits because the dust may have been acquired by a tidal interaction
without the galaxies necessarily merging, and some of the dust may
also have been produced inside the galaxy.

\section{Conclusions}
 \label{sec:conc}

 We present the FIR and submm observations with \Hersc~of 62
 ETGs, including 23 ellipticals and 39 \SO galaxies. We find the
 following results:

\begin{enumerate}
\item We detect 24\% of the ellipticals and 62\% of the S0s.  The
  optical and X-ray luminosities of the detected sources show we are
  detecting the most massive ellipticals, though the S0 sample is
  representative of the general population.  
  Of the ten `pure' ellipticals in our sample only one is detected by \Hersc\ (M86),
  while six of the eleven `impure' ellipticals are detected.
  The detection rate for
  the ETGs outside Virgo appears to be higher than for those inside
  the cluster but this is not statistically significant.

\item The mean dust masses for the detected galaxies are ${\rm log}M_d = 6.3
  \pm 0.1$ and $5.5 \pm 0.3\,M_{\sun}$ for the \SO and E
  population, with average dust-to-stellar-mass ratios of ${\rm
    log}(M_d/M_*)= -4.1 \pm 0.1$ and $-5.3 \pm 0.2$.     Including the upper limits, the average dust masses for the
  detected {\em and} undetected \SO and elliptical sources are ${\rm
    log}M_d (M_{\odot}) =  5.9 \pm 0.1$ and $5.2 \pm 0.1$, with dust-to-stellar-mass
  ratios of ${\rm log}(M_d/M_*)= -4.4 \pm 0.1$ and $-5.8 \pm 0.1$.

\item  The mean dust temperature for the ETGs {\em detected} by \Hersc~is
  $\sim 24\,\rm K$, warmer than the dust in
  late-type galaxies.    The {\em entire} early-type sample, including non-detections, has a dust mass of ${\rm log}M_d = 5.6
  \pm 0.1$ and ${\rm log}(M_d/M_*)= -5.1 \pm 0.1$.  In comparison, the average
  dust-to-stellar mass ratio for spiral galaxies in the HRS is $-2.59 \pm 0.03$.
   The dispersion in the dust-to-stellar mass ratio is much
  greater for ETGs than spirals.

\item The NUV--$r$ colors show that virtually all the ETGs lie close to
  the red sequence, but there is evidence from UV and optical colors,
  radio continuum observations, and literature searches that a $\sim
  20\%$ of the sample have had a recent star-forming epoch or have
  significant residual star formation.  However, the NUV colors
  have not been corrected for extinction or contamination from the old
  stellar population. Our ETGs are redder in NUV--$r$ with lower
  dust-to-stellar mass ratios than the ETGs detected in the H-ATLAS
  survey.

\item We show that the detection of cold dust, the ratio of
far-infrared to radio emission, and the gas-to-dust ratios all indicate
 that many ETGs contain a cool ISM in which stars are forming, similar
to that seen in spiral galaxies.

\item We find no evidence for a correlation between the FIR luminosity
  or dust mass of ETGs with the optical luminosity, suggesting that
  the main source of the dust in the galaxies detected by \Hersc~is
  not mass loss from evolved stars.  This, together with the large
  spread in the dust-to-stellar-mass ratio of ETGs, suggests a
  significant fraction of the dust in these galaxies is acquired
  externally via mergers or tidal interactions.

\item We use the results from the ATLAS$^{\rm 3D}$ survey to divide
  our sample into fast and slow rotators.  We show that the difference
  in dust properties between S0s and ellipticals is more obvious than
  the difference between slow and fast rotators. We suggest that this
  may be additional evidence that the dust in ETGs has been acquired
  by gravitational encounters and may not tell us much about
  the evolutionary state of the galaxy.

\item The low dust-to-stellar-mass ratios of S0s compared
  with early-type spirals cannot be explained by the larger
  bulge-to-disk ratios of S0s. The relative sizes of the
  dust sources in S0s are also smaller than seen in early-type spirals.
  These suggest that the
  disks in S0s contain much less dust (and presumably gas) than spiral
  disks with a similar size.  This effect is probably not
  being caused by current environmental processes adding to the weight
  of evidence that early-type spirals
  are not being transformed into S0s in significant numbers at the
  current epoch.

\item If the cool ISM in ETGs is acquired as the
result of a tidal interaction or merger, an upper limit on
the increase in the stellar mass of the ETG due to the
interaction is 1\% for the ellipticals and 10\% for the
S0s, suggesting that the interactions are not
significant galaxy-building events.

 \end{enumerate}

 Our sample of ETGs has lower dust-to-stellar-mass ratios and dust
 masses compared to previously published \Hersc~samples.  The results
 from our sample are important for interpreting chemical evolution
 models which model the production and destruction of dust in
 ellipticals as well as providing a low-redshift benchmark for
 understanding the evolution of dust and gas in high redshift surveys
 (e.g., Rowlands et al.\ 2012, Dunne et al.\ 2011). Finally, we note that to
 obtain a better understanding of some of these issues it will be
 important to follow-up our statistical study with detailed
 observational studies of individual ETGs, for example with ALMA.

 \acknowledgments We thank everyone involved with the
 \Hersc~Observatory.  PACS has been developed by a consortium of
 institutes led by MPE (Germany) and including UVIE (Austria); KU
 Leuven, CSL, IMEC (Belgium); CEA, LAM (France); MPIA (Germany); INAF-
 IFSI/OAA/OAP/OAT, LENS, SISSA (Italy); and IAC (Spain). This development
 has been supported by the funding agencies BMVIT (Austria),
 ESA-PRODEX (Belgium), CEA/CNES (France), DLR (Germany), ASI/INAF
 (Italy), and CICYT/MCYT (Spain).  SPIRE has been developed by a
 consortium of institutes led by Cardiff University (UK) and including
 University of Lethbridge (Canada); NAOC (China); CEA, OAMP (France); IFSI,
 University of Padua (Italy); IAC (Spain); Stockholm Observatory (Sweden);
 Imperial College London, RAL, UCL-MSSL, UKATC, University of Sussex (UK); and
 Caltech/JPL, IPAC, University of Colorado (USA). This development has been
 supported by national funding agencies: CSA (Canada); NAOC (China);
 CEA, CNES, CNRS (France); ASI (Italy); MCINN (Spain); Stockholm
 Observatory (Sweden); STFC and UKSA (UK); and NASA (USA). HIPE is a joint
 development by the \Hersc~Science Ground Segment Consortium,
 consisting of ESA, the NASA \Hersc~Science Center and the HIFI, PACS
 and SPIRE consortia.  This research made use of the NASA
 Extragalactic Database (NED: http://ned.ipac.caltech.edu/)
 which is operated by the Jet Propulsion Laboratory, California
 Institute of Technology, under contract with the National Aeronautics
 and Space Administration. The research leading to these results has 
received funding from the European Community's Seventh Framework Programme 
(/FP7/2007-2013/) under grant agreement no. 229517. This research made use of APLpy, an
 open-source plotting package for Python hosted at
   http://aplpy.github.com. We thank Edward Gomez for useful and
 informative discussions and H.L.G. acknowledges the support of Las
 Cumbres Observatory.

{\it Facilities:} \facility{{\it Herschel} (PACS and SPIRE)},
\facility{{\it Spitzer} (MIPS)}, \facility{{\it IRAS}}
  
\appendix

\section{A.  Notes on the Ellipticals detected by \textit{Herschel}}
\label{sec:desc}
\noindent {\it HRS3, NGC3226}. A semicircular lane of dust in NGC3226
is seen as absorption in the optical (Martel et al.\ 2004) with dusty
strands extending north--south.  The source is known to be interacting
with the nearby spiral NGC3227 (the bright source to the south of the
elliptical---Figure~\ref{fig:spireegals}), which has clear tidal gas
trails (Mundell et al.\ 2004). The elliptical appears slightly extended
in the 250\,$\mu$m image but more compact than in the
optical image (Figure~\ref{fig:spireegals}).  NGC3226 is part of the Leo Cloud.\\

\noindent {\it HRS138, NGC4374 (M84)}. M84 has a radio bright core and
radio jets. \textit{Hubble Space Telescope} (HST) images show dust lanes across the center of the galaxy
(Bower et al. 1997), which are approximately perpendicular to the
radio jet.  The galaxy was detected at 850\,$\mu$m with SCUBA by
Leeuw et al.\ (2000). They concluded that most of
the 850\,$\mu$m flux was synchrotron emission, but that the emission
at shorter wavelengths detected by \textit{IRAS} was produced by $\sim
10^{5}\,M_{\sun}$ of dust at a temperature of 35 K (see also
Boselli et al\ 2010a).  The 250\,$\mu$m source is unresolved and is
coincident with the
radio core (Figure\ref{fig:spireegals}). NGC4374 is a member of the Virgo Cluster.\\

\noindent {\it HRS150, NGC4406 (M86)}.
M86 is a well-known IR-bright giant elliptical.  Two dust features
were detected with {\it IRAS} and originally attributed to dust
stripped from M86 due to its motion through the cluster.  The
discovery of atomic gas offset from the center of M86 and decoupled
from its stellar disk supports a tidal interaction (Li \& van Gorkom
2001), confirmed when Kenney et al.\ (2008) detected strong $\rm
H{\alpha}$ features extending from M86 to the nearby spiral NGC4438.
The distribution and velocity of the ionized gas provide clear
evidence for tidal interaction between these two giants.  \Hersc~SPIRE
observations of M86 and NGC4438 showed that the dust emission is
spatially correlated with the ionized gas between the two galaxies
(Gomez et al.\ 2010; Cortese et al.\ 2010a) implying that the dust is
material stripped from the nearby spiral.  In
Figure~\ref{fig:spireegals}, the appearance of the
dust emission in M86 is clearly different to the other ellipticals in our sample, with faint filamentary features seen within $D(25)$. NGC4406 is a member of the Virgo Cluster.\\

\noindent {\it HRS183, NGC4486 (M87)}. M87 is the brightest galaxy in
the Virgo Cluster and well known for the jet extending from the
nucleus seen at radio, optical, and X-ray wavelengths (Junor
et al. 1999).  The presence of dust was inferred from \textit{HST} g -- z
color images (Ferrarese et al. 2006) and the strong FIR emission seen
with {\it Spitzer} (Perlman et al. 2007).  The latter result might be
explained by synchrotron emission from the central radio
source. Although the optical images show dust is present,
\Hersc~observations of M87 with PACS and SPIRE as part of \hevics~
(Baes et al.\ 2010; Boselli et al.\ 2010a) found no evidence of a dust
component in excess of the synchrotron emission in
the FIR and submm.  Baes et al. place an upper limit on the dust mass of $10^{5}\,M_{\odot}$.  In the \Hersc~image, M87 is a bright, extended source.  \\

\noindent {\it HRS186, NGC4494}. NGC4494 is often described as an
`ordinary elliptical' (Capaccioli et al. 1992). O'Sullivan
\& Ponman\ (2004) found that its X-ray luminosity was two order of
magnitudes lower than expected for its optical luminosity and there
are signs of a small dust disk in absorption (Tran et al. 2001). X-ray
faint galaxies such as NGC4494 may arise due to losing their hot X-ray
gas in outflows. Further evidence of interactions with nearby galaxies
or mergers is the low metallicity estimated from the X-ray gas
($<0.1Z_{\sun}$, O'Sullivan \& Ponman\ 2004), and this may indicate
dilution of interstellar material via an infall of unenriched (cold)
material. The galaxy is slightly
extended at 250\,$\mu$m.  NGC4494 is a member of the Coma I Cloud.\\

\noindent {\it HRS241, NGC4636}. NGC4636 has dust features
and an unusual X-ray
morphology (Temi et al.\ 2007b). The origin of this
morphology is thought to be recent outbursts from the central
AGN (Jones et al. 2002).  Temi et al. (2003)
used {\it ISO} observations to show that the dust mass for NGC4636 is
far in excess of that expected from stellar mass loss 
and proposed that the dust was accreted in a very recent
merger with a
dusty, gas-rich galaxy (similar to M86, Gomez et al.\ 2010).  NGC4636
is part of the Virgo Cluster.\\

\noindent {\it HRS258, NGC4697}. NGC4697 is an X-ray-faint galaxy
given its optical luminosity, which may be a result of severe loss of
interstellar gas via stripping or outflows (Sarazin et al.
2001). As in NGC4494, the metallicity of the gas is low
($<0.07Z_{\sun}$), pointing toward dilution of the interstellar
medium with unenriched cold gas.  The source is slightly extended at
250\,$\mu$m. NGC4697 is
in the outer regions
of the Virgo Cluster.\\

\section{B.  FIR emission versus optical and X-ray}
\label{sec:laws}

We investigated whether there is a correlation between $L_X$ and $L_B$
using the statistics in the ASURV package for dealing with censored
data. We find evidence for a correlation for the elliptical galaxies
($P=99.3\%$\footnote{\label{fn:kend}Using the Kendall $\tau$
  test, a statistic used to measure the association between $X$ and
  $Y$; this test is appropriate for small samples with upper limits
  where the underlying distributions of $X$ and $Y$ are not
  known.}). The relationship between the two parameters for
ellipticals\footnote{\label{fn:buck}Using the Buckley-James method, a
  standard linear regression estimator.  This test requires that the
  censoring distribution about the fitted line is random and does not
  require the residuals to be Gaussian as other regression
  methods do.}  is described by the following relationship with
deviation from the regression $\sigma_{\rm r}=0.43$:
\begin{equation} 
{\rm log} L_B = (2.95 \pm 1.03){\rm log}L_X +10.79
\label{eq:optlx}
\end{equation}
This relationship agrees with other IR samples dominated by massive
ETGs (see also Brown \& Bregman 1998), whereas X-ray studies indicative
of the entire population find are described by $L_X \propto
L_B^{2.3}$ (O'Sullivan et al.\ 2001; Temi et al.\ 2004).

If the dust produced in ETGs is due to stellar mass loss, the dust
mass would be roughly proportional to the mass of the stars, and hence
to $L_B$.  If the dust is then widely dispersed and heated by the
starlight as well as electron collisions in the hot, X-ray emitting
gas, the FIR emission will also depend on the density of the gas or on
the stellar density, which are both roughly proportional to
$L_{B}^{1/2}$.  Therefore, we would expect $L_{\rm FIR} \propto
L_{B}^{1.4-1.6}$. Bregman et al.\ (1998) indeed found this
relationship was true for {\it IRAS}-detected ETGs, though Temi et al.\ 2004
found no correlation for their {\it ISO}-detected sources.  In the
\Hersc~sample, we see that the FIR luminosity of \SO galaxies does
appear to increase with the optical luminosity, although this is not a statistical
correlation, with
Spearman rank\footnote{\label{fn:rho}Using the Spearman rho
  correlation test; this statistic is more appropriate than the
  Kendall $\tau$ used for the ellipticals since the S0s are a
  larger sample with more detections.  The correlation determines how
  well the relationship between $X$ and $Y$ can be described by a
  monotonic function.  This test is less sensitive to outliers than
  the similar Pearson correlation test.  We tested that the Kendall's
  $\tau$ and Spearman correlation tests give similar values.}
coefficient $r_S = 0.36$ and probability that a correlation is present
of $P=97\%$.  The \SO data can be fit by the following
equation\reffnmark{fn:buck} with $\sigma_{\rm
  r}=0.39$:
\begin{equation}
\noindent
{\rm log} L_{\rm FIR} ({\rm S0})  = (1.22 \pm 0.42){\rm
  log}\left(L_B\right)-3.53.
\label{eq:firso}
\end{equation}

The FIR luminosity of the elliptical galaxies increases with $L_B$
but again, we find that the correlation is not statistically significant, with $L_B$ ($\tau=0.36$ and $P = 91\%$\reffnmark{fn:kend}), the results of the
regression analysis produces the following relationship (with $\sigma_{\rm
  r}=0.33$\reffnmark{fn:buck}):

\begin{equation} 
\noindent {\rm log} L_{\rm FIR} ({\rm E})  = (1.00\pm 0.77){\rm log}\left(L_B\right)-2.44.
\label{eq:firel}
\end{equation}

The correlations in Equations~(\ref{eq:firso}) and (\ref{eq:firel}) are plotted in
Figure~\ref{fig:firtrends}. The lack of any strong correlation between
$L_{\rm FIR}$ and $L_B$ does not provide any evidence for the hypothesis
that the dust responsible for the FIR--submm emission is produced by
stellar mass loss, although the ETGs do at least fall in roughly the
right place in the figure. 

In Figure~\ref{fig:firtrends} (lower panel), we also compare the dust mass
with optical and X-ray luminosities but find no evidence for a
correlation due to the small numbers in the samples with both
\Hersc~detections and X-ray fluxes in the literature.

There is also no evidence for a significant correlation between dust mass and
$L_B$ for S0 galaxies ($r_S = 0.27$ and $P=90\%$), with a best-fit
relationship ($\sigma_{\rm r} = 0.58$\reffnmark{fn:rho}):
\begin{equation} 
\noindent {\rm log}M_d ({\rm S0}) = (1.11 \pm 0.62){\rm log} L_B -5.08
\label{eq:mdso}
\end{equation}

The dust mass and optical luminosity for the ellipticals are also not correlated ($\tau=0.30$ and
$P=90\%$\reffnmark{fn:kend}), with a best-fit relationship ($\sigma_{\rm
  r}=0.87$)\reffnmark{fn:buck}:
\begin{equation}
\noindent
{\rm log}M_d ({\rm E})= (1.89\pm 1.99){\rm log} L_B -14.40
\label{eq:mde}
\end{equation}

The remaining panel in Figure\ref{fig:firtrends} shows dust mass rather
than FIR luminosity plotted against X-ray luminosity, with both
quantities being normalized by $L_B$; the correlation is weaker when
FIR luminosity is converted into dust mass.

\end{document}